\documentclass[usenatbib]{mnras}
\pdfoutput=1

\usepackage{newtxtext,newtxmath}

\usepackage[T1]{fontenc}
\usepackage{ae,aecompl}

\usepackage{newtxtext,newtxmath}
\usepackage{graphicx}
\usepackage{color}
\usepackage{soul}
\usepackage{units}
\usepackage{url}
\usepackage{bm}         
\usepackage{times}
\usepackage{dcolumn}
\usepackage{bm}
\usepackage{epsf}
\usepackage{subcaption}
\usepackage[normalem]{ulem}
\usepackage{threeparttable}

\newcommand{\be}{\begin{equation}}
\newcommand{\ee}{\end{equation}}
\newcommand{\bs}{\boldsymbol}
\newcommand{\beq}{\begin{equation}}
\newcommand{\eeq}{\end{equation}}
\newcommand{\beqn}{\begin{eqnarray}}
\newcommand{\eeqn}{\end{eqnarray}}

\usepackage{color}

\title[Radio Burst Precursors of Neutron Star Mergers]{Shock-powered radio precursors of neutron star mergers from accelerating relativistic binary winds}

\author[N.~Sridhar, J.~Zrake, B.~D.~Metzger, L.~Sironi \& D.~Giannios]{
Navin~Sridhar$^{1}$, 
Jonathan~Zrake$^{1,2}$,
Brian~D.~Metzger$^{1,3}$,
Lorenzo~Sironi$^{1}$,
Dimitrios~Giannios$^{4}$
\\
$^{1}$Columbia Astrophysics Laboratory, Columbia University, 550 W 120th St, New York, NY 10027, USA\\
$^{2}$Department of Physics and Astronomy, Clemson University, SC 29634-0978, USA\\
$^{3}$Center for Computational Astrophysics, Flatiron Institute, New York, NY 10010, USA\\
$^{4}$Department of Physics and Astronomy, Purdue University, 525 Northwestern Avenue, West Lafayette, IN 47907, USA
}

\begin{document}

\label{firstpage}
\maketitle

\begin{abstract}
During the final stages of a compact object merger, if at least one of the binary components is a magnetized neutron star (NS), then its orbital motion substantially expands the NS's open magnetic flux---and hence increases its wind luminosity---relative to that of an isolated pulsar.  As the binary orbit shrinks due to gravitational radiation, the power and speed of this binary-induced inspiral wind may (depending on pair loading) secularly increase, leading to self-interaction and internal shocks in the outflow beyond the binary orbit.  The magnetized forward shock can generate coherent radio emission via the synchrotron maser process, resulting in an observable radio precursor to binary NS merger.  We perform 1D relativistic hydrodynamical simulations of shock interaction in the accelerating binary NS wind, assuming that the inspiral wind efficiently converts its Poynting flux into bulk kinetic energy prior to the shock radius. This is combined with the shock maser spectrum from particle-in-cell simulations, to generate synthetic radio light curves. The precursor burst with a fluence of $\sim1$ Jy$\cdot$ms at $\sim$GHz frequencies lasts $\sim 1-500$ ms following the merger for a source at $\sim3$ Gpc ($B_{\rm d}/10^{12}$ G)$^{8/9}$, where $B_{\rm d}$ is the dipole field strength of the more strongly-magnetized star. Given an outflow geometry concentrated along the binary equatorial plane, the signal may be preferentially observable for high-inclination systems, i.e. those least likely to produce a detectable gamma-ray burst. 
\end{abstract}

\begin{keywords}
Transients: neutron star mergers, fast radio bursts --- Software: simulations --- Physical data and processes: shock waves, radiation dynamics, plasmas
\end{keywords}

\section{Introduction} \label{sec:introduction}
Mergers of compact object binaries containing neutron stars (NS) and stellar mass black holes (BH) are the primary source class for current gravitational wave (GW) observatories, including Laser Interferometer Gravitational-Wave Observatory \citep[LIGO;][]{Harry+10}, Virgo \citep{Acernese+15}, and Kamioka Gravitational Wave Detector \citep[KAGRA;][]{Somiya12}.  One of the goals of multi-messeger astrophysics is the discovery of electromagnetic (EM) emission in coincidence with these GW sources (e.g.,~\citealt{Bartos+13,Fernandez&Metzger16}), which in the case of NS-NS and BH-NS mergers encodes key information on a number of topics, including the formation and properties of relativistic jets (e.g.,~\citealt{Zrake+18,Kathirgamaraju+19}), the equation of state of nuclear density matter (e.g.,~\citealt{Margalit&Metzger19}), and the origin of the heaviest elements (e.g.,~\citealt{Metzger+10}). 

The detection of the first NS-NS merger through GW emission, GW170817 \citep{LIGO+17DISCOVERY}, and the subsequent discovery of EM emission across the spectrum from radio to gamma-rays \citep{LIGO+17CAPSTONE}, was a watershed event.  It revealed the strongest evidence to date that NS mergers can produce relativistic collimated outflows consistent with those responsible for cosmological short-duration gamma-ray bursts (e.g.,~\citealt{Margutti+18,Beniamini+19}).  The origin of the gamma-ray emission \citep{LIGO+17FERMI}, which was detected despite our viewing the event at an angle $\gtrsim 20^{\circ}$ outside the core of the jet, remains a subject of ongoing debate \citep{Goldstein+17,Kasliwal+17,Nakar+18,Metzger+18,Beloborodov+18}.  Unfortunately, given the low luminosities of the gamma-ray and afterglow emission from events like GW170817, most future GW-detected mergers at greater distances will not produce detectable jetted emission \citep{Metzger&Berger12,Beniamini+19}.  No additional gamma-ray coincidences were reported during LIGO/Virgo's third observing run, despite the discovery of at least one additional NS-NS merger, GW190425 \citep{LIGO+20a}.  Consistent with this, an analysis by \citet{Dichiara+20} which cross-correlates short GRB sky positions with nearby galaxies, finds an all-sky rate of only $0.5-3$ short GRBs per year within the 200 Mpc horizon distance of Advanced LIGO at design sensitivity.

The kilonova and non-thermal emission which accompanied GW170817 chiefly result from processes---large mass ejection and jet formation---which take place {\it after} the merger is complete.  Compared to the post-merger phase, less theoretical work been dedicated to EM emission during the late inspiral phase prior to coalescence.  Nevertheless, the discovery and characterization of such {\it precursor} emission--observed prior to other EM counterparts (and potentially even before the end of the GW chirp)--would be of great importance.  It could provide unique information on the state of the binary and its constituent stars prior to their destruction \citep[e.g.,][]{Ramirez-Ruiz+19}, as well as offer a potential ``pre-warning'' signal to enable additional prompt EM follow-up (e.g.,~\citealt{Schnittman+18}).  Furthermore, in some cases, such as massive BH-NS binaries in which the NS is swallowed whole before being tidally disrupted \citep{McWilliams&Levin11} or massive NS-NS systems which undergo prompt collapse to a BH \citep{Paschalidis&Ruiz19}, the post-merger counterparts may be dim or non-existent; in such cases a precursor signal may be the best way to detect and localize this subset of GW events.

If at least one NS is magnetized, then the orbital motion of the companion NS or BH through its dipole magnetic field induces a strong voltage and current along the magnetic field lines connecting the two objects \citep{Lipunov&Panchenko96,Vietri96,Hansen&Lyutikov01,McWilliams&Levin11,Piro12,Lai12,DOrazio&Levin13,Palenzuela+13,Ponce+14, DOrazio+16,Wang+18,Paschalidis&Ruiz19,Lyutikov19,Crinquand_19,Gourgouliatos+19}, in analogy with the unipolar inductor model for the Jupiter-Io system \citep{Goldreich&LyndenBell69}.  A separate source of electromagnetic power arises due to the magnetic dipole radiation generated by the orbital acceleration of the magnetized primary, which is independent of the conducting properties of the companion \citep{Ioka&Taniguchi00,Carrasco&Shibata20}.  By tapping into the orbital energy of the binary, these interactions can in principle power EM emission that increases in strength as the orbital velocity increases and the binary separation decreases approaching merger. 

Resistive MHD \citep{Palenzuela+13} and force-free \citep{Most&Philippov20,Carrasco&Shibata20} numerical simulations show that the magnetic field geometry of the merging binary system can be complex and time-dependent, with the magnetic field lines connecting the neutron stars being periodically torn open to infinity.  Dissipation may also occur in spiral current sheets that develop beyond the binary orbit \citep{Carrasco&Shibata20}.  Averaged over many orbits, a fraction of the power dissipated in the circuit is dissipated as “heat” (particle acceleration) in the magnetosphere, while the remainder is carried to large distances by a Poynting flux along the open field lines.  
 
\citet{Metzger&Zivancev16} argue that if particle acceleration in the magnetosphere near the binary gives rise to non-thermal (e.g., synchrotron) emission, this radiation is unlikely to escape to the distant observer, because such a high density of > MeV photons within such a compact volume will result in copious electron-positron pair production via $\gamma-\gamma$ annihilation \citep{Usov92}.  They hypothesized that the final stages of the inspiral may generate a ``pair fireball'', similar to early models of GRB outflows \citep{Paczynski86,Goodman86}.  However, even making generous assumptions about the fraction of the power of magnetospheric interaction emerging in the MeV gamma-ray band, \citet{Metzger&Zivancev16} found that precursor gamma-ray emission is only detectable by {\it Fermi} GBM to a distance of $\lesssim 10$ Mpc.  

Recently, \cite{Beloborodov_20} put forth the prospects of observing X-ray precursors to NS mergers via quasi-periodic flaring---akin to magnetar bursts---of magnetized NS mergers. However, a coherent radio emission, if produced, provides a more promising precursor signal because of the comparatively greater sensitivity of radio telescopes (e.g.,~\citealt{Hansen&Lyutikov01}). Radio emission which emerges as a short  ($\sim$ millisecond duration) burst comparable to the final few orbits of the binary could in principle be observable as a cosmological fast radio burst (FRB) with properties similar to those recently discovered \citep{Lorimer+07,Thornton+13}.  Energy released by magnetospheric interaction during the very final stages of a BH-NS or NS-NS merger inspiral has long been speculated to give rise to a single (non-repeating) FRB-like signal \citep{Usov&Katz00,Hansen&Lyutikov01,Moortgat&Kuijpers03,Totani13,Zhang13,Mingarelli+15,Metzger&Zivancev16,Wang+16, Bhattacharyya_17,Zhang_2019}.  However, given the high-inferred volumetric rate of FRBs relative to NS-NS mergers, such ``one off'' bursts can account for at most only a small fraction $\sim 1\%$ of the total FRB population.  Furthermore, many FRB sources have now been observed to repeat \citep{Spitler+16,CHIME+19}, and the population of (thus far) non-repeaters may be consistent with those currently observed to recur (e.g.,~\citealt{Lu+20}).\footnote{Scenarios for generating recurring FRBs due to magnetospheric interaction from wide NS-NS binaries, on much longer timescales of decades to centuries before coalescence, are discussed by \citet{Gourgouliatos+19}, \citet{Zhang20}.}  

\begin{figure}
\centering
\includegraphics[width=0.5\textwidth]{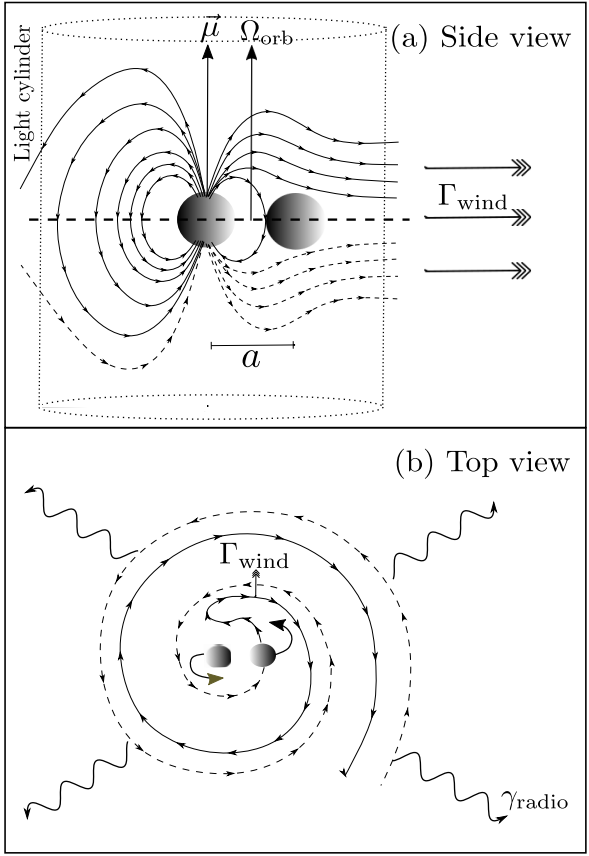}
\caption{A schematic diagram of the binary merger-induced accelerating pulsar wind, and the associated radio emission ($\gamma_{\rm radio}$). Top panel (a) shows the system from an equatorial plane (side view), where the magnetic field lines of the strongly magnetized pulsar (left sphere) are seen to be opened to infinity by the orbital motion of the primary or by interaction with the companion star (right sphere; NS or a BH) at a binary separation radius $a$, well within the light cylinder of the rotating magnetically-dominated pulsar. The binary wind is ejected opposite to the strongly magnetized star (at a given time), whose Lorentz factor ($\Gamma_{\rm wind}$) increases with decreasing binary separation, $a$. The binary wind, over several orbits traces a spiral pattern beyond $a$, that can be noticed from a top view of the system (panel b). It is the interaction of the smallest yet fastest spiral at the end of the binary inspiral phase with the earlier emanated larger yet slower spiral wind, that causes the formation of a forward shock at radius $\gg a$, and the associated coherent radio emission via synchrotron maser process. Note that the inspiral wind, and the shock emission are beamed along the orbital equatorial plane, and the direction of both the NS dipole moment ($\vec{\mu}$) and the orbital angular velocity ($\Omega_{\rm orb}$) are roughly normal to the plane of emission.}
\label{fig:cartoon}
\end{figure}

Even if the vast majority of the present FRB sample do not arise from NS-NS or NS-BH mergers, that does not exclude mergers from generating detectable radio emission.  Most FRB surveys cover only a fraction of the sky with a limited duty cycle.  By contrast, a targeted follow-up search with wide field of view observatories at low radio frequencies (few 100 MHz) is possible for mergers detected through their near-concomitant GW emissions because the latter offer a precise time window and a relatively precise sky position of $\lesssim 10-100$ deg$^{2}$ (when the merger is detected by three or more interferometers).  \citet{Callister+19} searched for transient radio emission within approximately one hour of the BH-BH merger event GW170104, offering a proof-of-concept of this technique.  Upper limits have also been placed on coherent radio signals from short GRBs (e.g.,~ \citealt{Palaniswamy+14,Kaplan+15,Anderson+18,Rowlinson&Anderson19,Gourdji+20,Rowlinson+20}), however the distances to typical short GRBs are much greater than those of the closest GW-detected events.  Furthermore, as we shall discuss, the beaming of the precursor radio emission may also differ from that of a GRB jet, the latter probably focused along the angular momentum axis of the binary.

Most models of merger radio precursors invoke particle acceleration or reconnection processes which are local to the inner magnetosphere, i.e. on spatial scales comparable to the binary separation or light cylinder. radius.  Magnetic reconnection can occur sporadically following the build-up of twist in the magnetic flux tubes connecting the stars (e.g., due to stellar spins or misaligned magnetospheres; \citealt{Most&Philippov20}) or in a spiral current sheet outside of the binary orbit \citep{Carrasco&Shibata20}.\footnote{In addition to magnetosphere interaction, tidal resonant excitation of modes in the NS crust provides an additional way to tap into the orbital energy of the merging binary \citep{Tsang+12,Tsang13}. If such modes shatter the crust, then $\sim 10^{46}-10^{47}$ erg may be released seconds prior to merger, some fraction of which will also couple to the magnetosphere in the form of outwardly-propagating Alfv\'en waves, which could generate coherent radio emission upon decaying at large radii \citep{Kumar&Bosnjak20,Lu+20,Yuan_2020}.}  Such reconnection events were argued to give rise to coherent radio emission, through either merging plasmoids beyond the light cylinder (\citealt{Lyubarsky19,Philippov+19}) or further out of the binary via a synchrotron maser emission as outgoing blobs collide with ambient plasma (e.g.,~\citealt{Most&Philippov20}).  However, given the highly dissipative environment and resulting high compactness in the inner magnetosphere, this region may become heavily loaded in pairs during the final stages of the inspiral \citep{Metzger&Zivancev16}; this could prevent the escape of radio waves, due to induced Compton scattering \citep{Lyubarsky08} or other plasma processes, from the immediate vicinity of the binary orbit. 

In this paper, we instead consider a new precursor emission mechanism that occurs on much larger radial scales well outside the binary orbit (Fig.~\ref{fig:cartoon}).  In the potentially common scenario that one NS is substantially more strongly magnetized than the other (e.g., if one pulsar is young with a strong dipole magnetic field $B_{\rm 1} \gtrsim 10^{12}$ G and the other ``recycled'' with a much weaker field $B_{\rm 2} \lesssim 10^{10}$ G; see the double pulsar; \citealt{Kramer&Stairs08}), the magnetic flux of the more strongly magnetized star which intersects the orbit of the companion will be open to infinity (e.g.,~\citealt{Palenzuela+13,Carrasco&Shibata20}).  In a time-averaged sense, this enhanced open flux will generate an equatorial outflow in the orbital plane with a power and bulk Lorentz factor which can (for a reasonable set of assumptions) increase in a secular manner approaching merger.  This accelerating binary-induced wind---akin to a pulsar ``spinning up'' instead of the usual magnetic braking---will generate relativistic internal shocks in the binary plane on scales well outside the binary orbit as the late portion of the wind interacts with the earlier wind.  These shocks will furthermore be magnetized, giving rise to synchrotron maser emission in the radio band (e.g.,~\citealt{Plotnikov&Sironi19,Metzger+19}) observable as a brief FRB-like signal.

This paper is organized as follows.  In \S\ref{sec:wind} we describe the merger-induced binary pulsar wind and present analytic estimates of the properties of the internal shocks.  In \S\ref{sec:simulations} we describe our numerical simulation setup of the shock interaction for various binary wind models.  In \S\ref{sec:results} we present the hydrodynamical results of shock interaction, and the associated FRB.  In \S\ref{sec:discussion} we discuss observational prospects for detecting the signal and the caveats.  In \S\ref{sec:conclusion} we summarize our conclusions.

\section{Binary-Induced Accelerating Pulsar Wind}
\label{sec:wind}
\subsection{Time-Dependent Wind Properties}
\label{sec:windproperties}

The spin-down Poynting luminosity of an isolated rotating magnetized neutron star can be expressed as (e.g.,~\citealt{Bucciantini+06})
\be
\dot{E}=\left(\frac{f_{\Phi}}{\tilde{f}_{\Phi}}\right)^{2}\frac{\mu^{2}\Omega^{4}}{c^{3}},
\label{eq:edot}
\ee 
where $\Omega$ is the angular rotation rate, $\mu \equiv B_{\rm d}R_{\rm ns}^{3}$ is the dipole moment of the neutron star of radius $R_{\rm ns}$, and $B_{\rm d}$ is the surface dipole field strength.  Here $f_{\Phi}$ is the fraction of the magnetic flux threading the NS surface which is open to infinity, normalized to its value $\tilde{f}_{\Phi} \approx R_{\rm ns}/2R_{\rm L}$ for an isolated dipole wind with aligned magnetic and rotation axes in the limit $R_{\rm L} \gg R_{\rm ns}$, where $R_{\rm L} \equiv c/\Omega$ is the light cylinder radius.

One effect of the close orbiting binary companion of radius $R_{\rm c} \sim R_{\rm ns}$ in a binary of semi-major axis $a \ll R_{\rm L}$ may be to open additional closed field lines (e.g.,~\citealt{Palenzuela+13}) for finite ranges of companion's resistivity \citep{Lai12}, such that the amount of magnetic flux is enhanced over that of an isolated pulsar by an amount :
\be
\left(\frac{f_{\Phi}}{\tilde{f}_{\Phi}}\right)_{\rm bin} \approx \frac{R_{\rm c}c}{4\pi \Omega_{\rm orb} a^{2}}, \label{eq:openfrac}
\ee
Compared to the case of an isolated pulsar, the open flux fraction has been reduced by a factor of $\sim 2R_{\rm c}/(4\pi a)$ to account for the azimuthal angle subtended by the binary companion, and increased by a factor of $\approx R_{\rm L}/a=c/(2\Omega_{\rm orb}a)$ because field lines crossing the equatorial plane exterior to the binary separation $a$ (instead of the light cylinder radius) are now open, where $\Omega_{\rm orb}=(GM_{\rm tot}/a^{3})^{1/2}$ is the orbital frequency of the binary of total mass $M_{\rm tot}$ (see \citealt{Fernandez&Metzger16} for further discussion).  This corresponds to the ``U/u'' case simulated by \citet{Palenzuela+13}, which as already mentioned is the most likely magnetic field configuration characterizing merging NS binary progenitors. A similar configuration should characterize BH-NS mergers because the BH is not magnetized.

Substituting Eq. (\ref{eq:openfrac}) into (\ref{eq:edot}), the power of the binary wind,
\begin{eqnarray}
\dot{E} &=& \left(\frac{f_{\Phi}}{\tilde{f}_{\Phi}}\right)_{\rm bin}^{2}\frac{\mu^{2}\Omega_{\rm orb}^{4}}{c^{3}} \approx \frac{\mu^{2}GM_{\rm tot}R_{\rm c}^{2}}{16 \pi c a^{7}} \nonumber \\
 &\approx& 2\times 10^{42}{\rm erg\,s^{-1}}\left(\frac{B_{\rm d}}{10^{12}{\rm G}}\right)^{2}\left(\frac{a}{2R_{\rm ns}}\right)^{-7}.
\label{eq:Edot1}
\end{eqnarray}
is an extremely sensitive function of the binary separation, $\dot{E} \propto a^{-7}$, where in the final line we have assumed an equal mass binary $M_{\rm tot}=2M_{\rm ns}$ with $M_{\rm ns}=1.4M_{\odot}$, and $R_{\rm ns}=12$ km.  \citet{Lai12} derive a similar expression similar to Eq. (\ref{eq:Edot1}) as the maximum power which can be dissipated in the closed ``binary circuit'' before the generation of a strong toroidal magnetic field from the poloidal current leads to magnetic field line inflation and reconnection.  Here we hypothesize that an order-unity fraction of this maximum power emerges as a quasi-steady magnetized wind from the binary equatorial plane.  

Even if the companion does not itself open the magnetic flux bundles, a separate source of electromagnetic power arises due to the magnetic dipole radiation generated by the orbital acceleration of the magnetized primary, which is independent of the properties of the companion.  This gives a contribution to the power of \citep{Ioka&Taniguchi00,Carrasco&Shibata20}
\begin{eqnarray}
\dot{E} &=& \frac{4\mu^{2}a^{2}}{15 c^{5}}\Omega_{\rm orb}^6=\frac{4}{15}\frac{\mu^{2}}{c^{5}}\frac{(GM_{\rm tot})^{3}}{a^{7}} \nonumber \\
&\approx& 4\times 10^{42}{\rm erg\,s^{-1}}\left(\frac{B_{\rm d}}{10^{12}{\rm G}}\right)^{2}\left(\frac{a}{2R_{\rm ns}}\right)^{-7},
\label{eq:Edot2}
\end{eqnarray}
similar in its normalization and scaling with binary separation as Eq. \ref{eq:Edot1}.

The power emitted by the ``binary wind'' (Eqs.~\ref{eq:Edot1},\ref{eq:Edot2}) eclipses that of the single more highly-magnetized pulsar ($\dot{E}$ from Eq.~\ref{eq:edot} with $f_{\Phi}=\tilde{f}_{\Phi}$ and where now $\Omega=2\pi/P$ for the pulsar spin-period $P$) for semi-major axes smaller than a critical value,
\be
a_{\rm bin} \approx 482\,{\rm km}\left(\frac{P}{\rm 0.1\,s}\right)^{1/7}.
\label{eq:abin}
\ee

The binary is driven to coalescence by gravitational waves, such that its semi-major axis shrinks with time $t$ according to,
\be
a=a_0(1 - t/t_{\rm m,0})^{1/4},
\ee
where $a(t=0)\equiv a_0$ is the initial binary separation, $t_{\rm m,0} \equiv t_{\rm m}(a_0)$ is the time to merge starting from $a_0$, and
\be
t_{\rm m}=\frac{5}{512}\frac{c^{5}a^{4}}{G^{3}M_{\rm ns}^{3}} \approx 1.2\times 10^{-3}{\rm s}\left(\frac{a}{24\,{\rm km}}\right)^{4},
\label{eq:tm}
\ee
where in the first line we have assumed equal bodies of mass $M_{\rm ns}=M_{\rm tot}/2$ and in the second line we have taken $M_{\rm ns}=1.4M_{\odot}$.  It will prove convenient to introduce the time until merger as an alternative variable,
\be\Delta t \equiv t_{\rm m,0}-t, \ee
in terms of which
\be
a=a_0\left(\frac{\Delta t}{t_{\rm m,0}}\right)^{1/4}
\ee
Thus, from Eq. (\ref{eq:Edot1}) we have $\dot{E} \propto a^{-7} \propto \Delta t^{-7/4}$ for times 
\be
\Delta t \lesssim t_{\rm m}(a_{\rm bin}) \approx 195\,{\rm s}\left(\frac{P}{\rm 0.1\,s}\right)^{4/7}.
\ee

The open magnetic flux lines will carry rest-mass as well as energy, likely in the form of electron/positron pairs.  
For ordinary pulsar winds, the wind mass-loss rate, $\dot{M}$, is typically thought to scale with the Goldreich-Julian (GJ) value,
\be
\dot{M}_{\rm GJ}=\mu_{\pm}m_{\rm e} \frac{I}{e} \approx \frac{\mu_{\pm} m_{\rm e}}e\frac{2\mu \Omega^{2}}{c} \underset{\Omega \rightarrow \Omega_{\rm orb}}= \frac{\mu_{\pm}}{e\cdot c}\frac{2\mu GM_{\rm tot}m_{\rm e} }{a^{3}} 
\label{eq:MdotGJ}
\ee
where $\mu_{\pm}$ is the pair multiplicity, and
\be
I \equiv 4\pi R_{\rm L}^{2}c \eta_{\rm GJ}|_{\rm R_{\rm L}} \approx \frac{2\mu \Omega^{2}}{ c} ;\,\,\,\eta_{\rm GJ}|_{\rm R_{\rm L}}=\left(\frac{\Omega B}{2\pi c}\right)_{\rm R_{\rm L}}=\frac{\Omega\mu}{2\pi c R_{\rm L}^{3}} 
\label{eq:etaGJ}
\ee
are the GJ current and charge density at the light cylinder, respectively.  

The dynamics of the binary-induced winds considered here, as well as the external electromagnetic and photon environment, are drastically different than isolated pulsar winds (e.g.,~\citealt{Wada+20}).  For example, the outflow may be loaded by plasma generated via $\gamma-\gamma$ annihilation due to high energy radiation from reconnection or other forms of dissipation in the magnetosphere (e.g.,~\citealt{Metzger&Zivancev16}).  Nevertheless, performing the na\"ive exercise of replacing the neutron star rotation rate with that of the binary orbit in Eq. (\ref{eq:MdotGJ}), i.e.~$\Omega \rightarrow \Omega_{\rm orb}$, then one finds $\dot{M}_{\rm GJ} \propto a^{-3} \propto \Delta t^{-3/4}$.  

Throughout this paper we parameterize the dependence of the wind mass-loss rate on binary separation as
\be
\dot{M} \propto a^{-m} \propto \Delta t^{-m/4},
\label{eq:Mdotparam}
\ee
where $m$ is a free parameter which takes on the value $m=3$ if $\dot{M} \propto \dot{M}_{\rm GJ}$ but is highly uncertain as it depends on the mechanism of wind pair-loading.

We define a wind ``mass loading'' parameter $\eta$ according to
\be
\eta \equiv \frac{\dot{E}}{\dot{M}c^{2}} \propto a^{m-7}.
\label{eq:sigma1}
\ee
A wind that fully converts its energy to kinetic form can achieve a bulk Lorentz factor $\Gamma \simeq \eta$ by large radii, but in general $\Gamma \ll \eta$ in cases where the bulk remains in Poynting flux or is otherwise lost (e.g., as radiation).  Substituting Eqs.~(\ref{eq:Edot1}), (\ref{eq:MdotGJ}) into Eq.~(\ref{eq:sigma1}), we obtain
\be
\eta=\frac{1}{32\pi}\frac{1}{\mu_{\pm}}\frac{e B_{\rm d}}{m_{\rm e} c^{2}}\frac{R_{\rm ns}^{3}R_{\rm c}^{2}}{ a^{4}} \underset{R_{\rm c}=R_{\rm ns}}\approx \frac{4.5\times 10^{11}}{\mu_{\pm}}\frac{B_{\rm d}}{10^{12}\,{\rm G}}\left(\frac{a}{2R_{\rm ns}}\right)^{-4},
\label{eq:eta}
\ee
for the case with $\dot{M}=\dot{M}_{\rm GJ}$.  For young, high-voltage pulsars such as the Crab pulsar, observations and theory show that $\mu_{\pm} \lesssim 10^{5}$ (e.g.,~\citealt{Kennel&Coroniti84,Timokhin&Harding19}), which applied to our case near the end of the inspiral (e.g.,~$a \lesssim 4 R_{\rm ns}$) would imply $\eta \gtrsim 10^{5}$ for $B_{\rm d}=10^{12}$ G.  However, this may substantially under-estimate the mass-loading in the NS merger case (e.g.,~due to $\gamma-\gamma$ pair creation in the inner magnetosphere), in which case $\eta$ would be substantially less.  

For $\dot{M}=\dot{M}_{\rm GJ}$ and $\mu_{\pm}={\rm constant}$ ($m=3$), Eq. (\ref{eq:eta}) predicts that $\eta \propto a^{-4}$ will increase approaching the merger.  Even in the more general case $m \ne 3$ (Eq.~\ref{eq:sigma1}), $\eta$ increases with time as long as $m < 7$.  In this case the binary-driven wind will grow in both power and speed, inevitably giving rise to self-interaction and the generation of internal shocks within the wind.

The strength of the internal shocks depends not just on the evolution of $\eta$, but also on the four-velocity attained by the wind at large radii $r$ well outside its launching point near the binary orbit.  In models of axisymmetric pulsar winds \citep{Goldreich&Julian70}, the bulk Lorentz factor achieved near the fast magnetosonic surface outside the light cylinder is $\Gamma=\eta^{1/3}$, while the asymptotic magnetization of the wind (ratio of Poynting flux to kinetic energy flux) is $\sigma=\eta^{2/3} \gg \Gamma$, i.e. the acceleration is highly inefficient and most of the energy remains trapped in magnetic form.  However, observations of pulsar wind nebulae reveal that---at least by the location of the wind termination shock---the wind has nearly fully converted its magnetic energy into bulk kinetic energy, i.e. $\Gamma \simeq \eta$; $\sigma \ll 1$ \citep{Kennel&Coroniti84}.  Several ideas have been proposed to explain this anomalous acceleration (the ``$\sigma$ problem''; \citealt{Lyubarsky&Kirk01,Sironi&Spitkovsky11,Porth+13,Zrake&Arons_17,Cerutti_20}), but none are universally agreed upon.  A similar inference of rapid and efficient acceleration of pulsar winds has been invoked to account for the gamma-ray flares of the Crab pulsar \citep{Aharonian+12} and of gamma-ray binaries \citep{Khangulyan+12}.  

In this paper, we assume the binary-driven wind manages to solve its ``$\sigma$ problem'' and hence to achieve a bulk Lorentz factor close to the maximal value, i.e.~
\be \label{eq:Gamma_deltat}
\Gamma \simeq \eta \propto a^{m-7} \propto \Delta t^{(m-7)/4},
\ee
by the radii $r \lesssim r_{\rm sh}$ at which internal shocks occur.  The wind self-interaction will then be mediated by strong shocks of moderate magnetization, which can be approximately modeled as hydrodynamical (i.e. neglecting magnetic fields). 
If the winds instead were to maintain $\sigma \gg 1$ to large radii, then the resulting strongly magnetized shocks could be considerably weaker. However, as we show below (eq.~\ref{eq:Granot}), under a reasonable assumptions for acceleration of the wind, we naturally expect $\sigma \sim 1$ by the radius of internal shocks. Nonetheless, the shock can still be strong for $\sigma\gg1$ if the post-shock region is weakly magnetized---likely as a result of shock-driven magnetic reconnection, which can be faithfully probed with resistive magnetohydrodynamical simulations. We leave a detailed study of the magnetized shock case to future work, though we comment on how it would effect our predictions in $\S\ref{sec:discussion}$.

\subsection{Internal Shocks: Analytic Estimates}
\label{sec:shocks}

The accelerating binary wind will collide with the earlier, slower phases of the wind ejecta, generating magnetized ultra-relativistic shocks.  In this section we provide analytic estimates of this interaction and the resulting radio emission in particular, tractable limits, which we later test with numerical simulations (\S\ref{sec:results}).

Given the rapid rise of the outflow power $\dot{E} \propto a^{-7}$, most of the total wind energy, 
\begin{eqnarray}
E_{\rm f} &=& \int_{\rm a_{\rm f}}^{\infty} \dot{E}\left|\frac{dt}{da}\right| da=\frac{4}{3}\dot{E}_{\rm f}t_{\rm f} \approx \frac{5}{24\pi}\frac{\mu^{2}R_{\rm ns}^{2}c^{4}}{(GM)^{2}a_{\rm f}^{3}}  \nonumber \\
&\approx& 5\times 10^{41}{\rm erg}\,\,\left(\frac{B_{\rm d}}{10^{12}{\rm G}}\right)^{2}\left(\frac{a_{\rm f}}{2R_{\rm ns}}\right)^{-3},
\label{eq:Etot}
\end{eqnarray}
is released near the end of the inspiral, where $a_{\rm f}$ is the final separation (e.g., as defined by the physical collision between the NSs) and $\dot{E}_{\rm f} \equiv \dot{E}[t_{\rm f}]$.  This energy will emerge in a ``fast shell'' of width $t_{\rm f} \equiv t_{\rm m}[a_{\rm f}] \sim 1$ ms and Lorentz factor $\Gamma_{\rm f} \equiv \Gamma[t_{\rm f}]$.  The shell will initially expand ballistically into the pre-existing wind released by the earlier phases of the inspiral, before sweeping up enough mass to decelerate (depending on the time evolution of $\dot{M}$).  In what follows, it is convenient to express the energy-loss rate and Lorentz factor of the binary wind as
\be \dot{E}=\dot{E}_{\rm f}(\Delta t/t_{\rm f})^{-7/4}; \,\,\, \Gamma=\Gamma_{\rm f}(\Delta t/t_{\rm f})^{(m-7)/4}.
\label{eq:windproperties}
\ee
The fast shell released at the end of the merger initially expands with $\Gamma_{\rm f} \gg 1$, reaching a radius $r_{\rm sh} \simeq ct$ by a time $t$ after the merger.  Throughout the remainder of this section we ignore deceleration of the fast shell; the conditions for this ``freely coasting'' evolution to be valid are estimated below.  

A question immediately arises: Did the inspiral wind material met by the fast shell at radius $r_{\rm sh}$ expand freely prior to that point, or had it earlier collided/shocked with the even slower material ahead of it?  Let us assume the former and check the conditions under which this assumption is violated.  If the upstream inspiral wind has not decelerated appreciably since injection, then the slow wind released $\Delta t$ prior to the merger will reach $r_{\rm sh}$ by a time $t$ after the merger given by
\be
t \simeq 2\Gamma^{2}[\Delta t]\Delta t \approx 2 \Gamma_{\rm f}^{2}t_{\rm f}(\Delta t/t_{\rm f})^{(m-5)/2} ,
\label{eq:collision}
\ee
where the second equality makes use of Eq. (\ref{eq:windproperties}). This expression also assumes that $\Gamma[\Delta t]$ is much smaller than the fast shell $\Gamma_{\rm f}$ and hence is not accurate for small $\Delta t$.  For $m \gtrsim 5$, the wind met by the fast shell at $r_{\rm sh}$ is ``pristine'' and corresponds to that ejected at the time $\Delta t$ satisfying Eq. (\ref{eq:collision}).  By contrast, for $m \lesssim 5$ we obtain the seemingly unphysical result that $\Delta t$ increases with increasing $t$, which instead shows that the inspiral wind will hit earlier ejecta prior to interacting with the fast shell (and hence Eq.~\ref{eq:collision} cannot be used).

As discussed near the end of the previous section, we have assumed that the wind has largely converted its Poynting flux into bulk kinetic energy by the radius of internal shock interaction.  We now check whether such efficient acceleration is possible under a specific assumption regarding the acceleration mechanism of the wind.  Following \citet{Granot+11}, the Lorentz factor of the impulsively ejected fast shell will grow with radius from the wind launching point $r_0 \simeq 2a$ (near the fast magnetosonic point, where $\Gamma \simeq \eta^{1/3}$; \citealt{Goldreich&Julian70}) as
\be
\Gamma \simeq \eta^{1/3}\left(\frac{r}{r_0}\right)^{1/3},
\label{eq:Granot}
\ee
thus reaching its maximal value ($\Gamma \simeq \eta; \sigma \lesssim 1)$ by the radius
\be
r = r_{\Gamma} \gtrsim 2a \eta^{2} 
\ee
Assuming efficient acceleration, the radius of the shock generated by the fast shell corresponding to an observer time $\sim t_{\rm f}$ is given by (eq.~\ref{eq:collision}),
\be
r_{\rm sh} \simeq c t \simeq 2 \Gamma_{\rm f}^{2}ct_{\rm f}.
\ee
Thus,
\be
\frac{r_{\Gamma}}{r_{\rm sh}} \gtrsim \frac{a/c}{t_{\rm f}} \sim 0.1, 
\ee 
where in the final line we have taken typical values $a \sim 12$ km, $t_{\rm f} \sim 1$ ms.  Thus, we generally expect that the shell will have accelerated to nearly its maximal Lorentz factor ($\Gamma \sim \eta$) but also that the residual magnetization of the wind material will not be too low $\sigma \sim 1$.

We now check the condition for the fast shell to appreciably decelerate by sweeping up pristine (previously unshocked) inspiral wind.  The rest mass swept up by the fast shell as it meets material released at times $\lesssim \Delta t$ prior to the merger grows as $M_{\rm rest} \sim \dot{M}\Delta t \propto \Delta t^{(4-m)/4}$ and hence is dominated by the matter already present in the fast shell for $ m > 4$.  On the other hand, what matters for deceleration is the swept-up comoving {\it inertial} mass $M_{\rm th} \propto M_{\rm rest}\gamma_{\rm th}$ which accounts also for the thermal Lorentz factor $\gamma_{\rm th} \gg 1$ of relativistically hot gas behind the shock.  The latter should scale with the Lorentz factor of the forward shock $\Gamma_{\rm sh}$, i.e. $\gamma_{\rm th} \propto \Gamma_{\rm sh} \simeq \Gamma_{\rm f}/2\Gamma \propto \Delta t^{(7-m)/4}$, where we have made use of Eq. (\ref{eq:collision}).  Thus, $M_{\rm th} \propto \Delta t^{(11-m)/2}$ will grow with increasing $\Delta t$ for $m < 11/2=5.5$, leading to a violation of our assumption of constant velocity of the fast shell and hence also a violation of the assumptions entering Eq. (\ref{eq:collision}).  The regimes of wind interaction are summarized in Table \ref{tab:regimes}.  

We now estimate the properties of the maser emission in the regime of freely-coasting evolution of the fast shell ($5.5 < m < 7$).  The upstream rest-frame pair density met by the fast shell, as it meets inspiral wind released at time $\Delta t$, is given by
\begin{eqnarray}
n_{\pm} &=& \left.\frac{\dot{M}}{4\pi  \Gamma r_{\rm sh}^{2}m_{\rm e} c }\right|_{\Delta t} \underset{\dot{M} \approx \frac{\dot{E}}{\Gamma c^{2}}}= \frac{\dot{E}[\Delta t]\Delta t}{2\pi m_{\rm e} c^{5}t^{3}}=  \frac{3E_{\rm f}(\Delta t/t_{\rm f})^{-3/4}}{8\pi m_{\rm e} c^{5}t^{3}} \nonumber \\
&\approx& \frac{3E_{\rm f}}{8\pi m_{\rm e} c^{5}t^{3}}\left(\frac{t}{2 \Gamma_{\rm f}^{2}t_{\rm f}}\right)^{-3/[2(m-5)]},
\label{eq:npm}
\end{eqnarray}
where in the final line we have used Eq. (\ref{eq:collision}).

Observer time is connected to lab-frame time according to the standard expression,
\be
t_{\rm obs} \simeq \frac{r_{\rm sh}}{2\Gamma_{\rm f}^{2}c} \approx \frac{t}{2\Gamma_{\rm f}^{2}} \underset{\rm Eq.~\ref{eq:collision}}\approx t_{\rm f}\left(\frac{\Delta t}{t_{\rm f}}\right)^{(m-5)/2}  \Rightarrow \frac{\Delta t}{t_{\rm f}} \approx \left(\frac{t_{\rm obs}}{t_{\rm f}}\right)^{2/(m-5)} ,
\label{eq:t_obs}
\ee
such that Eq. (\ref{eq:npm}) becomes,
\begin{eqnarray}
n_{\pm} \approx \frac{3E_{\rm f}}{64\pi m_{\rm e} c^{5}\Gamma_{\rm f}^{6} t_{\rm obs}^{3}}\left(\frac{t_{\rm obs}}{t_{\rm f}}\right)^{-\frac{3}{2(m-5)}}
\approx \frac{3E_{\rm f}}{64\pi m_{\rm e} c^{5}\Gamma_{\rm f}^{6} t_{\rm f}^{3}}\left(\frac{t_{\rm obs}}{t_{\rm f}}\right)^{\frac{27-6m}{2(m-5)}}.
\end{eqnarray}

For moderate upstream magnetization $\sigma \lesssim 1$, the spectral energy distribution of the maser peaks at a few times the plasma frequency of the upstream medium $\nu_{\rm p}=(4\pi n_{\pm} e^{2}/m_{\rm e})^{1/2}$ \citep{Plotnikov&Sironi19}, which boosted into the observer frame corresponds to a peak emission frequency
\begin{eqnarray}
\nu_{\rm pk} \approx 3\nu_{\rm p}\Gamma_{\rm f} \approx \nu_{\rm pk,f}\left(\frac{t_{\rm obs}}{t_{\rm f}}\right)^{\frac{27-6m}{4(m-5)}},
\label{eq:nuobs}
\end{eqnarray}
where
\be
\nu_{\rm pk,f} \equiv \left(\frac{27e^{2} E_{\rm f}}{16 m_{\rm e}^{2} c^{5}\Gamma_{\rm f}^{4} t_{\rm f}^{3}}\right)^{1/2} \approx 75\,{\rm GHz}\left(\frac{\Gamma_{\rm f}}{10^{3}}\right)^{-2}\left(\frac{B_{\rm d}}{10^{12}{\rm G}}\right),
\label{eq:nupkf}
\ee
where we have used Eq. (\ref{eq:Etot}) and take $t_{\rm f}=t_{\rm m}[a_f] \approx 1.2$ ms (Eq.~\ref{eq:tm}).  The spectral peak of the radio emission will thus start at high frequencies at $t_{\rm f}$ and drift lower after that time.

The luminosity of the maser emission is given by
\be
(\nu L_{\nu})|_{\nu_{\rm pk}} \approx f_{\xi}f_{\rm b}^{-1}L_{\rm e},
\ee
where $L_{\rm e}$ is the electron kinetic luminosity entering the forward shock seen by the external observer, $f_{\xi} \sim 10^{-3}-10^{-2}$ is the maser radiative efficiency defined relative to the kinetic energy of the electrons \citep{Plotnikov&Sironi19}, and $f_{\rm b}<1$ is geometric beaming factor due to the fraction of the total solid angle subtended by the binary outflow (e.g.,~the wedge near the binary equatorial plane; Fig.~\ref{fig:cartoon}).  

The shock luminosity in the observer frame is given by
\begin{eqnarray} \label{eq:luminosity_1}
L_{\rm e} &\approx& 4\pi r_{\rm sh}^{2} m_{\rm e} n_{\pm} c^{3}\frac{\Gamma_{\rm f}^{4}}{\Gamma^{2}[\Delta t]} \underset{\rm Eq.~\ref{eq:npm}}\approx \dot{E}\left(\frac{\Gamma[\Delta t]}{\Gamma_{\rm f}}\right)^{-4} \nonumber \\
 &\approx& \dot{E}_{\rm f}\left(\frac{\Delta t}{t_{\rm f}}\right)^{(21-4m)/4} 
\underset{\rm Eq.~\ref{eq:t_obs}}
\approx \dot{E}_{\rm f}\left(\frac{t_{\rm obs}}{t_{\rm f}}\right)^{\frac{21-4m}{2(m-5)}}.
\label{eq:Le}
\end{eqnarray}
Even though the fast shell is not significantly decelerated for $m > 5.5$, the total energy dissipated by the shock, $\int_{t_{\rm f}} L_{\rm e} {\rm d}t_{\rm obs} \sim \dot{E}_{\rm f}t_{\rm f}$, is of the same order of magnitude as the total energy of the fast shell.   

\subsubsection{Example ($m=6$)} \label{sec:example_m6}

As an example, for $m=6$ we have $\nu_{\rm pk} \propto t_{\rm obs}^{-9/4}$.  The spectral peak of the maser emission will cross down through observer bandpass (central frequency $\nu_{\rm obs}$) on a timescale
\be
t_{\rm FRB} \approx 6.8\,{\rm ms}\left(\frac{B_{\rm d}}{10^{12}{\rm G}}\right)^{4/9}\left(\frac{\Gamma_{\rm f}}{10^{3}}\right)^{-8/9}\left(\frac{\nu_{\rm obs}}{\rm 1\,GHz}\right)^{-4/9},
\label{eq:tFRB}
\ee
where again we take $t_{\rm f}=1.2$ ms.  We identify $t_{\rm FRB}$ as the characteristic duration of the observed radio burst as seen by a telescope with bandwidth $\Delta \nu \sim \nu_{\rm obs}$.

Note that to produce a detectable burst we require that $\nu_{\rm obs} > \nu_{\rm pk,f}$ ($t_{\rm FRB} > t_{\rm f}$), thus placing an upper limit on the final Lorentz factor
\be \label{eq:LF_limit}
\Gamma_{\rm f} \lesssim 8650 \left(\frac{B_{\rm d}}{10^{12}{\rm G}}\right)^{1/2}\left(\frac{\nu_{\rm obs}}{\rm 1\,\rm GHz}\right)^{-4/9}
\ee
Likewise, there is a maximum burst duration 
\be \label{eq:t_limit}
t_{\rm FRB,max} \approx 500\,{\rm ms}\,\left(\frac{P}{0.1{\rm s}}\right)^{2/7},
\ee
set by the time after which the fast shell will reach wind material released at large binary separations $a \gtrsim a_{\rm bin}$ (Eq.~\ref{eq:abin}) when the wind of the isolated pulsar dominates that of the binary. 

The isotropic radio luminosity reaches a peak value 
\be \label{eq:iso_luminosity}
L_{\rm r}[t_{\rm f}] \approx \left(\frac{f_{\xi}}{f_{\rm b}}\right)\dot{E}_{\rm f} \approx 4\times10^{42}\,{\rm erg\,s^{-1}}\,\left(\frac{f_{\xi,-3}}{f_{\rm b,-1}}\right)\left(\frac{B_{\rm d}}{10^{12}{\rm G}}\right)^{2}, \ee 
and decays thereafter as $L_{\rm r} \propto t_{\rm obs}^{-3/2}$, where $f_{\xi,-3}=f_{\xi}/10^{-3}$ and $f_{\rm b,-1}=f_{\rm b}/0.1$.  

The isotropic radiated energy of the burst $\mathcal{E}_{\rm FRB} \approx L_{\rm r} t_{\rm obs} \propto t_{\rm obs}^{-1/2}$ over a timescale $\sim t_{\rm obs}$ is a weak function of time.  Its value measured by an observer at frequency $\sim \nu_{\rm obs}$ on the timescale $t_{\rm FRB}$ is given by
\begin{eqnarray}
\mathcal{E}_{\rm FRB}&\approx&L_{\rm r}[t_{\rm FRB}] t_{\rm FRB} \nonumber \\
&\approx&1.2\times 10^{37}{\rm erg\,} \left(\frac{f_{\xi,-3}}{f_{\rm b,-1}}\right)\left(\frac{\Gamma_{\rm f}}{10^{3}}\right)^{4/9}\left(\frac{\nu_{\rm obs}}{\rm 1\,GHz}\right)^{2/9}\left(\frac{B_{\rm d}}{10^{12}{\rm G}}\right)^{16/9}
\label{eq:EFRB}
\end{eqnarray}

\begin{table}
  \begin{center}
    \caption{Critical regimes of $m$, where $\dot{M} \propto a^{-m}$.}
    \label{tab:regimes}
    \begin{tabular}{c|c} 
      $m$ & Description\\
      \hline
\hline
	$> 7$ & Binary wind decelerates approaching merger; no shocks \\
    $< 7$ & Binary wind accelerates approaching merger; internal shocks \\
	$> 5.5$ & Coasting fast shell \\
	$< 5.5$ & Fast shell sweeps up more than its own inertial mass, decelerates \\ 
	$> 5$ & Coasting fast shell meets unshocked binary wind \\
	$< 5$ & Coasting fast shell meets earlier-shocked binary wind \\ 
	$<4$ & Coasting fast shell sweeps up more than its own rest mass \\ 
	$3$ & ``Goldreich-Julian'' scaling (Eq.~\ref{eq:MdotGJ})
    \end{tabular}
  \end{center}
\end{table}

\section{Simulations of Shock Interaction}
\label{sec:simulations}

For the magnetic field configuration of the two stars most likely to be realized in nature, the binary-induced outflow should be concentrated in the orbital plane (Fig.~\ref{fig:cartoon}).  Furthermore, although the outflow will emerge opposite of the weakly-magnetized star on the radial scale of the binary separation $\sim a$, upon averaging over many orbits the outflow will become approximately azimuthally-symmetric upon reaching large radii $\gg a$ where the internal shock interaction we are modeling takes place.  

For simplicity, we perform 1D calculations of the wind interaction along the radial direction corresponding to the equatorial plane.  We model the outflow under the assumption of spherical symmetry, with the understanding that due to relativistic beaming the emission may only be observable for observers within a limited solid angle of binary equatorial plane subtended by the outflow (i.e. all outflow properties and observed luminosities are understood to be {\it isotropic equivalent} quantities).  This section describes the simulation setup and the code used in our numerical calculations.

\subsection{Numerical Set-Up}

We consider the wind launched from the binary during the final stages of the inspiral, as the semi-major axis shrinks from some specified initial distance $a_{\rm 0}$ to a final size $a_{\rm f}$.  We take $a_{\rm f}=2 R_{\rm ns}=24$ km in our simulations, appropriate to NS-NS mergers.  For yet smaller separations $a \lesssim a_{\rm f}$, additional complications arise (e.g.,~deviations from point-mass inspiral picture, tidal mass-loss from the star(s)---baryon loading in the wind, non-dipolar contributions to the magnetic field, etc.), and therefore we ignore the very final phase. 

The binary separation at the beginning of the simulation, $a_{\rm 0}$, must be chosen large enough so as to generate a sufficient upstream medium for the subsequent wind (the bulk of the energy of which is released near the end of the inspiral) to interact with and, potentially, be decelerated by.  On the other hand, larger values of $a_{\rm 0}$ require a longer simulation time prior to and after the merger to follow the full gravitational wave inspiral and shock interaction, respectively.  In our numerical simulations, we take $a_{\rm 0}/a_{\rm f}=3-10=6-20~R_{\rm ns}$ as a compromise between these considerations.  This is still a factor $\gtrsim$ 2 smaller than the total range of binary separation over which the binary wind dominates that of the single pulsar (Eq.~\ref{eq:abin}).  Defining the start of the calculation as $t=0$, the merger takes place at $t=t_{\rm m}[a_0] \equiv t_{\rm m,0} \approx 0.1-12$ s (for $a_{0}/a_{\rm f}=3-10$), and the physical part of the wind injected from the inner boundary of the simulation is assumed to terminate just prior to the merger at $t \simeq t_{\rm m,0} - t_{\rm f} \approx t_{\rm m,0}$, where $t_{\rm f} \approx 1.2$ ms.

Since the shock interaction takes place at radial distances $\gg a_{\rm f}$, we fix the inner boundary of our simulation grid at $r_{\rm in}=a_{\rm f}$, rather than following the binary contraction self-consistently.  In general, the injected wind power, $\dot{E}_{\rm in}$, can be divided into kinetic and magnetic components,
\be
\dot{E}_{\rm in}=\dot{E}_{\rm kin} + \dot{E}_{\rm mag}=\dot{E}_{\rm kin}(1+\sigma_{\rm in}),
\label{eq:Edotin}
\ee
where $\sigma_{\rm in} \equiv \dot{E}_{\rm mag}/\dot{E}_{\rm kin}$ is the magnetization, and
\be
\dot{E}_{\rm mag}=4\pi r_{\rm in}^{2}c \left(\frac{B_{\rm in}^{2}}{4\pi}\right);\,\,\,
\dot{E}_{\rm kin}=4\pi r_{\rm in}^{2} \rho_{\rm in} \Gamma_{\rm in} c^{3},
\ee
where $B_{\rm in}(t)$ and $\rho_{\rm in}(t)$ are the lab-frame magnetic field strength and rest-frame mass density injected at the inner boundary, and $\Gamma_{\rm in}$ is the bulk Lorentz factor of the injected wind.  

As discussed in \S\ref{sec:windproperties}, we assume that the wind is able to accelerate over a short radial baseline, effectively immediately converting its energy entirely into kinetic form.  Hence we take $\sigma_{\rm in}=0$ and perform purely hydrodynamical simulations of the shock interaction (however, note that a moderate value of $\sigma=0.1$ is assumed in calculating the properties of the shock maser emission; \S\ref{sec:m6_emission}).  Future work will perform MHD simulations exploring the more general case of highly magnetized winds $\sigma_{\rm in} \gg 1$.

\begin{figure}
    \centering
    \includegraphics[width=80mm]{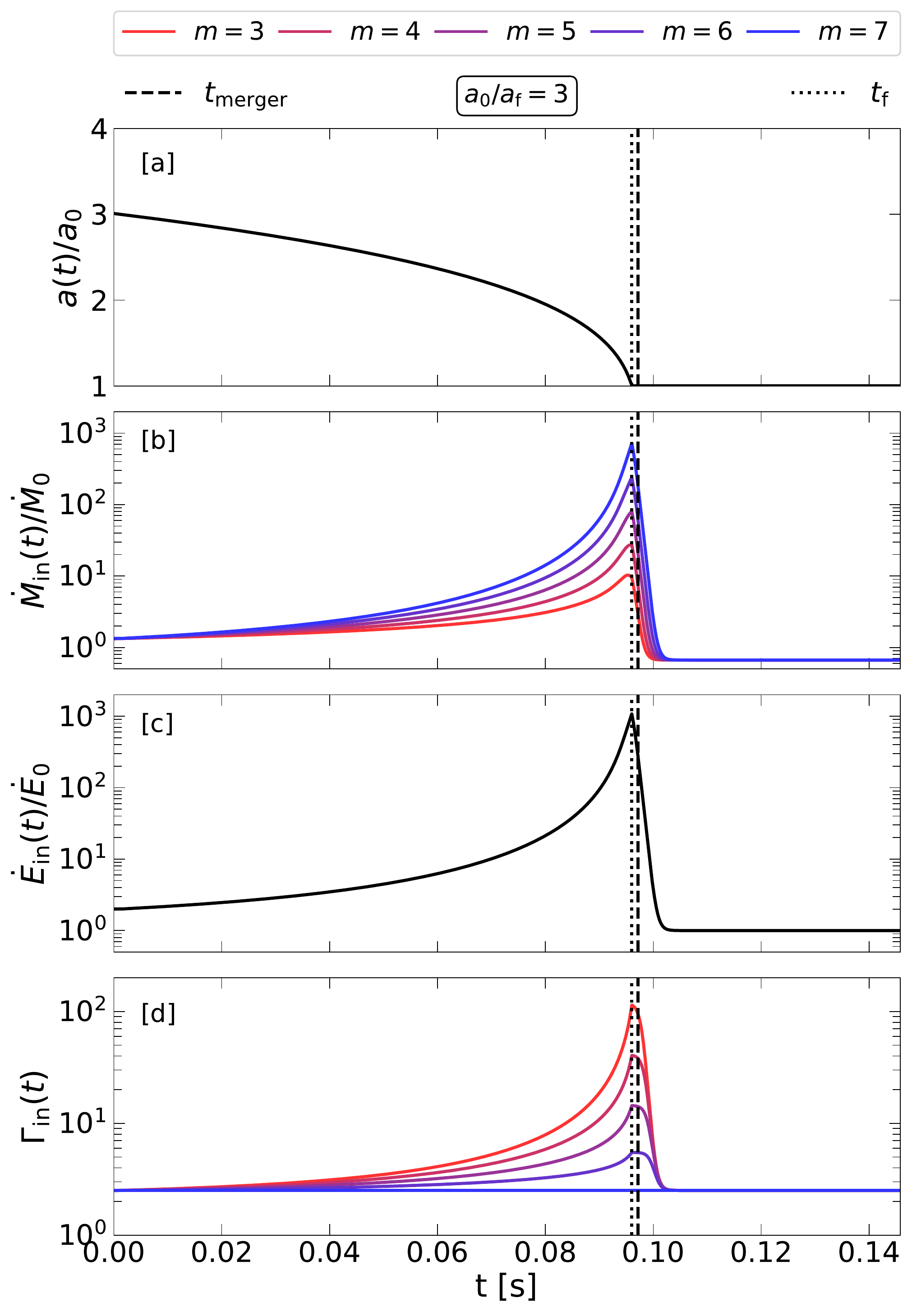}
    \caption{Time evolution of the properties of the injected binary wind in our simulations, meant to mimick the final stages of a neutron star merger.  Quantities shown include the [a] binary semi-major axis; [b] mass-loss rate, [c] wind power, and [d] bulk Lorentz factor of the inspiral wind. The quantities in panels [a], [b] and [c] are normalized by their initial values---defined at the binary separation of $a_{\rm 0}=3a_{\rm f}=6R_{\rm ns}$. The vertical dashed line denotes the time for the merger ($t_{\rm m,0}$) from an initial separation of $a_{\rm 0}$, and the dotted black vertical lines denote the time when the central engine is turned off $(t_{\rm m,0}-t_{\rm f})$. The colors (red -- blue) depict different assumptions for the wind mass-loss power law index $m \in [3,7]$ (see Table \ref{tab:models})}
    \label{fig:wind_profile}
\end{figure}

As summarized in Fig.~\ref{fig:wind_profile} and Table~\ref{tab:models}, the time-dependent wind properties are specified via boundary conditions at the inner domain of the simulation grid.  Following Eq. (\ref{eq:windproperties}), the kinetic luminosity of the injected wind $\dot{E}=\dot{E}_{\rm kin}$ evolves according to
\begin{equation}\label{eq:edotin}
    \dot{E}_{\rm in}(t)=
    \begin{cases}
\dot{E}_{0} & (t < 0) \\
          \frac{\dot{E}_0}{(1 - t/t_{\rm m,0})^{7/4}}f_{\rm taper}(t) & (0 \le t \lesssim t_{\rm m,0} - t_{\rm f})\\
          \dot{E}_{0} & (t \gtrsim t_{\rm m,0}),
    \end{cases}
\end{equation}
where the initial wind power $\dot{E}_0 \approx \dot{E}_{\rm in}[t_{\rm f}](a_{\rm 0}/a_{\rm f})^{-7}$ is chosen such that the final wind power $\dot{E}_{\rm in}[t_{\rm f}] \equiv \dot{E}_{\rm f} \approx (10^{3}-10^{7})\dot{E}_0$ following Eq. (\ref{eq:Edot1}), for different values of $m$.  Likewise, the mass-loss rate of the wind injected from the inner boundary evolves as (Eq.~\ref{eq:windproperties})
\begin{equation}\label{eq:mdotin}
    \dot{M}_{\rm in}(t)=
    \begin{cases}
\dot{M}_{\rm 0} & (t < 0) \\
          \frac{\dot{M}_{\rm 0}}{(1-t/t_{\rm m,0})^{m/4}}f_{\rm taper}(t) & (0 \le t \lesssim t_{\rm m,0} - t_{\rm f})\\
          \dot{M}_{\rm 0} & (t \gtrsim t_{\rm m,0})
    \end{cases}
\end{equation}
where $\dot{M}_{\rm 0}=\dot{E}_{\rm 0}/\Gamma_{\rm 0}c^2$ (Eq.~\ref{eq:sigma1}) and $\Gamma_{0}=1.5$.  For numerical stability, we gradually taper the wind power and mass-loss rate just after the merger ($t \gtrsim t_{\rm m,0}$), instead of abruptly shutting the engine off.  This is achieved in Eqs. (\ref{eq:edotin}, \ref{eq:mdotin}) by multiplying the wind power by a tapering function,
\be \label{eq:taper}
f_{\rm taper}(t) \equiv \frac{1}{2}\bigg\{ 1 + \tanh \left({\frac{t - t_{\rm m,0} - t_{\rm f}}{t_{\rm f}}}\right) \bigg\},
\ee
We have checked that the precise form of the tapering does not significantly affect the overall shock dynamics relevant to the radio emission.  

We vary the power-law index $m \in [3,7]$ in order to explore a diversity of wind mass-loading behavior (\S\ref{sec:shocks}).  The values of $\dot{E}_0$ and $\dot{M}_0$ are chosen such that the initial outflow Lorentz factor $\Gamma_{\rm in}(t=0)=\Gamma_{\rm 0} + 1.01\simeq2.5$ for all models.  For our chosen set-up, different values of $m$ and initial inspiral radius $a_{\rm 0}/a_{\rm f}$ result in different values for the peak $\Gamma_{\rm in}[t_{\rm f}] \simeq \Gamma_{\rm f} \sim 3-115$ of the outflow achieved near the end of the inspiral (Fig.~\ref{fig:wind_profile}; Table \ref{tab:models}).  These relatively modest peak Lorentz factors were chosen for purposes of numerical stability ($\S\ref{sec:code}$), despite being much smaller than are likely achieved in the actual wind (e.g., Eq.~\ref{eq:eta} and surrounding discussion).  However, insofar as we are able to capture the shock dynamics in the ultra-relativistic regime, in our final analysis we can simply rescale the shock properties and radio emission to the physical values of $\dot{E}_{\rm f}$ and $\Gamma_{\rm f}$ ($\S\ref{sec:discussion}$).

The medium exterior to the binary, into which the inspiral wind enters at the start of the simulation ($t=0$), is taken to be that of a steady wind of power $\dot{E}_0$, mass-loss rate $\dot{M}_{\rm 0}$ and Lorentz factor $\Gamma_0$.  Physically, this wind could represent the ordinary spin-down powered pulsar wind prior to the binary orbit entering the light cylinder, albeit for larger values of $a_0 \approx a_{\rm bin}$ (Eq.~\ref{eq:abin}) than adopted in our numerical simulations.  We have checked that the pre-existing wind does not affect the shock interaction during the times we simulate.  All of the wind material is ejected with a low thermal pressure ($p_{\rm gas}/\rho c^2=10^{-3}$, where $p_{\rm gas}$ is the gas pressure as defined in the upcoming section).

After the binary wind is shut-off at $t_{\rm m,0} - t_{\rm f}$, we continue to run the simulations to later times, $t_{\rm sim} \gg \Gamma_{\rm f}^{2}t_{\rm f}(a_{\rm 0}/a_{\rm f})^{4/3}$, to capture the interaction between the fast shell released at the end of the inspiral with the slowest material released at the beginning (Table \ref{tab:models} compiles the simulation run time).  For numerical stability, the same small residual wind power $\dot{E}_{0}$ and mass-loss rate $\dot{M}_0$ of the pre-inspiral wind are also injected at late times after the merger $\gg t_{\rm m,0}$; however, this slow late-time wind does not immediately catch up to the fast shell and has no appreciable impact on the forward shock properties at larger radii of greatest interest.

\begin{table*}
\begin{center}
\renewcommand{\arraystretch}{1.25}
\caption{ Parameters of the inspiral wind models simulated in this work. }
\begin{tabular}{ |c|c c c c c c c| } 
\hline 
 Sim. ID & m$^{[\rm a]}$ & $a_{\rm 0}/a_{\rm f} ^{[\rm b]}$ & $t_{\rm m,0}^{[\rm c]}$ [s] & max$(\dot{E}_{\rm in})/\dot{E}_{\rm 0}^{[\rm d]}$ & max$(\dot{M}_{\rm in})/\dot{M}_{\rm 0}^{[\rm e]}$ & max$(\Gamma_{\rm in})^{[\rm f]}$ & $t_{\rm sim}^{[\rm g]}$ [s] \\[0.25em]
\hline \hline
 m3a3  & 3 &   &       &                & 10.3             & 114 &   \\
 m4a3  & 4 &   &       &                & 27.3             & 41  &   \\
 m5a3  & 5 & 3 & 0.097 & $1\times 10^3$ & 80.6             & 14.4  &  10 \\
 m6a3  & 6 &   &       &                & $2.4\times 10^2$ & 5.5   &   \\
 m7a3  & 7 &   &       &                & $7.2\times 10^2$ & 2.5   &   \\
 \hline 
 m5a10 & 5 & 10 & 12.0 & $5\times 10^6$ & $5\times 10^4$ & 151  & 72  \\
 m6a10 & 6 & 10 & 12.0 & $5\times 10^6$ & $3.3\times 10^5$ & 16  & 72  \\
 \hline
\end{tabular}
      \begin{tablenotes}
      \footnotesize
      \item \textit{Note.} A brief description of the defining parameters of each performed simulation (identifier in first column) is as follows. $^{[\rm a]}$Power-law index of the mass-loss rate (Eq.~\ref{eq:Mdotparam}); $^{[\rm b]}$Ratio of the initial ($a_{\rm 0}$) to final binary semi-major axis ($a_{\rm f}=2R_{\rm ns}$); $^{[\rm c]}$Time until merger from an initial separation of $a_{\rm 0}$; $^{[\rm d,e]}$Ratio of the maximum wind power ($3\times 10^{44}$ erg s$^{-1}$) and mass-loss rate achieved near the end of the inspiral phase to their initial values; $^{[\rm f]}$Lorentz factor of the fastest shell ejected at the end of the inspiral phase; $^{[\rm g]}$Simulation duration in lab frame $t_{\rm sim}$.
      \end{tablenotes}
\renewcommand{\arraystretch}{1}
\label{tab:models}
\end{center}
\end{table*}

\subsection{Numerical Code and Simulation Scheme}
\label{sec:code}

Our numerical simulations are performed using {\sc Mara3}, an open-source higher-order Godunov code first described in \citet{Zrake_MacFadyen_12}. The code has been extended with an Eulerian-Lagrangian moving mesh scheme that evolves the mesh vertices along with the outflowing plasma, similar to the scheme described in \citet{Duffell_MacFadyen_13}. This approach automatically concentrates numerical resolution in regions of high compression naturally arising around the forward shock. As mentioned above, we neglect the dynamical importance of the magnetic fields, assuming that the wind converts its Poynting flux to bulk kinetic energy at smaller radii than we simulate.

The code solves the special relativistic hydrodynamical equations in flux-conservative form,
\begin{equation}\label{eq:conserved_form}
	\frac{\partial \mathbf{U}}{\partial t} + \bs{\nabla} \cdot \mathbf{F}=\mathbf{S}_{\rm geom} \, ,
\end{equation}
where the vector of conserved quantities $\mathbf{U}=(D, S, \tau)^T$ consists of the lab-frame mass density $D=\Gamma \rho$, radial momentum density $S=\rho h \Gamma^2 v / c^2$, and energy density (excluding rest-mass) $\tau=\rho h \Gamma^2 - p_{\rm gas} - D c^2$. $\Gamma=(1 - v^2/c^2)^{-1}$ is the gas Lorentz factor, and $h=c^2 + \varepsilon + p_{\rm gas} / \rho$ is specific enthalpy. $\varepsilon$, $p_{\rm gas}$, $\rho$ denote the specific internal energy, gas pressure, and proper mass density, respectively. We adopt a gamma-law equation of state $p_{\rm gas}=(\gamma_{\rm ad} - 1)\varepsilon \rho c^2$, with adiabatic index $\gamma_{\rm ad}=4 / 3$.  In making this choice, we are neglecting the effects of radiative cooling of the post shock, i.e. we take the evolution to be adiabatic.  The validity of this assumption will depend on the detailed composition of the binary wind (e.g., the presence of ions, which cool less effectively than pairs) and the synchrotron cooling timescale of these particles relative to the expansion rate.  A more thorough exploration of the dynamical effects of cooling is left to future work (see also a discussion in \S\ref{sec:caveats}).

The vector of fluxes is given by
\begin{equation}\label{eq:flux}
\mathbf{F}=\begin{pmatrix}
D v \\
\tau v + p_{\rm gas} v \\
\mathbf{S} v + p_{\rm gas}
\end{pmatrix} \, ,
\end{equation}
and the geometrical source term is $\mathbf{S}_{\rm geom}=(0, 0, 2 p_{\rm gas}/r)^T$.

Eq. (\ref{eq:conserved_form}) is discretized in terms of the extrinsic conserved quantities, $\mathbf{Q}_i=\int_{i} \mathbf{U} {\rm d}V$ integrated over each control volume. The control volumes are spherical shells whose inner and outer boundaries are moved to enforce a zero-mass-flux condition.  The discretized solution is updated in time using the method of lines and an explicit second-order Runge-Kutta integration, whose base scheme is given by
 \begin{equation} \nonumber
 \delta \mathbf{Q}_i=\delta t \left(\mathbf{\hat F}_{i-1/2} A_{i-1/2} - \mathbf{\hat F}_{i+1/2} A_{i+1/2} + \mathbf{S}_{i,\rm geom} V_i \right) \, .
 \end{equation}
Here $\mathbf{\hat F}_{i \pm 1/2}$ are the Godunov fluxes through the inner and outer cell boundaries, $A_{i+1/2}=4 \pi r_{i+1/2}^2$ is the area of the cell boundary at $i+1/2$, $V_i$ is the cell volume, and $\mathbf{S}_{i,\rm geom}$ is the geometrical source term sampled half way between $r_{i+1/2}$ and $r_{i-1/2}$.
The Godunov fluxes are given by $\mathbf{\hat F}=\mathbf{F}^* - \mathbf{U}^* v^*$, where $v^*$ is the speed of the contact discontinuity, and $\mathbf{U}^*$ and $\mathbf{F}^*$ are the intrinsic conserved quantities and associated fluxes at the $i+1/2$ cell interface. The starred quantities are obtained by sampling the solution (which is self-similar in $r/t$) of the Riemann problem $(\mathbf{U}_i, \mathbf{U}_{i+1})$ at $r/t=v^*$. This solution is approximated using the Harten–Lax–van Leer contact (HLLC) solver of \cite{Mignone_Bodo_06}. The cell interfaces are advanced in time according to the same Runge-Kutta time-integration scheme, where
 \begin{equation}
 \delta r_{i+1/2}= v^*_{i+1/2} \delta t \, .
\end{equation}

New cells are inserted at the radial inner boundary $r=r_{\rm in}$ by adding an additional face at $r_{\rm in}$ when the innermost face has advanced past $r_{\rm in} + \delta r$, where $\delta r$ is chosen based on the desired grid resolution. The fluid state of the inserted zone is set according to the wind inner boundary condition described in Eqs. (\ref{eq:Edotin}, \ref{eq:mdotin}). Due to the high compression factor arising at high-$\Gamma$ shocks, cells can become very narrow, requiring a prohibitively short time step to satisfy the Courant Friedrichs-Lewy condition. Cells that are compressed below a threshold value $\delta r_{\rm min}$ are joined with an adjacent cell in a conservative fashion, by erasing the interface between the two cells and combining their extrinsic conserved quantities. To minimize the size disparity of adjacent zones, the removed interface is always the one between the too-small cell, and the smaller of its two neighbors. We also employ a conservative cell-splitting technique to keep from under-resolving rarified regions where the cells grow larger than $\delta r_{\rm max}$. The parameters $\delta r_{\rm min}$ and $\delta r_{\rm max}$ are typically kept within an order of magnitude of one another.

The outer edge of the simulation domain is placed at $L_{\rm box}/c=10^3$ s ($L_{\rm box}\approx \unit[3\times 10^{13}]{cm} \gtrsim 10^{7} a_{\rm f}$) and the initial grid has a resolution of 512 cells per decade. The 1D nature of our special relativistic hydrodynamical simulations enables us to study the evolution of the ejecta at high resolution for longer durations---from the initial acceleration until late time deceleration phase.
The duration of each simulated model, $t_{\rm sim}$, are listed in Table \ref{tab:models}.

Two-dimensional simulations of decelerating relativistic ejecta by \cite{Duffell_MacFadyen_13} have shown that the onset of Rayleigh-Taylor (RT) instability can disrupt the contact discontinuity, and generate turbulence in the shocked gas. The mixing causes the shocked gas to acquire more entropy and internal energy. This results in the reverse shock being pushed to smaller radii, and delays its observed emission relative to equivalent 1D models which are artificially stabilized against RT. We do not expect this phenomenon to change our overall results, as the radio emission in our model primarily arises from the forward shock (\S\ref{sec:m6_emission}), which is generally unaffected by the development of RT. 

\section{Simulation results} \label{sec:results}

\begin{figure*}
    \centering
    \includegraphics[width=170mm]{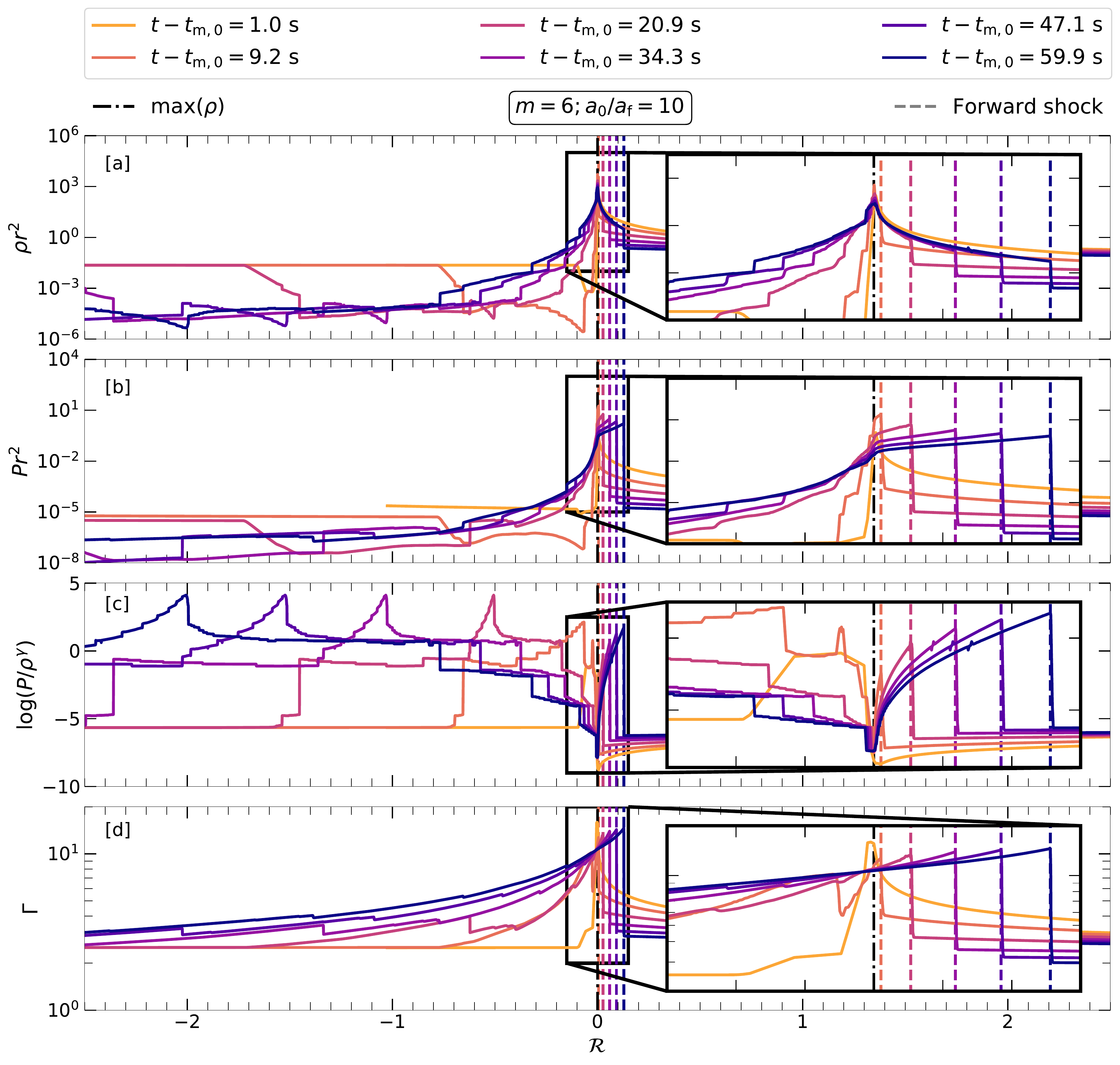}
    \caption{Snapshots of the radial profiles of hydrodynamical quantities demonstrating the self-interaction of the inspiral-driven wind from the \textit{m6a10} model ($m=6$; $a_{\rm 0}/a_{\rm f}=10$).  To follow the narrow shock structure, we employ an Eulerian radial coordinate ${\cal R}\equiv(r-r_{\rho})/r_{\rm in}$, where $r_{\rho}$ is the location of the peak density, and $r_{\rm in}$ is the inner edge of the simulation box.  Panel [a]: Comoving density ($\rho$); Panel [b]: Pressure ($P$); Panel [c]: Entropy $\log(P/\rho^{\gamma}$) where $\gamma=4/3$ is the adiabatic index.  Panel [d]: Bulk Lorentz factor.  $\rho$ and $P$ are multiplied by $r^2$ in order to compensate for their secular radial evolution in spherical coordinates. Vertical black dash-dotted lines (at ${\cal R}=0$) denotes the location of the peak density, while dashed vertical lines denote the location of the forward shock.  The density and pressure are shown in code units (but are scaled to physical units when calculating radio light curves).  Different colored lines denote different snapshots in time $t - t_{\rm m,0}$ after the merger.  The inset in each panel shows a zoomed-in view around $r_{\rho}$ $(-0.15\le{\cal R}\le0.15)$.}
    \label{fig:radial_evol_m6}
\end{figure*}

This section describes the results of our hydrodynamical simulations of the binary wind interaction and our method of post-processing the inferred forward shock properties to obtain light curves of the radio maser emission.  As described in $\S\ref{sec:shocks}$, the final stage of the inspiral generates a highly energetic relativistic shell which then propagates into the density field laid down by the earlier inspiral wind.  The dynamics of this shell---and hence of the dominant forward shock it generates---depends on the properties of the upstream binary wind, in particular how quickly the mass-loss rate (parametrized by the power-law index, $m$) rises with decreasing binary separation (Table \ref{tab:models}). We start in \S\ref{sec:m6} by analyzing first the \textit{m6a10} model in detail (Table \ref{tab:models}), as the $m=6$ case falls in the regime $m > 5.5$ for which the final shell is expected to coast freely (Table \ref{tab:regimes}), and for which analytic results were obtained in $\S\ref{sec:shocks}$.  Then in \S\ref{sec:lowerm} we explore the \textit{m5a10} model as a representative case which samples the $m<5.5$ regime, for which stronger deceleration of the fast shell may be expected. We do not analyze the $m=7$ model, corresponding to a constant wind velocity, because as expected no shocks were seen to develop.

\subsection{Coasting Fast Shell (m=6)}
\label{sec:m6}

\subsubsection{Hydrodynamics of the Shock Interaction}

Figure \ref{fig:wind_profile} shows the time evolution of the injected wind properties for all the critical regimes of wind interaction ($3\le m \le 7$; Table \ref{tab:regimes}), with a small initial separation of $a_{\rm 0}/a_{\rm f}=3$.  Compared to most of our other models with $a_{\rm 0}/a_{\rm f}=3$ (Table \ref{tab:models}), the large initial binary separation $a_{\rm 0}/a_{\rm f}=10$ extends the start of the simulation prior to the merger to $\approx$ 12 s.  Though still not in the expected physical range, this relatively longer lead-in allows us to follow the dynamical interaction of the fast shell with the earlier inspiral wind material to larger radii and for a longer duration, before the FS enters regions influenced by the steady pulsar wind (the initial outflow assumed to exist on the grid prior to the inspiral phase simulated).

Self-interaction within the inspiral wind gives rise to an outwardly propagating shock complex.  In general, the complex is expected to include a forward shock (FS) propagating into the unshocked inspiral wind, a reverse shock (RS) propagating back through fast shell (and ultimately the post-merger wind), and a contact discontinuity separating the two regions of shocked gas.  There furthermore exists a complex series of shocks and other features in the post-merger wind region---caused due to the intereaction of RS with the post-merger wind---which we largely neglect in what follows as they are neither physical nor germane to the synchrotron maser emission, which originates from the energetically-dominant FS generated by the fast shell. Figure \ref{fig:radial_evol_m6} illustrates the radial profiles of the gas density, pressure, entropy, and velocity at various snapshots in time (color coded with respect to the time after merger, $t-t_{\rm m,0}$).  In order to better highlight the different structures and discontinuities of the hydrodynamical parameters associated with the shock complex, we employ an Eulerian radial coordinate ${\cal R}\equiv(r-r_{\rho})/r_{\rm in}$, where $r_{\rho}$ is the location of the peak density on the grid, and $r_{\rm in}$ is the inner edge of the simulation box.

The yellow line in Fig. \ref{fig:radial_evol_m6} shows a snapshot taken soon after the merger ($t-t_{\rm m,0}\sim$ 1 s). Although the inspiral ejecta is spatially extended ($\Delta\simeq ct_{\rm m,0}\sim10^{11}$ cm; ${\cal R}\gtrsim4$), most of its energy---estimated to be $\sim5\times10^{41}$ erg (Eq. \ref{eq:Etot})---is carried by the narrow shell at ${\cal R}\sim0$.  This narrow shell has a width of $\Delta_{\rm f}\simeq ct_{\rm f}\sim3\times10^7$ cm, with a bulk Lorentz factor of $\Gamma_{\rm f}\sim16$ (see Table \ref{tab:models})---being the fastest portion of the entire ejecta. The peak seen at ${\cal R}\sim0$ in the spatial extent of $\rho r^2$ (Fig. \ref{fig:radial_evol_m6}a) also indicates that most of the total ejected wind mass is located in the same shell.  At this early time after the merger, strong interaction between the fastest and slower parts of the inspiral wind have not yet developed.  

However, by the second snapshot (orange line in Fig. \ref{fig:radial_evol_m6}; $t-t_{\rm m,0}\sim9$ s), a high density spike appears ahead of the fast shell.  This is accompanied by a discontinuous jump in entropy (at ${\cal R}\sim0+\epsilon$ of Fig. \ref{fig:radial_evol_m6}c), indicating the formation of the nascent FS, as the fast shell interacts with the upstream inspiral wind.  The shocked downstream material in the posterior tail of the fast shell starts exhibiting semblance of a reverse shock (RS). This feature is seen at ${\cal R}\sim-0.03$ (see inset of Fig. \ref{fig:radial_evol_m6}) with a peak in entropy (panel c) and a drop in the Lorentz factor ($\Gamma_{\rm f}\sim6$; panel d) with respect to the fast shell ahead of it ($\Gamma_{\rm f}\sim16$), and propagates behind into the post-merger pulsar wind at later times. 
Given the coasting shell evolution expected in the $m=6$ case, the prompt RS---formed as a result of the interaction of the FS with the upstream inspiral wind---is too weak to influence the dynamics of the fast shell.  The kinetic energy of the fast shell is efficiently transferred to gas behind the FS (denoted by a vertical dashed line in Fig.~\ref{fig:radial_evol_m6}) without any significant deceleration of the FS itself, due to the RS.  

In the following stage, more wind material is swept up by the narrow fast shell, which then begins to broaden it.  This can be observed at $t-t_{\rm m,0}\gtrsim21$ s (magenta curves in Fig. \ref{fig:radial_evol_m6}) by a modest drop in the density (at ${\cal R}\sim0.0$ of panel a), and a broadening of the fast shell behind the FS. The onset of the stratification of the fast shell into a distinct FS and a RS as seen in the previous stage ($t-t_{\rm m,0}\sim9$ s) is accompanied by the formation of the internal reverse shocks that continue to propagate back into material released after the merger (e.g., the discontinuous entropy jumps seen at ${\cal R}<0$).  The latter evolution is complex but is not physical (we are not attempting to model the post-merger phase) and, since the post-merger outflow carries little energy, has little impact on the FS.   
 
\begin{figure}
    \centering 
        \centering
        \includegraphics[width=85mm]{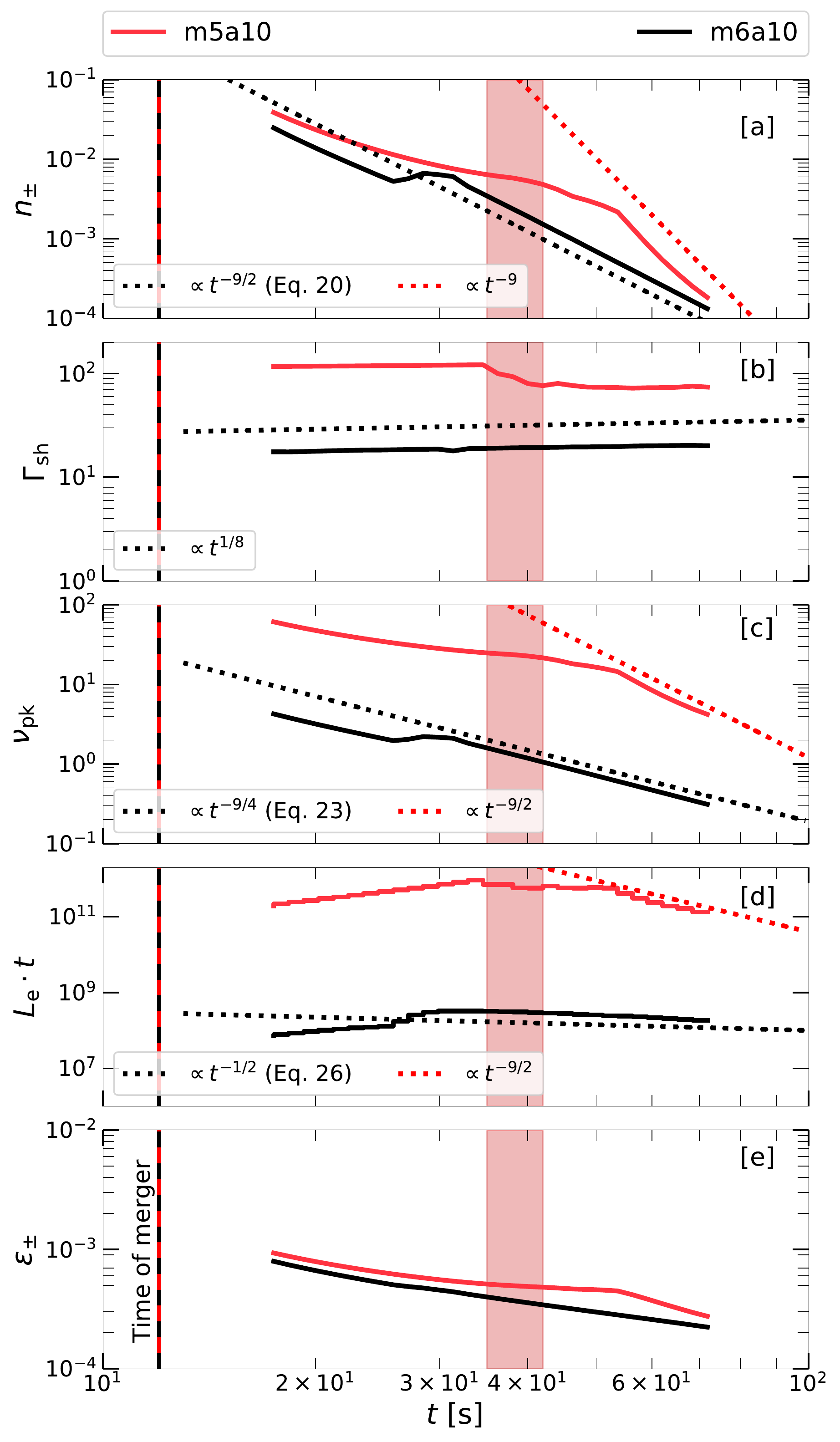}
        \caption{Evolution of shock properties for different models of the inspiral wind (see Table \ref{tab:models}), with respect to time in lab frame. Shock properties corresponding to fast shell coasting into pristine upstream medium ($m=6$) are denoted by the black curves, and the red curves ($m=5$) are representative of the $m<5.5$ regime wherein a decelerating fast shell drives through a pre-shocked upstream medium (see Table \ref{tab:regimes}).  The simulations are performed for an initial inspiral separation of $a_{\rm 0}/a_{\rm f}=10$. The top panel [a] denotes the upstream rest-frame mass density---corresponding to the earlier discharged inspiral wind---as intercepted by the forward shock; panel [b] shows the shock Lorentz factor; panel [c] shows the evolution of the peak of the synchrotron maser emission in observer frame; panel [d] depicts the fluence evolution of the burst---measured from the shock luminosity $L_{\rm e}$, and panel [e] illustrates the specific internal energy of the upstream medium entering the forward shock. $n_{\pm},\nu_{\rm pk}$ and $L_{\rm e}\cdot t$ are denoted in arbitrary units, and $\varepsilon_{\pm}$ in units of $c^2$, as defined in \S\ref{sec:code}. In panels [a], [c] and [d], the black dotted lines represent the agreement of the ($m=6$) simulation results with the analytical estimates---corresponding to the labelled equations in \S \ref{sec:shocks}. Red-dotted lines in all the panels, and the black dotted lines in panel [b] denote an empirically fitted powerlaw slope for respective shock properties. The moment of merger is denoted by the vertical black-red dashed line (for both the models, \textit{m5a10} and \textit{m6a10}). The red band at $35 \lesssim t~[s] \lesssim 42$ denotes the reverse shock crossing phase in the \textit{m5a10} model.}
\label{fig:time_evolution}
\end{figure}

Solid black curves in Fig. \ref{fig:time_evolution} show the time evolution of properties relevant to the upstream inspiral wind, FS, and its radio emission extracted from \textit{m6a10} simulation.  These include the upstream density just ahead of the shock $n_{\pm}$, the Lorentz factor of the FS $\Gamma_{\rm sh}$---well approximated to an accuracy of ${\cal O}(\Gamma_{\rm sh}^{-1})$ by $\sqrt{2}$ times the Lorentz factor of the immediate post-shock gas \citep{Blandford_McKee_76}, the peak frequency of the maser emission $\nu_{\rm pk}$ (Eq.~\ref{eq:nuobs}), and the (bolometric) kinetic fluence $L_{\rm e} \cdot t$ (kinetic shock luminosity $L_{\rm e}$ times lab time $t$).\footnote{Since $\Gamma_{\rm sh}$ is approximately constant, lab time $t$ and observer time $t_{\rm obs}=t/(2\Gamma_{\rm sh}^{2})$ are roughly proportional, and hence $L_{\rm e} \cdot t$ shows the time evolution of the observed fluence.} After an initial transient phase during which the FS is still developing (and hence cannot be properly identified on the grid), $\Gamma_{\rm sh}$ evolves only weakly with time ($\propto t^{\alpha}$, where $\alpha=1/8$), roughly consistent with the expectation of coasting evolution for $m=6$.  The evolution of $n_{\pm}$, $L_{\rm e} \cdot t$ and $\nu_{\rm pk}$, also achieve power-law-like behavior with decay indices that agree well with the analytic expectations (Eq.~\ref{eq:npm}, \ref{eq:nuobs} and \ref{eq:Le}; shown as grey dotted lines in Fig.~ \ref{fig:time_evolution}). 

In summary, the $m=6$ case is characterized by a rapid transfer of the kinetic energy of the fast shell to the FS, with no deceleration of the Lorentz factor of the fastest material throughout the evolution, consistent with the expected coasting behavior for $m > 5.5$ (\S\ref{sec:shocks}).  The properties of the FS agree well with analytic predictions made under the assumption of free expansion and that the material met by the fast shell was not shocked at a prior stage.

\subsubsection{Synchrotron Maser Emission} \label{sec:m6_emission}

\begin{figure}
    \centering
    \includegraphics[width=85mm]{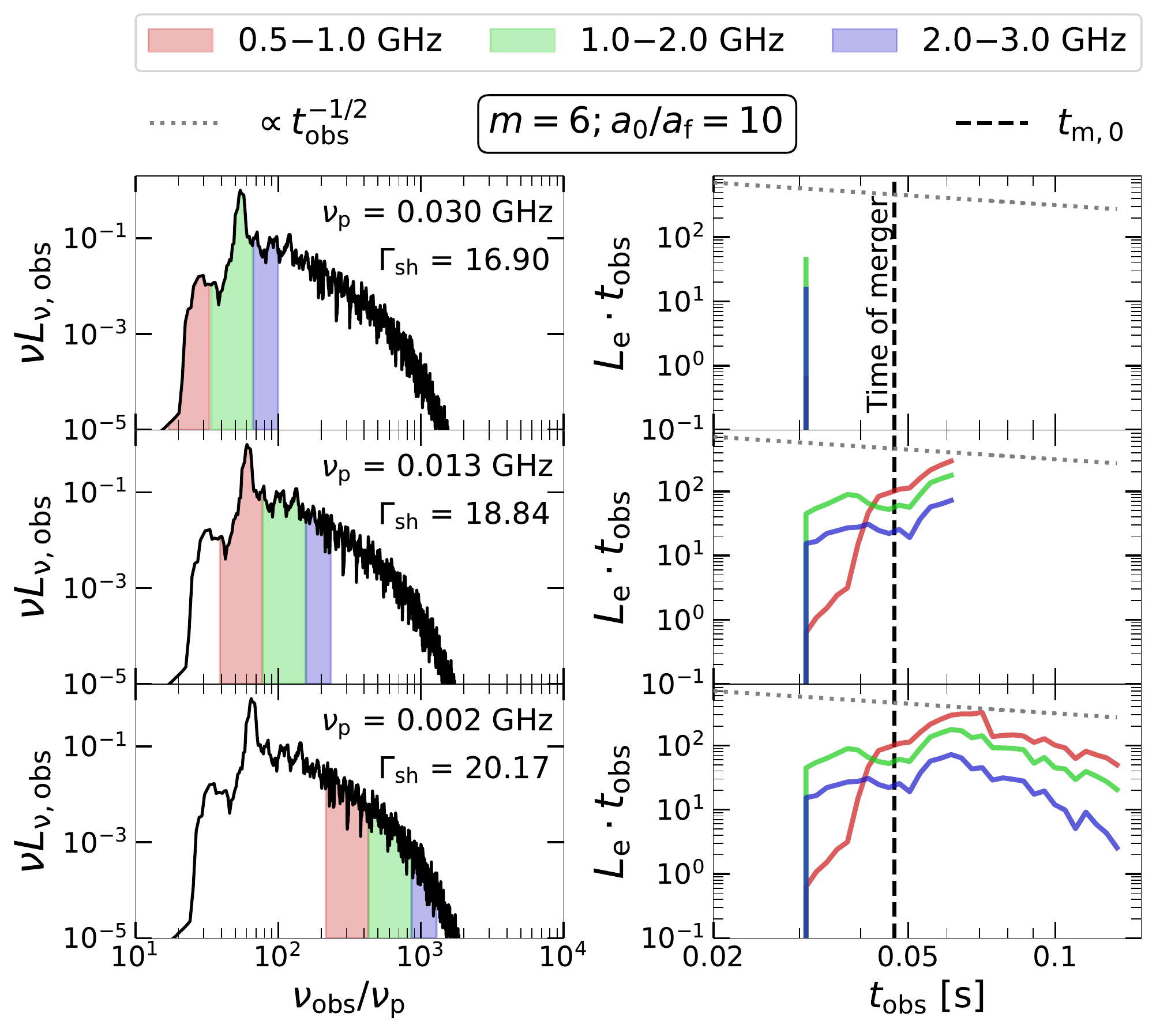}
    \caption{Schematic illustration of how light curves in different radio frequency bands are calculated by post-processing the time-dependent shock properties of our \textit{m6a10} simulation (Fig.~\ref{fig:time_evolution}).  Each row corresponds to a different snapshot in time (top: $t_{\rm obs}\sim0.03$ s, middle: $t_{\rm obs}\sim0.06$ s, bottom: $t_{\rm obs}\sim1.5$ s). \textit{Left column:} The spectral energy distribution $\nu L_{\nu}$ (in arbitrary units) of the synchrotron maser emission from magnetized shocks based on particle-in-cell (PIC) simulations \protect\citep{Plotnikov&Sironi19}, calculated assuming magnetization $\sigma=0.1$ in the inspiral wind.  The frequency axis is normalized to the upstream plasma frequency $\nu_{\rm p}$, whose values at different times are indicated at the top right corner of each panel.  Colored regions denote the contribution of particular frequency bands to the overall spectrum (red: 0.5--1.0 GHz, green: 1.0--2.0 GHz, blue: 2.0--3.0 GHz).  As the shock moves outwards through lower density material, $\nu_{\rm p}$ decreases and the $\nu_{\rm obs}$ moves up the high-frequency tail of the maser SED.  \textit{Right column:} Radio light curve in each frequency band shown on the left (same color coding).  The time of merger is denoted by vertical dashed lines, while the dotted line shows the analytically-predicted power-law decay of the bolometric luminosity (Eq.~\ref{eq:luminosity_1}).}
    \label{fig:lc_PICspectra}
\end{figure}

\begin{figure*}
     \begin{subfigure}[b]{0.49\textwidth}
        \includegraphics[width=0.9\textwidth]{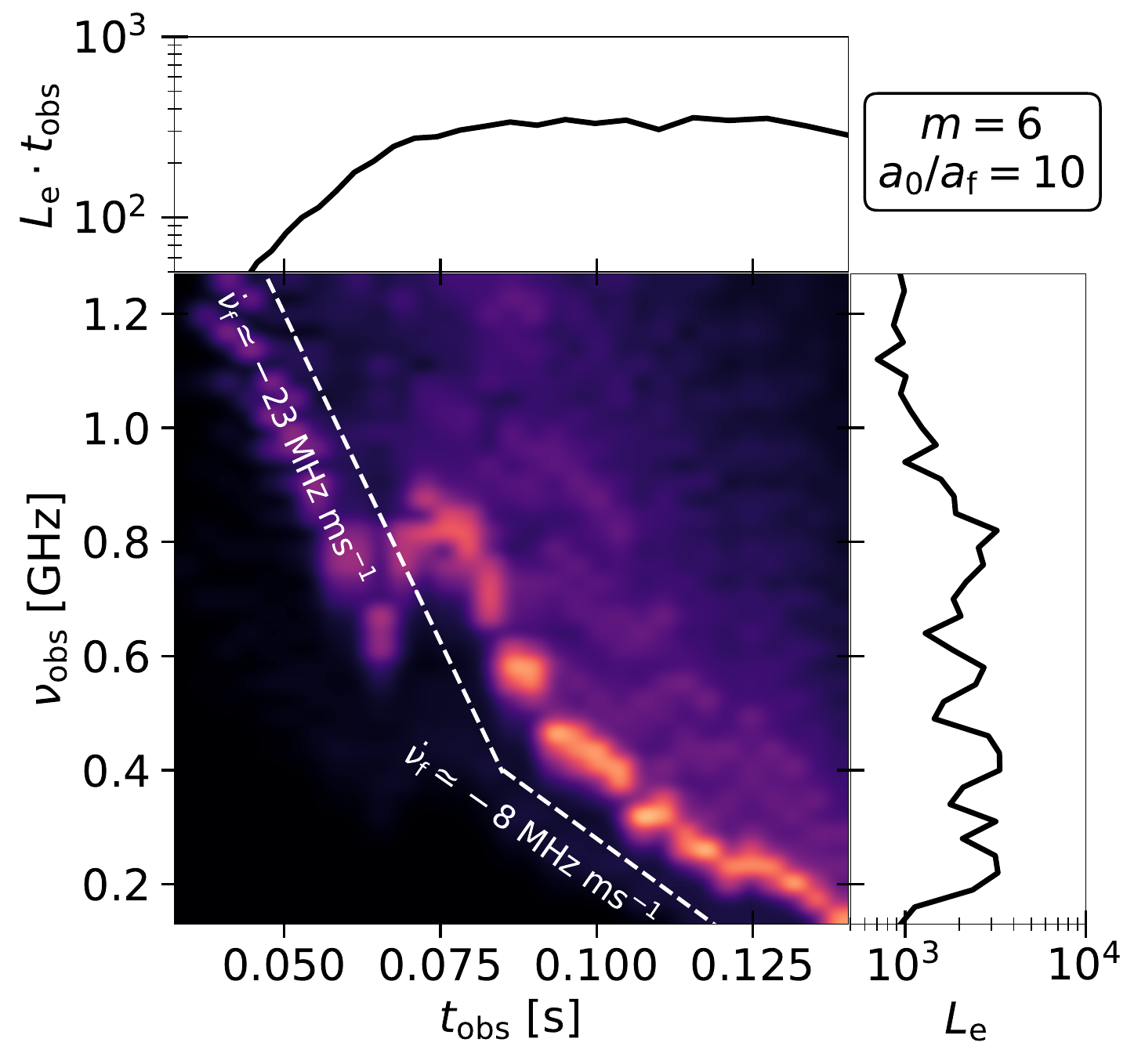}
        \caption{}
        \label{fig:waterfall_m6}
     \end{subfigure}
     \hfill
     \begin{subfigure}[b]{0.49\textwidth}
        \includegraphics[width=0.9\textwidth]{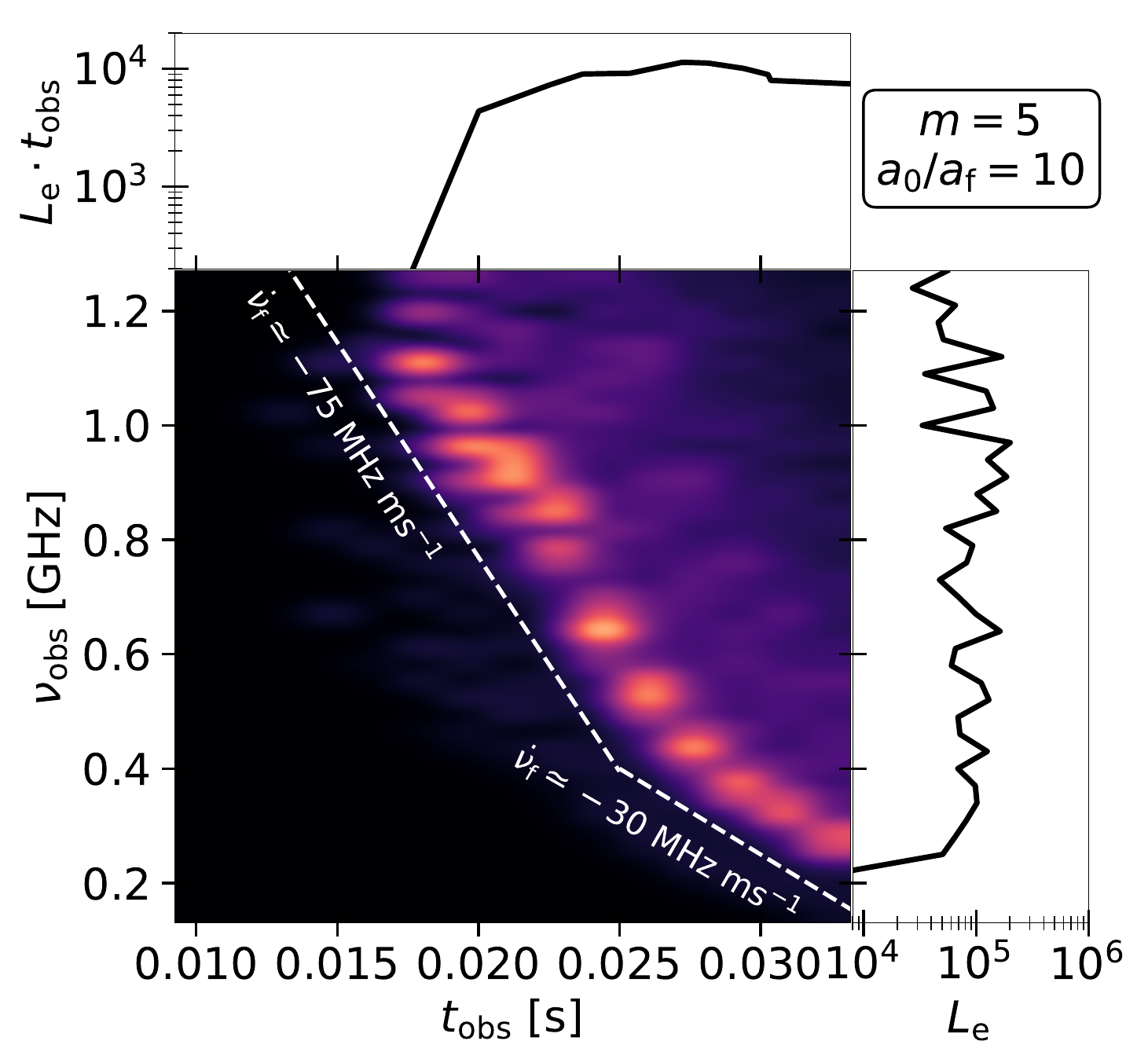}
        \caption{}
        \label{fig:waterfall_m5}
    \end{subfigure}           
    \caption{Synthetic dynamic spectra (`waterfall plots') of FRBs, showing radio flux (in arbitrary units) as a function of observer frame frequency and time for the \textit{m6a10} model---as in  Fig.~\ref{fig:lc_PICspectra}---in the left column (a), and \textit{m5a10} model in the right column (b).  All times indicated are in the post-merger phase.  In columns (a) and (b), the upper panel shows the 0.1--1.3 GHz frequency-integrated radio light curve, and the bottom right panel shows the time-integrated energy spectra.  The white dashed lines represent an approximate frequency drift rate ($\dot{\nu}_{\rm f}$) at different times. Our model does not account for time-of-arrival effects due to different emission across the shock front, which can smear out some of the finer sub-burst features.}
    \label{fig:waterfall}
\end{figure*}

\begin{figure*}
    \centering
    \includegraphics[width=170mm]{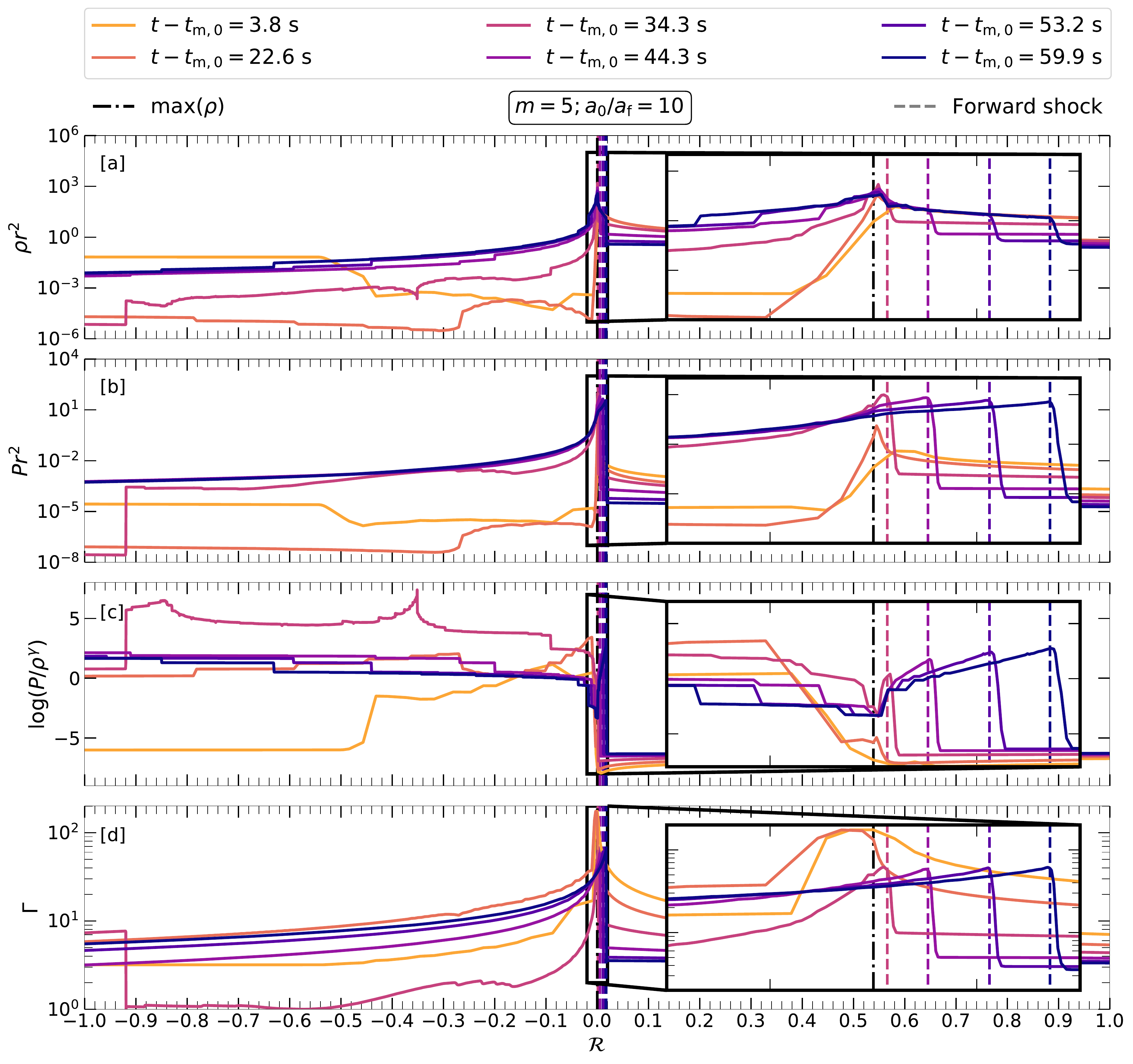}
    \caption{Snapshots of the radial profiles of hydrodynamical quantities demonstrating the self-interaction of the inspiral-driven wind, calculated for the \textit{m5a10} model.  The inset in each panel shows a zoomed-in view around $r_{\rho}$ $(-0.02\le{\cal R}\le0.02)$. A description of all the other quantities illustrated here can be found in Fig. \ref{fig:radial_evol_m6}.}
    \label{fig:radial_evol_m5}
\end{figure*}

Given the FS properties extracted from our simulations, we now describe how this information is used to calculate synthetic radio light curves using the results of particle-in-cell (PIC) simulations of the synchrotron maser emission from magnetized shocks \citep{Plotnikov&Sironi19}.  The efficiency, $f_{\xi}$, and spectral energy distribution (SED) of the maser emission depend on the magnetization of the upstream wind $\sigma$.  We take $\sigma=0.1$ in our calculations to follow, for which $f_{\xi} \sim 10^{-3}$ based on 2D/3D calculations (see Fig.~B2 of \citealt{Plotnikov&Sironi19}).  Neither the maser spectral energy distribution (SED) nor $f_{\xi}$ depend sensitively on $\sigma \lesssim 1$ insofar as its value is not so low ($\sigma \ll 1$) that the shock is still mediated by Larmor gyration of the incoming particles \citep{Iwamoto_17}, as opposed to the Weibel instability \citep{Weibel_59}. 

The radiative efficiency and spectrum of the maser emission can also depend on the temperature of the upstream medium \citep{Babul_Sironi_20}.  In particular, $f_{\xi}$ is drastically suppressed if the upstream medium is relativistically hot, with an internal energy density $\varepsilon \gtrsim 0.1 c^{2}$ \citep{Babul_Sironi_20}.  However, as shown in the bottom panels of Fig. \ref{fig:time_evolution}, we find a cold upstream $\varepsilon \lesssim 10^{-3} c^{2}$ ahead of the forward shock in all of our simulations.   This is not surprising in the $m=6$ case, because the upstream medium is expected to be ``pristine'' (unshocked prior to the arrival of the fast shell) for $m > 5.5$ (Table \ref{tab:regimes}).  However, we find a similarly cold upstream even in our $m < 6$ simulations (\S\ref{sec:lowerm}), as a result of the upstream shocked gas experiencing adiabatic losses prior to the arrival of the fast shell.  We are thus justified in using calculations of the maser emission which assume a cold upstream \citep{Plotnikov&Sironi19}.

The maser SED typically peaks at a Doppler boosted observer frequency $\nu_{\rm pk} \approx 3\Gamma_{\rm sh}\nu_{\rm p}$ where $\nu_{\rm p}$ is the rest-frame plasma frequency of the upstream medium (Eq.~\ref{eq:nuobs} and surrounding discussion) and falls off rapidly at higher frequencies $\gg \nu_{\rm pk}$.  From the time evolution of $n_{\pm}$ and $\Gamma_{\rm sh}$ derived from our simulations (Fig. \ref{fig:time_evolution}), we calculate $\nu_{\rm pk}(t)$ and then use the maser spectrum $\nu L_{\nu}[\nu/\nu_{\rm pk}]$ from \citet{Plotnikov&Sironi19} to calculate the observed luminosity in different frequency bands, where the spectrum is normalized such that $\int_{0}^{\infty} L_{\nu}{\rm d}\nu=f_{\xi}L_{\rm e}$.  Finally, we convert from lab-frame time $t$ to observer time $t_{\rm obs}$ using the Eq. (\ref{eq:t_obs}) in order to generate light curves in each band. This methodology is depicted in the left column of Fig.~\ref{fig:lc_PICspectra}, where different colors denote the position of different radio frequency bands (red: 0.5--1 GHz, green: 1.0--2.0 GHz, blue: 2.0--3.0 GHz) at three snapshots in the shock evolution.  At early times, the higher frequency radio bands overlap  $\nu_{\rm pk}$.  However, as the shock propagates into lower density material, $\nu_{\rm pk} \simeq 3\Gamma_{\rm sh}\nu_{\rm p}$ decreases and crosses lower frequency bands.   The light curve (right column of Fig. \ref{fig:lc_PICspectra}) in a given observing band centered at $\nu_{\rm obs}$ peaks on the timescale $t_{\rm FRB}$ (Eq.~\ref{eq:tFRB}) when $\nu_{\rm obs} \sim \nu_{\rm pk}$ (Eq.~\ref{eq:nuobs}).  The radio light curve peaks earlier at higher frequencies than at lower ones due to the downward sweeping nature of the signal.  

As noted earlier, the wind Lorentz factor in our numerical simulations is limited to values $\lesssim {\cal O}(10^2)$ for numerical reasons.  This would result in values of $\nu_{\rm pk}$ near the end of the inspiral (Eq.~\ref{eq:nupkf}) close to $\sim 10^{15}$ Hz, i.e.~at optical/UV frequencies instead of the radio band.  Thus, to account for more realistic wind Lorentz factors $\Gamma \gg 10$ which make the maser emission accessible to radio telescopes, in Fig.~\ref{fig:lc_PICspectra} we have re-scaled $\Gamma_{\rm f}$ from its normalization in the simulations up a value $\sim 10^{4}$. This scaled-up $\Gamma_{\rm f}$ is chosen such that $\nu_{\rm pk, f} \simeq 1.2$ GHz (see Fig.~\ref{fig:waterfall_m6}). 

Some FRBs are observed to exhibit temporal sub-structure within a given burst, with distinct spectral properties \citep{Farah_18, Gajjar_18}. A useful tool for visualizing such features is the dynamic energy spectrum (`waterfall plot').  The bottom left panel of Fig.~\ref{fig:waterfall_m6} shows a synthetic dynamic spectrum for the example from Fig.~\ref{fig:lc_PICspectra}, which we have calculated using light curves from 40 different frequency bins spanning the 0.1--1.3 GHz range.  Bright (dark) regions denote the moments in time or frequency with enhanced (diminished) radio power.  Due to frequency structure in the maser SED (Fig.~\ref{fig:lc_PICspectra}), the dynamic spectrum likewise shows sub-burst features, seen as narrow pockets of enhanced emission. The time integrated spectrum, illuminating the presence of burst sub-structures is depicted in the bottom-right panel of Fig \ref{fig:waterfall_m6}. The dynamic spectrum is seen to exhibit downward drifting frequency behavior in the burst sub-structure---with a frequency drift rate of $\dot{\nu}_{\rm f}$ of $\approx -23$ MHz ms$^{-1}$ in the 0.4--1.0 GHz range, and  $\dot{\nu}_{\rm f}$ of $\approx -8$ MHz ms$^{-1}$ in the 0.1--0.4 GHz range. We note that this is qualitatively similar to that observed in several repeating FRBs \citep{Hessels+19,CHIME+19,Caleb_20}.  Although the binary neutron star merger precursors studied here are clearly not repeating events, similar physics of an outwardly-propagating shock wave may operate in other central engine models \citep{Metzger+19}.

An important caveat (see \S\ref{sec:caveats} for more) to our above calculations is that we do not include the effects of differential arrival time of emission across the relativistic shock front, which will act to smooth out short temporal structure in the burst \citep{Beniamini&Kumar20}.  
We have also neglected the effects of attenuation of the radio signal by induced Compton scattering in the wind material ahead of the shock \citep{Lyubarsky08}.  Although induced scattering can play an important role in shaping the maser emission in the case of an effectively stationary upstream medium \citep{Metzger+19}, its effects are less severe when the upstream is itself a relativistic wind \citep{Beloborodov20}.

\subsection{Decelerating Fast Shell ($3 \le m \le 5$)}
\label{sec:lowerm}

While the $m=6$ models provide an ideal way to test the results of our numerical simulations against analytic predictions, the coasting nature of the fast shell and a pristine unshocked upstream medium are assumptions that are expected to break down for $m \le 5.5$.  For $m \le 5.5$ the fast shell sweeps up more than its own inertial mass from an upstream medium (which itself has been earlier shocked; Table \ref{tab:regimes}), resulting in a stronger reverse shock and more significant deceleration of the fast shell.  Having established the reliability of our numerical results in the $m > 5.5$ case---with the matching of analytical estimates, we now rely on hydrodynamical simulations alone to explore the shock evolution in the $m < 5.5$ regime.

We choose the $m=5$ case to explore in detail.  Akin to our simulation of the \textit{m6a10} model, we choose an initial binary separation of $a_{\rm 0}/a_{\rm f}=10$ for simulating the inspiral wind interaction in the $m=5$ case, to sufficiently probe the wind interaction for a meaningfully long duration (before it interacts with the initial pulsar wind).  We find qualitatively similar behavior of the wind interaction during the early phases in the $m=3, 4$ models with a shorter initial separation ($a_{\rm 0}/a_{\rm f}=3$; Table \ref{tab:models}), but do not explore these models in depth because they could not be run long enough (large enough $a_{\rm 0}/a_{\rm f}$) to capture the asymptotic behavior of the shock deceleration.\footnote{ 
This is because in models with low mass loadings ($m=3,4$), the maximum value of $\Gamma_{\rm f}\gg {\cal O}(10^2)$ that would be attained can cause the numerical scheme to suffer from a failure to recover the primitive variables from the conserved quantities \citep{Marti_03}.
} 

Figure \ref{fig:radial_evol_m5} shows snapshots in the evolution of the inspiral wind for the \textit{m5a10} model, similar to that previously shown in Fig.~\ref{fig:radial_evol_m6} for the $m=6$ model.  In addition to the FS generated by the fastest shell released at the end of the inspiral (\S\ref{sec:m6}), the upstream wind in the $m=5$ case now contains an additional, slower shock complex generated by the interaction of the inspiral wind with the pre-inspiral (constant power) ``pulsar'' wind.  However, this pre-shock region is too far ahead of the fast shell (${\cal R}>4$) to have any direct influence on the evolution of the dominant FS at early times when most of the shock generated power is being released.  We hereafter focus on the evolution of the FS generated by the fast shell (solid red curves in Fig. \ref{fig:time_evolution}), as this generates the greatest kinetic luminosity.  Less luminous radio emission could occur prior to the end of the inspiral from weaker internal shocks ahead of the final shock, but we do not explore that here.

Unlike the $m=6$ case, where the fast shell rapidly transfers its kinetic energy to the FS, the FS in the $m=5$ case only develops by $t-t_{\rm m,0}\sim23$ s after merger\footnote{This delay is simply a consequence of the larger $\Gamma_{\rm f}\simeq150$ which increases the lab-frame time for shock interaction, $t_{\rm int}\propto\Gamma_{\rm f}^2(a_{\rm 0}/a_{\rm f})^{4/3}$.}, as seen by a small discontinuous peak in the entropy at ${\cal R}=0+\epsilon$ in Fig. \ref{fig:radial_evol_m5} (orange curve; dashed vertical lines denoting FS are not drawn for this snapshot in time due to the proximity of the FS to the black dash-dotted line at ${\cal R}=0$). The radial profile of the gas Lorentz factor reveals the presence of the uniform fast shell prior to $t-t_{\rm m,0} \lesssim 23$ s (yellow curve), with most of it still intact at ${\cal R}<0$, behind the nascent FS.  However, by $t-t_{\rm m,0}=34$ s (light magenta curve), the RS has passed through the head of the fast shell towards its tail, as seen prominently at ${\cal R}\sim-0.9$.  By the next snapshot at $t-t_{\rm m,0}=44$ s, the RS has already reached post-merger wind (${\cal R}<-1$), at which point both the fast shell and FS have been decelerated.  This phase is accompanied by heating of gas behind the FS, as revealed by the increase in the entropy (panel c).  The decelerating shock phase due to RS crossing through the fast shell is shown as a red shaded region in Fig.~\ref{fig:radial_evol_m5}.

The RS crossing phase is followed by a phase of approximately coasting behavior of the shock velocity ($\Gamma_{\rm sh}\sim 80$ at $t\gtrsim 43$ s; red curve in Fig. \ref{fig:time_evolution}b).  However, within this phase, the slope of the upstream density (${\rm d}n_{\pm}/{\rm d}t$) encountered by the FS is seen to change, with a break at $t\sim53$ s. Until this break point ($44\lesssim t \lesssim 53$), the slopes of the peak synchrotron frequency (${\rm d}\nu_{\rm pk}/{\rm d}t$) and the shock fluence (${\rm d}(L_{\rm e}\cdot t)/{\rm d}t$) of the \textit{m5a10} model resembles that of \textit{m6a10} model, albeit with a consistently higher value of both the observables. After the break point (at $t \gtrsim53$ s), upstream density follows a power-law profile, $n_{\pm}\propto t^\alpha$, where the slope is empirically measured to be $\alpha=-9$ (red dotted line in Fig. \ref{fig:time_evolution}a).  This precipitous decrease in $n_{\pm}$ is accompanied by a steeper decline in $\nu_{\rm pk}$ and $L_{\rm e}\cdot t$, each following a power-law profile with a slope of -4.5.  Note that observable properties like $\nu_{\rm pk}$ and $L_{\rm e}\cdot t$ in $m=5$ drop faster than for $m=6$ case.

Finally at times $t-t_{\rm m,0}\gtrsim80$ s, the FS is being influenced by upstream gas which has been shocked by interacting with the pre-inspiral pulsar wind, resulting in an abrupt decline in the upstream density and shock fluence.  
However, as already mentioned, this phase (or at least the timescale on which it occurs in the simulation) is not physical because the shocked material being encountered by the FS is an artifact of the steady pulsar wind assumed as an initial condition prior to the simulated inspiral phase. Therefore, the shock properties from this phase are not reported in our results, and not used for the calculation of the radio emission.

The shock properties extracted for the \textit{m5a10} model from the relativistic hydrodynamical simulation are then convolved with the synchrotron maser SED, to obtain the radio light curves and spectrum, following the procedures detailed in \S\ref{sec:m6_emission}. The dynamical spectrum of the FRB produced by a decelerating shock encountering a pre-shocked inspiral wind is displayed in Fig. \ref{fig:waterfall_m5}.  Here, we see a $\dot{\nu}_{\rm f}$ of $\approx-75$ MHz ms$^{-1}$ in the 0.4--1.0 GHz range, and a $\dot{\nu}_{\rm f}$ of $\approx-30$ MHz ms$^{-1}$ in the 0.1--0.4 GHz range. This presence of a break in the drift frequency rate is qualitatively similar to that seen in $m=6$ case. The faster speed of the FS in $m=5$ case and a steeper decline in the observable quantities ($\nu_{\rm pk}$ and $L_{\rm e}\cdot t$) conspire together to yield a relatively shorter duration FRB, with the peak frequency drifting down at a rate $\gtrsim3\times$ than that of the $m=6$ case.

In summary, the properties of the dominant FS in the $m=5$ case are qualitatively similar to that of the $m=6$ case, with the notable exception of a strong RS which passes through the ejecta (on an observer time $\sim t_{\rm f}$) and abruptly decreases (at least temporarily) the Lorentz factor and kinetic luminosity of the shock.  However, because the shocks are adiabatic in our set-up, this energy is not ``lost'' but may instead be eventually transferred back to the FS; indeed, this may be seen by the mild acceleration of $\Gamma_{\rm sh}$ at $t\gtrsim43$ s (see Fig. \ref{fig:time_evolution}).  After the RS crossing phase, the FS shock properties (relevant to the synchrotron maser emission) appear to decay as power-laws which are steeper than the $m=6$ case.  In a more realistic set-up, we expect that this self-similar phase of shock evolution would last considerably longer than in our simulation, before the fast shell eventually reaches the region of the wind dominated by the single pulsar wind.   

\section{Observational Prospects} \label{sec:discussion}

\begin{figure*}
        \begin{subfigure}[b]{0.5\textwidth}
                \centering
                \includegraphics[width=0.9\textwidth]{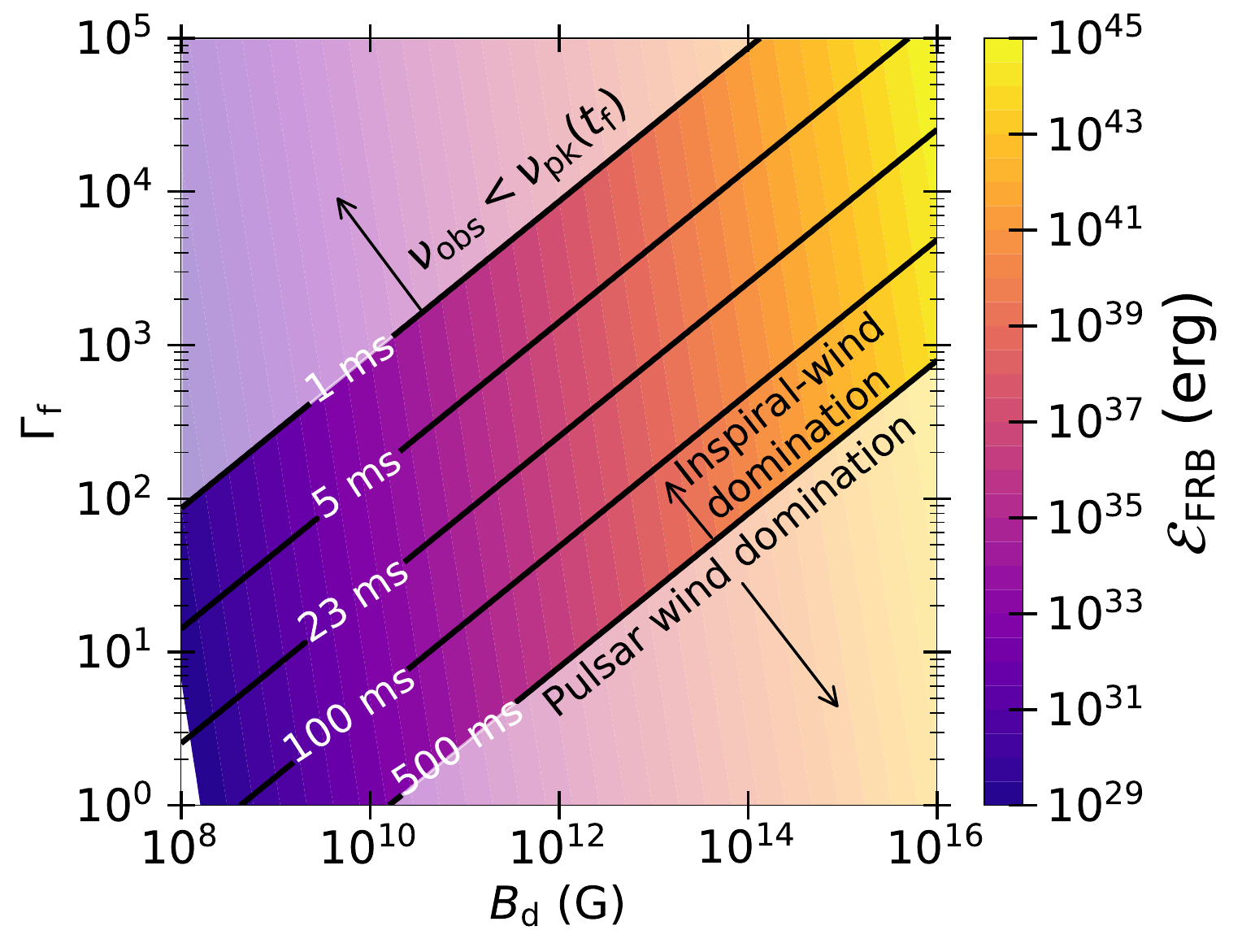}
                \caption{$\nu_{\rm obs}=1.0$ GHz}
        \end{subfigure}%
        \begin{subfigure}[b]{0.5\textwidth}
                \centering
                \includegraphics[width=0.9\textwidth]{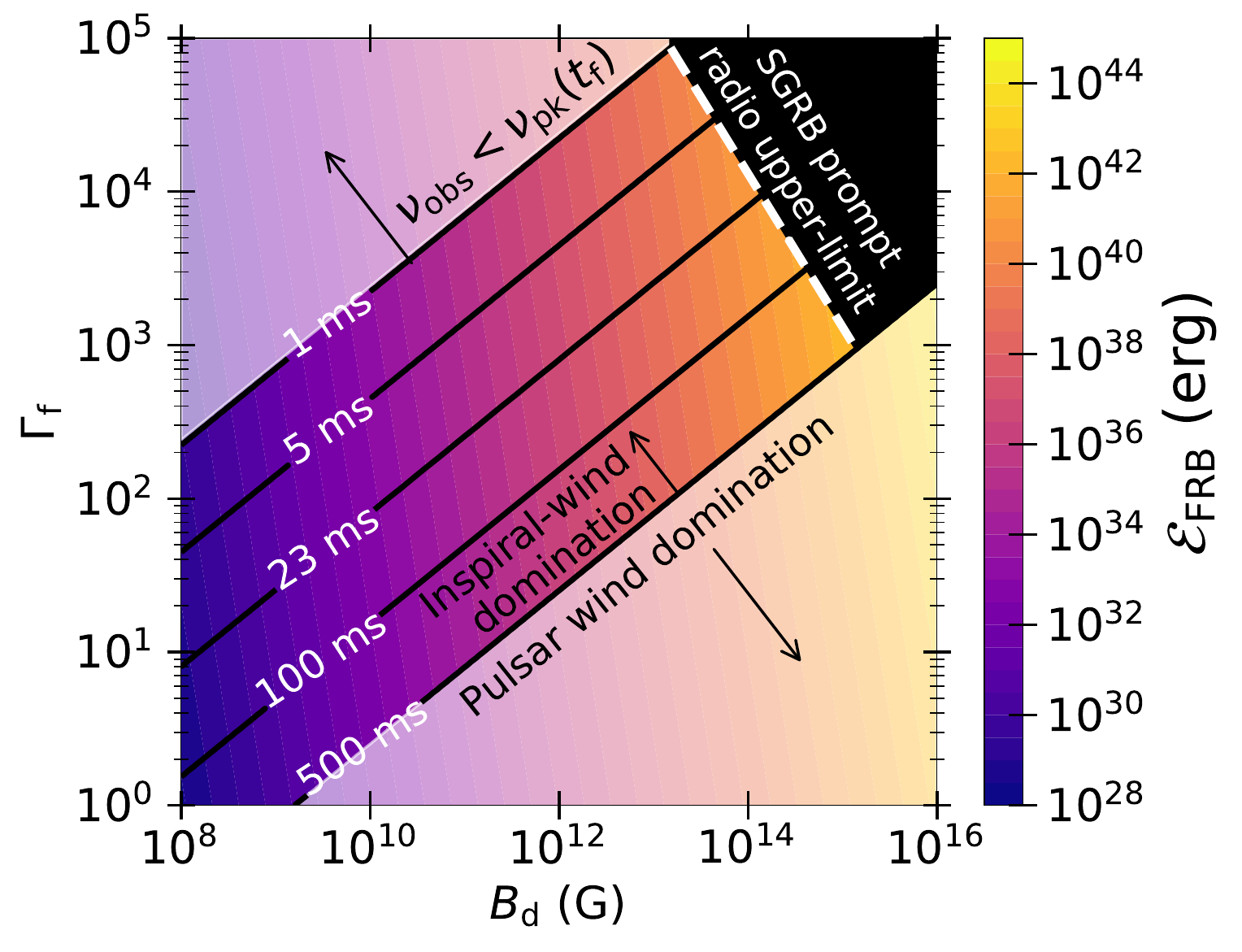}
                \caption{$\nu_{\rm obs}=0.1$ GHz}
        \end{subfigure}%
        \caption{Properties of the radio burst precursor as a function of the surface magnetic dipole field of the neutron star ($B_{\rm d}$; abscissa) and the peak Lorentz factor of the inspiral wind ejecta ($\Gamma_{\rm f}$; ordinate). The energy (${\cal E}_{\rm FRB}$) and the duration of the radio burst ($t_{\rm FRB}$) are depicted by the colored bands and black solid line contours, respectively (Eqs. \ref{eq:EFRB} and \ref{eq:tFRB}). The color-faded regions in the top-left and bottom-right corners of the plot constitute those values of $\Gamma_{\rm f}$ and $B_{\rm d}$, for which no observable FRB emission is possible. The emission with $t_{\rm FRB}\lesssim t_{\rm f}\sim1$ ms is suppressed due to the observing frequency $\nu_{\rm obs}$ being smaller than the peak frequency of the maser spectrum $\nu_{\rm pk}$ (Eq.~\ref{eq:LF_limit}), and at $t_{\rm FRB}\gtrsim500$ ms, the assumptions of our model may break down due to the potential interaction of the fast shell with the isolated pulsar wind (of assumed spin period $P=0.1$ s; Eq.~\ref{eq:t_limit}). The ${\cal E}_{\rm FRB}$ and $t_{\rm FRB}$ contours are calculated assuming a maser efficiency $f_{\xi}=10^{-3}$, geometric beaming fraction $f_{\rm b}=0.1$ and observing frequency $\nu_{\rm obs}=1.0$ GHz (left figure), and $\nu_{\rm obs}=0.1$ GHz (right figure).  The wind mass-loss rate is modelled by a power-law with an index $m=6$, corresponding to a coasting fast shell of ejecta meeting unshocked inspiral wind. The white dashed line in the right panel (b) is set by the observed fluence upper limits to prompt radio emission that accompanied the short gamma ray bursts GRB150424A \protect\citep{Kaplan+15} and GRB170112A \protect\citep{Anderson+18}.}
\label{fig:limitations}
\end{figure*}

This section discusses observational prospects for detecting the shock-powered radio precursor emission from neutron star mergers and addresses some caveats of our results.

\subsection{Parameter space of radio transients}
\label{discussion_wind}

One of the biggest uncertainties in our proposed scenario is the dependence of the wind mass-loading on binary separation (or, equivalently, time until merger).  It is fortunate then that our numerical results for the time-dependent forward shock properties (Fig.~\ref{fig:time_evolution}) show broadly similar evolution in both the freely-expanding ($5.5 < M < 7$) and decelerating ($3 \le m < 5.5$) cases.  In what follows, we apply our analytic results for the $m=6$ case (\S\ref{sec:shocks}) to explore the range of shock-powered emission.  Qualitatively similar results should apply to the more general $m < 7$ cases.

Figure \ref{fig:limitations} shows the energy $\mathcal{E}_{\rm FRB}$ (Eq.~\ref{eq:EFRB}) and duration $t_{\rm FRB}$ (Eq.~\ref{eq:tFRB}) of the predicted radio emission (at fiducial observing frequencies $\nu_{\rm obs}$ = 0.1, 1 GHz) as a function of two unknowns: the surface magnetic field $B_{\rm d}$ of the more strongly magnetized neutron star and the final Lorentz factor $\Gamma_{\rm f}$ of the binary wind (related to the uncertain wind mass-loading).  We also show the constraints imposed on $\Gamma_{\rm f}$ and the emission duration (1 ms $\lesssim t_{\rm FRB} \lesssim$ 500 ms) by the conditions needed to produce an observable burst ($\nu_{\rm obs} > \nu_{\rm pk,f}$; Eq.~\ref{eq:LF_limit}) and for the binary wind to interact with itself instead of the single pulsar wind (Eq.~\ref{eq:t_limit}).

The burst energy depends most sensitively on the neutron star magnetic field, ranging from $\mathcal{E}_{\rm FRB} \sim 10^{34}$ erg for a weakly magnetized pulsar ($B_{\rm d} \sim 10^{11}$ G) to $\mathcal{E}_{\rm FRB} \gtrsim 10^{41}$ erg for a magnetar primary ($B_{\rm d} \gtrsim 10^{14}$ G).  These energies, as well as the range of burst durations $t_{\rm FRB} \sim 1-500$ ms, broadly overlap those of observed FRBs (e.g.,~\citealt{Thornton+13,Bochenek+20,CHIME+20}).  The required range of Lorentz factors ($\Gamma_{\rm f} \sim 10^{2}-10^{5}$ for $B_{\rm d} \sim 10^{11}-10^{15}$ G) to produce an observable signal, requires mass-loading of the binary wind corresponding to values of the effective pair multiplicity $\mu_{\pm} \sim 10^{6}-10^{11}$ (Eq.~\ref{eq:eta}), which is higher than in rotational-powered pulsar winds (e.g.,~\citealt{Timokhin&Harding19}). However, note that this pair multiplicity is calculated during the final moments before merger ($a\lesssim2R_{\rm ns}$), whence the existence of additional dissipation processes and high compactness of the binary interaction region near the final inspiral could substantially enhance pair creation (e.g.,~\citealt{Metzger&Zivancev16}).

\subsection{Detection rates and strategies} \label{discussion_detection}

There are several strategies for detecting the radio counterparts we have described.  Firstly, they could be discovered ``blindly'' by existing surveys (and, indeed, may already be present in FRB samples).  A radio telescope with a fluence sensitivity of $F_{\rm lim}$ at $\nu_{\rm obs}=1$ GHz can detect a burst of energy $\mathcal{E}_{\rm FRB}$ (Eq.~\ref{eq:EFRB}) out to a distance
\begin{eqnarray}
D_{\rm rad} &\simeq& \left(\frac{\mathcal{E}_{\rm FRB}}{4\pi F_{\rm lim}}\right)^{1/2} \nonumber \\
&\approx& 3.2 \,{\rm Gpc}\left(\frac{f_{\xi,-3}^{1/2}}{f_{\rm b,-1}^{1/2}}\right)\left(\frac{F_{\rm lim}}{1\,\rm Jy\cdot ms}\right)^{-1/2}\left(\frac{B_{\rm d}}{10^{12}{\rm G}}\right)^{8/9}\left(\frac{\Gamma_{\rm f}}{10^{3}}\right)^{2/9},
\end{eqnarray}
where we have used Eq. (\ref{eq:EFRB}).  Given the local volumetric rate of NS-NS mergers inferred by LIGO/Virgo of $\mathcal{R} \approx 320_{-240}^{+490}$ Gpc$^{-3}$ yr$^{-1}$ \citep{LVC2020population}, the all-sky rate of NS-NS merger-associated FRBs above $F_{\rm lim}={1\,\rm Jy\cdot ms}$ is estimated to be (for $f_{\xi}=10^{-3}$)
\begin{eqnarray}
\dot{N}_{\rm rad} &\approx& \frac{4\pi}{3}f_{\rm b}D_{\rm rad}^{3}\mathcal{R} \nonumber \\
&\approx& 12\,{\rm day^{-1}}\left(\frac{f_{\rm b}}{0.1}\right)^{-1/2}\left(\frac{\cal R}{320{\rm Gpc^{-3}yr^{-1}}}\right)\left(\frac{B_{\rm d}}{10^{12}{\rm G}}\right)^{8/3}\left(\frac{\Gamma_{\rm f}}{10^{3}}\right)^{2/3},
\label{eq:Ndot}
\end{eqnarray}
where we have assumed Euclidean cosmology, neglected redshift effects on the FRB properties, and $B_{\rm d}/\Gamma_{\rm f}$ should be interpreted as average values.  A similar rate estimate can be made for radio emission through BH-NS merger channel, for which we would predict similar emission properties to the NS-NS case.  Making similar assumptions, e.g., for $B_{\rm d}$ of the NS and considering a BH-NS merger rate upper limit of $\mathcal{R}_{\rm BHNS} \lesssim 610$ Gpc$^{-3}$ yr$^{-1}$ provided by the LIGO/Virgo O2/O3 observations \citep{LIGO+19}, we estimate $\dot{N}_{\rm rad}\lesssim 23~{\rm day}^{-1}$ via BH-NS merger channel. The total FRB detection rate---through the mechanism proposed in this work---via BH-NS and NS-NS merger channel of $\dot{N}_{\rm rad}\lesssim 35~{\rm day}^{-1}$ is still a small fraction ($\sim 1\%$) of the total all-sky FRB rate above a few Jy $\cdot$ ms \citep{Ravi_19} of $\sim 10^{3}-10^{4}$ day$^{-1}$ \citep{Champion+16,Bhandari+18}.

Due to their different beaming patterns, mergers that produce detectable FRB emission may not be accompanied by a GRB (see below).  However, other post-merger counterparts which generate more isotropic emission---such as an optical/infrared kilonova (e.g., \citealt{Metzger+10}) or a radio afterglow (e.g., \citealt{Nakar&Piran11})---are potentially more promising.  It is not currently feasible to follow-up a large enough sample of FRBs at optical or radio wavelengths to discover such counterparts, given the large distances $D_{\rm rad} \gtrsim $ Gpc of the sources and the small fraction of FRBs which may be associated with this formation channel.  However, the sky positions of many FRBs will be followed up automatically by future deep surveys, such as the Vera C.~Rubin Observatory and the Square Kilometer Array, enabling a more systematic search. 

In addition to producing coherent radio emission, the shocks we have described will heat pairs to relativistic temperatures, simultaneously generating (incoherent) synchrotron radiation (e.g.,~\citealt{Metzger+19}).  For an $e^{\pm}$-pair upstream this emission will peak at optical or X-ray wavelengths \citep{Beloborodov20}, though can occur in the gamma-ray range if the upstream is baryon loaded.  However, even assuming the entire energy of the inspiral wind $\sim E_{\rm f}$ is converted into gamma-rays ($\sim10^{-8}-10^{-9}$ erg cm${^{-2}}$), such emission is only detectable by wide-field facilities like {\it Fermi} Gamma-ray Burst Monitor (GBM) to a few tens of Mpc \citep{Metzger&Zivancev16}.  The narrow fields of view of sensitive X-ray and optical telescope facilities also likely preclude them detecting thermal synchrotron emission from the FRB-generating shocks we have described.

Radio precursor bursts could also be discovered via targeted follow-up of NS-NS or BH-NS mergers which are first detected through their GW emission. The sky-position and orientation-averaged NS-NS detection range during the upcoming LIGO/Virgo O4 run is expected to be $D_{\rm GW} \simeq 150$ Mpc \citep{LVK_2018}.  However, due to the nature of the GW beaming pattern, the range for the approximately edge-on systems (those predicted to give rise to the brightest emission in our scenario; Fig.~\ref{fig:cartoon}) is smaller than average, 
\be \label{eq:DGW_edge}
D_{\rm GW,edge} \approx f_{\rm b,GW}D_{\rm GW} \sim 75 {\rm Mpc}\left(\frac{f_{\rm b,GW}}{0.5}\right),
\ee
where $f_{\rm b,GW}$ factors-in the effect of inspiral orientation on the GW beaming pattern in determining the NS-NS detection range, and ranges from 0.5--1.5 for inspiral orientations between edge-on and face-on, respectively \citep{Schutz11, Metzger&Berger12}.  Given that $D_{\rm rad} > D_{\rm GW,edge}$ for a range of plausible magnetic fields strengths $B_{\rm d} \sim 10^{11}-10^{12}$ G, the rate of GW events accompanied by an FRB signal could be as high as \footnote{We do not include the rate of GW events from BH-NS merger channel accompanying FRBs due to the uncertain nature of GW190814, and the low significance of GW190426\_152155 and other BH-NS candidates.}
\begin{eqnarray} \label{eq:N_GW}
\dot{N}_{\rm GW} &\approx& \frac{4\pi}{3}D_{\rm GW, edge}^{3}\mathcal{R} \nonumber \\
&\lesssim& 1\,{\rm year^{-1}}\left(\frac{D_{\rm GW,edge}}{75 {\rm Mpc}}\right)^3\left(\frac{\cal R}{320{\rm Gpc^{-3}yr^{-1}}}\right).
\end{eqnarray}

The predicted FRB signal (of duration $t_{\rm FRB}$) starts roughly simultaneously with the end of the GW inspiral  (Figs.~\ref{fig:time_evolution}, \ref{fig:lc_PICspectra}) (though some radio emission can start before the merger due to stronger internal shocks between earlier stages of the inspiral wind).  However, the FRB signal will arrive on Earth delayed relative to the GWs by an additional amount, $\Delta t_{\rm GW}$, due to the finite propagation speed of radio waves through ionized plasma,
\be
\Delta t_{\rm GW} \approx \left(\frac{e^2}{m_{\rm e}c\nu_{\rm obs}^2}\right)\int_{\rm 0}^{\rm d}n_{\rm e}{\rm d}s
\simeq 0.42\,{\rm s}\left(\frac{\nu_{\rm obs}}{1{\rm GHz}}\right)^{-2}\left(\frac{{\rm DM}}{10^2 {\rm pc~cm}^{-3}}\right),
\ee
where DM $=\int_{\rm 0}^{\rm d}n_{\rm e}{\rm d}s$ is total dispersion measure of the burst along the path of propagation, $n_{\rm e}$ is the free electron density and $d$ the source distance.  The DM in general includes contributions from the Milky Way ISM \citep[$\sim$40 pc cm$^{-3}$ at high Galactic latitudes;][]{Cordes_Lazio_02}, its halo \citep[50--80 pc cm$^{-3}$;][]{Prochaska_Zheng_2019}, inter-galactic medium \citep[24 pc cm$^{-3}\left(z/0.02\right)$, where $z$ is the redshift corresponding to $D_{\rm GW,edge}\sim75$ pc cm$^{-3}$;][]{Ioka_2003,Inoue_2004,Lorimer+07,Karastergiou_2015}, and the local host ISM and halo.  These local DM contributions including from the inspiral wind itself is considered negligible.  Given that ${\rm DM}[D_{\rm GW,edge}]\gtrsim 115$ pc cm$^{-3}$, we expect the propagation delay to be $\Delta t_{\rm GW}[D_{\rm GW,edge}] \gtrsim 0.48$ s $\gtrsim t_{\rm FRB}$ in some circumstances. 

Radio wavelength coverage of GW sky regions on timescales of seconds or less after the end of the GW inspiral is currently possible only at low frequencies $\sim 100$ MHz due to the very large field of dipole antenna arrays (such as OVRO-LWA; \citealt{Anderson+19}) which cover a large fraction of the sky and hence can catch FRB counterparts to GW events serendipitously.  However, rapid GW-triggered follow-up is not yet feasible at higher radio frequencies $\sim 1$ GHz where telescopes are significantly more sensitive.  In the future this technique may be more promising, particularly given improvements in technology and planned improvements in GW detection pipelines to provide shorter---or even negative---latency GW detections (e.g., \citealt{James+19}). 

Finally, one could also target short gamma-ray bursts (GRB) to search for potential early radio counterparts of NS-NS or NS-BH merger.  The right panel of Fig.~\ref{fig:limitations} shows upper limits on the energy of the putative radio burst $\mathcal{E}_{\rm FRB}$ associated with two short GRBs for which prompt radio follow-up observations were conducted (\citealt{Kaplan+15,Anderson+18}; see also \citealt{Rowlinson&Anderson19,Gourdji+20,Rowlinson+20}).  The observed upper-limit on radio fluence, with a range of ad-hoc burst durations, can be used alongside Eqs. (\ref{eq:tFRB},\ref{eq:EFRB}) to obtain the upper limits on the magnetic field of the neutron star $B_{\rm d}$, and the Lorentz factor of the fastest shell $\Gamma_{\rm f}$, as per our model.  The dark region delimited by the white dashed line in Fig. \ref{fig:limitations}(ii) corresponds to those values of the intrinsic parameters $B_{\rm d}$ and $\Gamma_{\rm f}$, that are not allowed by the observed radio fluence upper limits, for different burst durations. Depending on the calculated value of $\Gamma_{\rm f}$, these limits appear to rule out NS-NS mergers containing magnetars ($B_{\rm d} \gtrsim 10^{14}-10^{15}$ G) for the considered observations.  However, due to geometric effects, the FRB emission we predict is unlikely to be detectable in coincidence with a GRB.  The binary wind---and hence its shock-generated radio emission---is expected to be focused along the binary orbital plane (Fig.~\ref{fig:cartoon}), well outside the direction of the GRB jet (typically $\lesssim 10^{\circ}$ of the binary rotational axis; e.g., \citealt{Berger14}). 

\subsection{Caveats and Uncertainties}
\label{sec:caveats}

We now discuss several caveats and uncertainties that potentially affect our conclusions.  

\begin{itemize}

    \item \textit{Energy dissipation:} One uncertainty in our assumption is that a large fraction of the total energy dissipated by magnetic interaction between the binary stars emerges in the form of an ordered outflow, as opposed to dissipation close to the stars.  \cite{Palenzuela+13} find that the total dissipated energy is nearly equally partitioned in between the radiated Poynting energy and Joule heating (acceleration) of particles. Note that this is in the most probable configuration (considered here) of a neutron star binary with individual dipole moments along the same direction with widely different magnitudes.  However, the constituents of the binary can as well have misaligned dipole moments, in which case, a larger fraction of the ejected energy will be in the form of Poynting energy compared to heating up of particles.  The energy which is dissipated in the form of charged particle acceleration in the magnetosphere may generate an initially opaque pair fireball, which expands to several times the binary radius before becoming transparent, releasing a prompt burst of $\sim$ MeV gamma-rays \citep{Metzger&Zivancev16}. In this work, we have not considered any potential impact of photon-loading at small scales on the relativistic wind which generates internal shocks at much larger radii.  

    \item \textit{Dimensionality:} The relativistic hydrodynamical simulations of interacting inspiral wind in this work are performed in 1D (along the radial outflow direction).  The emission from the forward shock can be expected to be visible from a range of viewing angles, which depends upon the beaming angle of the outflow.  This corresponds to the solid angle of the fast shell ejected at the end of the inspiral, which requires multi-dimensional simulations to accurately capture the subtended solid angle.  In this work, we parametrize the uncertainty associated with the emission solid angle by a geometric beaming factor, $f_{\rm b}$ set to a constant 10\% (Eqs. \ref{eq:iso_luminosity}, \ref{eq:DGW_edge}).  However, we note that the outflow can, in principle, subtend a different solid angle at earlier times (smaller binary separation), and therefore the isotropic luminosity of the wind (along the direction of the FRB set by the fast shell) might be changing in a way different than just $\dot{E}$---potentially due to evolving $f_{\rm b}$.
   
    \item \textit{Magnetic fields:} We have neglected the dynamical impact of magnetic fields on the shock properties. Although this is a reasonable assumption if the wind has solved its ``$\sigma$ problem'' (see \S\ref{sec:introduction}) by the radius of the shock interaction, this is not guaranteed to be the case.  If instead the magnetization of the wind remains $\sigma \gg 1$ then the shocks will be substantially weaker \citep{Zhang_Kobayashi_05, Giannios_08, Mizuno_09, Mimica_10} and the efficiency of the maser emission will drop \citep{Plotnikov&Sironi19}.  It is still possible in this case that forced reconnection would occur in the outflow, liberating the magnetic energy and generating coherent radio emission through another mechanism (e.g.,~\citealt{Lyubarsky20,Most&Philippov20}).
   
    \item \textit{Mass loading:} We have assumed that the mass-loading of the binary wind increases with decreasing binary separation $\dot{M} \propto a^{-m}$, and increases slower than the wind power $\dot{E} \propto a^{-7}$, i.e., $m < 7$, such that the wind Lorentz factor increases in time, giving rise to shocks.  Although the na\"ive Goldreich-Julian scaling ($m=3$) satisfies this constraints, other plausible assumptions (e.g., $m\ge7$, where a fixed or growing fraction of $\dot{E}$ goes into pair rest-mass) would not. However, we note that the issue of wind acceleration is potentially coupled to the above issue of the wind magnetization.  If the wind is only partially able to accelerate by the radius of internal shocks (e.g.,~via reconnection between ``stripes'' of alternating magnetic polarity;  \citealt{Lyubarsky&Kirk01,Drenkhahn&Spruit02}), then the 4-velocity should scale inversely with the radial length scale of the stripes.  In ordinary pulsar winds (with misaligned rotational and magnetic axes), the stripe length scales with the light cylinder radius.  However, in the binary wind case the stripes may instead scale with the size of the binary orbit $\sim a$.  Since the latter decreases approaching the merger, all else being equal, the velocity achieved by the wind at a fixed distance from the binary should grow in time.

    \item \textit{Radiative cooling and instabilities:} We neglect the effects of radiative cooling of the shock-heated electron/positrons on the dynamics of the shock interaction in the inspiral wind, despite the fact that synchrotron cooling is likely to be rapid on the dynamical timescale \citep{Metzger+19,Beloborodov20}.  Given that the FS dynamics is largely driven by ballistic motion of the fast shell, the presence of cooling should not qualitatively affect its evolution relative to the predictions of the adiabatic case.  However, we cannot exclude that fast cooling generates instabilities that negatively impact the production of the maser emission. For example, \cite{Duffell_14} demonstrate that a shock front experiences Rayleigh-Taylor corrugations and significant turbulence behind it in the presence of cooling---which they approximate with a softer equation of state ($\gamma < 4/3$).  While we leave the full-fledged relativistic magnetohydrodynamic modelling of inspiral winds with radiative cooling effects to a future work, we speculate that such instabilities at the shock interface can generate short timescale variability in the orientation of the upstream magnetic field and hence in the polarization \citep{Nimmo_20}. 
   
    \item  \textit{Additional emission:} We have assumed that secular timescale evolution of the wind properties gives rise to the shock emission.  If the wind properties are instead highly variable on timescales much shorter than the GW inspiral time (e.g.,~\citealt{Most&Philippov20}), then shock interaction between the ejecta from discrete ``flares'' (or from discrete flare ejecta propagating into the otherwise quasi-steady binary wind) would provide an additional source of radio emission.

\end{itemize}

\section{Conclusions} \label{sec:conclusion}

We have proposed a concrete mechanism to transform a significant fraction (up to $f_{\xi} \sim 10^{-3}$) of the total Poynting energy released by the gravitational wave inspiral of a magnetized compact object binary (NS-NS or BH-NS) into a burst of coherent radio emission.  The predicted radio emission---with a typical duration of $1\sim500$ ms---exhibits properties similar to the observed FRBs.  Several past works have proposed mechanisms for generating coherent radio emission from NS mergers.  Here we go beyond most previous efforts by simulating the first dynamic spectrum (and radio light curves) of an FRB to our knowledge from first principles, and go on to make specific and quantitative predictions for the luminosity, temporal-frequency evolution, geometric beaming of the signal relative to the binary orientation, and prospects of observing them.  

In the paradigm proposed here, the power and the speed of a magnetically dominated pulsar's wind is enhanced secularly during the final stages of merger (Eq. \ref{eq:Edot2}).  This enhancement gives rise to shocks via self-interaction of the inspiral wind in the outflowing ejecta---whose properties are contingent on the uncertain mass-loading in the wind (\S\ref{sec:windproperties}), $\dot{M} \propto a^{-m}$ that we parametrize by $m$.  The wind pair loading is challenging to calculate from first principles, even in the case of ordinary pulsar winds.  Nevertheless, insofar as $m < 7$, the qualitative features of the fast shell generated at the end of the inspiral, and the resulting synchrotron maser emission, are relatively robust.  We have identified different regimes of $m$ which result in qualitatively different strength of the reverse shock and behavior during the deceleration process (Table \ref{tab:regimes}).  However, all of these regimes ultimately result in the transfer of the majority of the energy released by the end of the inspiral into a forward shock, which is then available to generate synchrotron emission sweeping across a relatively wide range of radio frequencies.  

We analytically derive (\S\ref{sec:shocks})---and confirm via hydrodynamical simulations (\S\ref{sec:m6})---the properties of the forward shock and the radio emission for a wind model that gives raise to a coasting fast shell.  For coasting shock case, the properties of the radio burst (e.g., energy ${\cal E}_{\rm FRB}$, duration $t_{\rm FRB}$) are also derived as a function of the central engine's intrinsic parameters---the dipole field ($B_{\rm d}$) of the NS and the ejecta speed $\Gamma_{\rm f}$ (\S\ref{sec:example_m6}).  The wind models that give rise to decelerating FS, on the other hand, are explored solely via hydrodynamical simulations (\S\ref{sec:lowerm}).  The properties of the FS extracted from the hydrodynamical simulations are used alongside the spectra of synchrotron maser emission obtained from particle-in-cell simulations, in order to simulate the generated FRB in various radio bands, for the case of a coasting fast shell model (\S\ref{sec:m6_emission}), and a decelerating forward shock model (\S\ref{sec:lowerm}) independently.

For a strong---but not unreasonable---primary magnetic field strength $B_{\rm d} \gtrsim 10^{12}$ G, we predict the burst fluence at $\sim$GHz band to be sufficiently bright to be detected to $\sim$Gpc distances by existing radio survey telescopes. Indeed, such merger counterparts could already be lurking in existing FRB samples, as mergers can account for $\sim$1\% ($\lesssim 40~{\rm day}^{-1}$) of the total all-sky FRB rate (Eq.~\ref{eq:Ndot}).  Out of these bursts, a few per year are predicted to be contemporaneous with gravitational wave detections (Eq. \ref{eq:N_GW}).  FRBs from this channel could be identified as a subset of non-repeating class of FRBs which exhibit downward-drifting frequency structure (Fig.~\ref{fig:waterfall}), similar to that seen thus far exclusively from recurring FRBs \citep{Hessels+19}.  This is not a coincidence because the mechanism for generating coherent radio emission proposed here for NS mergers from magnetized relativistic shocks is closely related to that which are proposed in magnetar flares as an explanation for repeating FRBs (e.g.,~\citealt{Beloborodov+18,Metzger+19}).  Host galaxy demographics may also help to distinguish an FRB subpopulation arising from NS-NS or BH-NS mergers.  While growing evidence shows that the majority of FRBs trace star-formation (e.g.~\citealt{Bhandari+20,Heintz+20}), neutron star mergers---due to natal birth kicks and a long delay time until merger---can be traced to various galaxy types (e.g., \citealt{Belczynski_06,Margalit+19}) with a wide range of expected offsets---both spatial and temporal---from star-formation.

\section*{Acknowledgements}
We thank Sasha Philippov for helpful comments and conversations. NS acknowledges support from Columbia University Dean's fellowship.  BDM acknowledges support from the NSF (grant AST-2002577) and the Simons Foundation (grant 606260). LS acknowledges support from the Cottrell Scholar Award, the Sloan Fellowship, and NASA ATP 80NSSC18K1104.

\section*{Data availability}
The data underlying this article will be shared on reasonable request to the corresponding author.

\bibliographystyle{mnras}
\bibliography{references}

\begin{thebibliography}{}
\makeatletter
\relax
\def\mn@urlcharsother{\let\do\@makeother \do\$\do\&\do\#\do\^\do\_\do\%\do\~}
\def\mn@doi{\begingroup\mn@urlcharsother \@ifnextchar [ {\mn@doi@}
  {\mn@doi@[]}}
\def\mn@doi@[#1]#2{\def\@tempa{#1}\ifx\@tempa\@empty \href
  {http://dx.doi.org/#2} {doi:#2}\else \href {http://dx.doi.org/#2} {#1}\fi
  \endgroup}
\def\mn@eprint#1#2{\mn@eprint@#1:#2::\@nil}
\def\mn@eprint@arXiv#1{\href {http://arxiv.org/abs/#1} {{\tt arXiv:#1}}}
\def\mn@eprint@dblp#1{\href {http://dblp.uni-trier.de/rec/bibtex/#1.xml}
  {dblp:#1}}
\def\mn@eprint@#1:#2:#3:#4\@nil{\def\@tempa {#1}\def\@tempb {#2}\def\@tempc
  {#3}\ifx \@tempc \@empty \let \@tempc \@tempb \let \@tempb \@tempa \fi \ifx
  \@tempb \@empty \def\@tempb {arXiv}\fi \@ifundefined
  {mn@eprint@\@tempb}{\@tempb:\@tempc}{\expandafter \expandafter \csname
  mn@eprint@\@tempb\endcsname \expandafter{\@tempc}}}

\bibitem[\protect\citeauthoryear{{Abbott} et~al.,}{{Abbott}
  et~al.}{2017a}]{LIGO+17DISCOVERY}
{Abbott} B.~P.,  et~al., 2017a, \mn@doi [Physical Review Letters]
  {10.1103/PhysRevLett.119.161101}, \href
  {http://adsabs.harvard.edu/abs/2017PhRvL.119p1101A} {119, 161101}

\bibitem[\protect\citeauthoryear{{Abbott} et~al.,}{{Abbott}
  et~al.}{2017b}]{LIGO+17CAPSTONE}
{Abbott} B.~P.,  et~al., 2017b, \mn@doi [\apjl] {10.3847/2041-8213/aa91c9},
  \href {https://ui.adsabs.harvard.edu/abs/2017ApJ...848L..12A} {848, L12}

\bibitem[\protect\citeauthoryear{{Abbott} et~al.}{{Abbott}
  et~al.}{2017c}]{LIGO+17FERMI}
{Abbott} B.~P.,  et~al., 2017c, \mn@doi [\apjl] {10.3847/2041-8213/aa920c},
  \href {https://ui.adsabs.harvard.edu/abs/2017ApJ...848L..13A} {848, L13}

\bibitem[\protect\citeauthoryear{{Abbott} et~al.,}{{Abbott}
  et~al.}{2018}]{LVK_2018}
{Abbott} B.~P.,  et~al., 2018, \mn@doi [Living Reviews in Relativity]
  {10.1007/s41114-018-0012-9}, \href
  {https://ui.adsabs.harvard.edu/abs/2018LRR....21....3A} {21, 3}

\bibitem[\protect\citeauthoryear{{Abbott} et~al.,}{{Abbott}
  et~al.}{2019}]{LIGO+19}
{Abbott} B.~P.,  et~al., 2019, \mn@doi [\apjl] {10.3847/2041-8213/ab3800},
  \href {https://ui.adsabs.harvard.edu/abs/2019ApJ...882L..24A} {882, L24}

\bibitem[\protect\citeauthoryear{{Acernese}, {Agathos}, {Agatsuma}
  et~al.}{{Acernese} et~al.}{2015}]{Acernese+15}
{Acernese} F.,  {Agathos} M.,  {Agatsuma} K.,   et~al., 2015, \mn@doi
  [Classical and Quantum Gravity] {10.1088/0264-9381/32/2/024001}, \href
  {https://ui.adsabs.harvard.edu/abs/2015CQGra..32b4001A} {32, 024001}

\bibitem[\protect\citeauthoryear{{Aharonian}, {Bogovalov}  \&
  {Khangulyan}}{{Aharonian} et~al.}{2012}]{Aharonian+12}
{Aharonian} F.~A.,  {Bogovalov} S.~V.,   {Khangulyan} D.,  2012, \mn@doi [\nat]
  {10.1038/nature10793}, \href
  {https://ui.adsabs.harvard.edu/abs/2012Natur.482..507A} {482, 507}

\bibitem[\protect\citeauthoryear{{Anderson} et~al.,}{{Anderson}
  et~al.}{2018}]{Anderson+18}
{Anderson} M.~M.,  et~al., 2018, \mn@doi [\apj] {10.3847/1538-4357/aad2d7},
  \href {https://ui.adsabs.harvard.edu/abs/2018ApJ...864...22A} {864, 22}

\bibitem[\protect\citeauthoryear{{Anderson} et~al.,}{{Anderson}
  et~al.}{2019}]{Anderson+19}
{Anderson} M.~M.,  et~al., 2019, \mn@doi [\apj] {10.3847/1538-4357/ab4f87},
  \href {https://ui.adsabs.harvard.edu/abs/2019ApJ...886..123A} {886, 123}

\bibitem[\protect\citeauthoryear{{Babul} \& {Sironi}}{{Babul} \&
  {Sironi}}{2020}]{Babul_Sironi_20}
{Babul} A.-N.,  {Sironi} L.,  2020, arXiv e-prints, \href
  {https://ui.adsabs.harvard.edu/abs/2020arXiv200603081B} {p. arXiv:2006.03081}

\bibitem[\protect\citeauthoryear{{Bartos}, {Brady}  \& {M{\'a}rka}}{{Bartos}
  et~al.}{2013}]{Bartos+13}
{Bartos} I.,  {Brady} P.,   {M{\'a}rka} S.,  2013, \mn@doi [Classical and
  Quantum Gravity] {10.1088/0264-9381/30/12/123001}, \href
  {https://ui.adsabs.harvard.edu/abs/2013CQGra..30l3001B} {30, 123001}

\bibitem[\protect\citeauthoryear{{Belczynski}, {Perna}, {Bulik}, {Kalogera},
  {Ivanova}  \& {Lamb}}{{Belczynski} et~al.}{2006}]{Belczynski_06}
{Belczynski} K.,  {Perna} R.,  {Bulik} T.,  {Kalogera} V.,  {Ivanova} N.,
  {Lamb} D.~Q.,  2006, \mn@doi [\apj] {10.1086/505169}, \href
  {https://ui.adsabs.harvard.edu/abs/2006ApJ...648.1110B} {648, 1110}

\bibitem[\protect\citeauthoryear{{Beloborodov}}{{Beloborodov}}{2020a}]{Beloborodov_20}
{Beloborodov} A.~M.,  2020a, arXiv e-prints, \href
  {https://ui.adsabs.harvard.edu/abs/2020arXiv201107310B} {p. arXiv:2011.07310}

\bibitem[\protect\citeauthoryear{{Beloborodov}}{{Beloborodov}}{2020b}]{Beloborodov20}
{Beloborodov} A.~M.,  2020b, \mn@doi [\apj] {10.3847/1538-4357/ab83eb}, \href
  {https://ui.adsabs.harvard.edu/abs/2020ApJ...896..142B} {896, 142}

\bibitem[\protect\citeauthoryear{{Beloborodov}, {Lundman}  \&
  {Levin}}{{Beloborodov} et~al.}{2018}]{Beloborodov+18}
{Beloborodov} A.~M.,  {Lundman} C.,   {Levin} Y.,  2018, arXiv e-prints, \href
  {https://ui.adsabs.harvard.edu/abs/2018arXiv181211247B} {p. arXiv:1812.11247}

\bibitem[\protect\citeauthoryear{{Beniamini} \& {Kumar}}{{Beniamini} \&
  {Kumar}}{2020}]{Beniamini&Kumar20}
{Beniamini} P.,  {Kumar} P.,  2020, arXiv e-prints, \href
  {https://ui.adsabs.harvard.edu/abs/2020arXiv200707265B} {p. arXiv:2007.07265}

\bibitem[\protect\citeauthoryear{{Beniamini}, {Petropoulou}, {Barniol Duran}
  \& {Giannios}}{{Beniamini} et~al.}{2019}]{Beniamini+19}
{Beniamini} P.,  {Petropoulou} M.,  {Barniol Duran} R.,   {Giannios} D.,  2019,
  \mn@doi [\mnras] {10.1093/mnras/sty3093}, \href
  {https://ui.adsabs.harvard.edu/abs/2019MNRAS.483..840B} {483, 840}

\bibitem[\protect\citeauthoryear{{Berger}}{{Berger}}{2014}]{Berger14}
{Berger} E.,  2014, \mn@doi [Annu. Rev. Astron. Astrophys.]
  {10.1146/annurev-astro-081913-035926}, \href
  {http://adsabs.harvard.edu/abs/2014ARA&A..52...43B} {52, 43}

\bibitem[\protect\citeauthoryear{{Bhandari} et~al.,}{{Bhandari}
  et~al.}{2018}]{Bhandari+18}
{Bhandari} S.,  et~al., 2018, \mn@doi [\mnras] {10.1093/mnras/stx3074}, \href
  {https://ui.adsabs.harvard.edu/abs/2018MNRAS.475.1427B} {475, 1427}

\bibitem[\protect\citeauthoryear{{Bhandari} et~al.,}{{Bhandari}
  et~al.}{2020}]{Bhandari+20}
{Bhandari} S.,  et~al., 2020, \mn@doi [\apjl] {10.3847/2041-8213/ab672e}, \href
  {https://ui.adsabs.harvard.edu/abs/2020ApJ...895L..37B} {895, L37}

\bibitem[\protect\citeauthoryear{{Bhattacharyya}}{{Bhattacharyya}}{2017}]{Bhattacharyya_17}
{Bhattacharyya} S.,  2017, arXiv e-prints, \href
  {https://ui.adsabs.harvard.edu/abs/2017arXiv171109083B} {p. arXiv:1711.09083}

\bibitem[\protect\citeauthoryear{{Blandford} \& {McKee}}{{Blandford} \&
  {McKee}}{1976}]{Blandford_McKee_76}
{Blandford} R.~D.,  {McKee} C.~F.,  1976, \mn@doi [Physics of Fluids]
  {10.1063/1.861619}, \href
  {https://ui.adsabs.harvard.edu/abs/1976PhFl...19.1130B} {19, 1130}

\bibitem[\protect\citeauthoryear{{Bochenek}, {Ravi}, {Belov}, {Hallinan},
  {Kocz}, {Kulkarni}  \& {McKenna}}{{Bochenek} et~al.}{2020}]{Bochenek+20}
{Bochenek} C.~D.,  {Ravi} V.,  {Belov} K.~V.,  {Hallinan} G.,  {Kocz} J.,
  {Kulkarni} S.~R.,   {McKenna} D.~L.,  2020, arXiv e-prints, \href
  {https://ui.adsabs.harvard.edu/abs/2020arXiv200510828B} {p. arXiv:2005.10828}

\bibitem[\protect\citeauthoryear{{Bucciantini}, {Thompson}, {Arons}, {Quataert}
   \& {Del Zanna}}{{Bucciantini} et~al.}{2006}]{Bucciantini+06}
{Bucciantini} N.,  {Thompson} T.~A.,  {Arons} J.,  {Quataert} E.,   {Del Zanna}
  L.,  2006, \mn@doi [\mnras] {10.1111/j.1365-2966.2006.10217.x}, \href
  {https://ui.adsabs.harvard.edu/abs/2006MNRAS.368.1717B} {368, 1717}

\bibitem[\protect\citeauthoryear{{CHIME/FRB Collaboration} et~al.,}{{CHIME/FRB
  Collaboration} et~al.}{2019}]{CHIME+19}
{CHIME/FRB Collaboration} et~al., 2019, \mn@doi [\apjl]
  {10.3847/2041-8213/ab4a80}, \href
  {https://ui.adsabs.harvard.edu/abs/2019ApJ...885L..24C} {885, L24}

\bibitem[\protect\citeauthoryear{{Caleb} et~al.,}{{Caleb}
  et~al.}{2020}]{Caleb_20}
{Caleb} M.,  et~al., 2020, \mn@doi [\mnras] {10.1093/mnras/staa1791}, \href
  {https://ui.adsabs.harvard.edu/abs/2020MNRAS.496.4565C} {496, 4565}

\bibitem[\protect\citeauthoryear{{Callister} et~al.,}{{Callister}
  et~al.}{2019}]{Callister+19}
{Callister} T.~A.,  et~al., 2019, \mn@doi [\apjl] {10.3847/2041-8213/ab2248},
  \href {https://ui.adsabs.harvard.edu/abs/2019ApJ...877L..39C} {877, L39}

\bibitem[\protect\citeauthoryear{{Carrasco} \& {Shibata}}{{Carrasco} \&
  {Shibata}}{2020}]{Carrasco&Shibata20}
{Carrasco} F.,  {Shibata} M.,  2020, \mn@doi [\prd]
  {10.1103/PhysRevD.101.063017}, \href
  {https://ui.adsabs.harvard.edu/abs/2020PhRvD.101f3017C} {101, 063017}

\bibitem[\protect\citeauthoryear{{Cerutti}, {Philippov}  \& {Dubus}}{{Cerutti}
  et~al.}{2020}]{Cerutti_20}
{Cerutti} B.,  {Philippov} A.,   {Dubus} G.,  2020, arXiv e-prints, \href
  {https://ui.adsabs.harvard.edu/abs/2020arXiv200811462C} {p. arXiv:2008.11462}

\bibitem[\protect\citeauthoryear{{Champion} et~al.,}{{Champion}
  et~al.}{2016}]{Champion+16}
{Champion} D.~J.,  et~al., 2016, \mn@doi [\mnras] {10.1093/mnrasl/slw069},
  \href {https://ui.adsabs.harvard.edu/abs/2016MNRAS.460L..30C} {460, L30}

\bibitem[\protect\citeauthoryear{{Cordes} \& {Lazio}}{{Cordes} \&
  {Lazio}}{2002}]{Cordes_Lazio_02}
{Cordes} J.~M.,  {Lazio} T.~J.~W.,  2002, arXiv e-prints, \href
  {https://ui.adsabs.harvard.edu/abs/2002astro.ph..7156C} {pp
  astro--ph/0207156}

\bibitem[\protect\citeauthoryear{{Crinquand}, {Cerutti}  \&
  {Dubus}}{{Crinquand} et~al.}{2019}]{Crinquand_19}
{Crinquand} B.,  {Cerutti} B.,   {Dubus} G.,  2019, \mn@doi [\aap]
  {10.1051/0004-6361/201834610}, \href
  {https://ui.adsabs.harvard.edu/abs/2019A&A...622A.161C} {622, A161}

\bibitem[\protect\citeauthoryear{{D'Orazio} \& {Levin}}{{D'Orazio} \&
  {Levin}}{2013}]{DOrazio&Levin13}
{D'Orazio} D.~J.,  {Levin} J.,  2013, \mn@doi [\prd]
  {10.1103/PhysRevD.88.064059}, \href
  {http://adsabs.harvard.edu/abs/2013PhRvD..88f4059D} {88, 064059}

\bibitem[\protect\citeauthoryear{{D'Orazio}, {Levin}, {Murray}  \&
  {Price}}{{D'Orazio} et~al.}{2016}]{DOrazio+16}
{D'Orazio} D.~J.,  {Levin} J.,  {Murray} N.~W.,   {Price} L.,  2016, \mn@doi
  [Phys. Rev. D] {10.1103/PhysRevD.94.023001}, \href
  {http://adsabs.harvard.edu/abs/2016PhRvD..94b3001D} {94, 023001}

\bibitem[\protect\citeauthoryear{{Dichiara}, {Troja}, {O'Connor}, {Marshall},
  {Beniamini}, {Cannizzo}, {Lien}  \& {Sakamoto}}{{Dichiara}
  et~al.}{2020}]{Dichiara+20}
{Dichiara} S.,  {Troja} E.,  {O'Connor} B.,  {Marshall} F.~E.,  {Beniamini} P.,
   {Cannizzo} J.~K.,  {Lien} A.~Y.,   {Sakamoto} T.,  2020, \mn@doi [\mnras]
  {10.1093/mnras/staa124}, \href
  {https://ui.adsabs.harvard.edu/abs/2020MNRAS.492.5011D} {492, 5011}

\bibitem[\protect\citeauthoryear{{Drenkhahn} \& {Spruit}}{{Drenkhahn} \&
  {Spruit}}{2002}]{Drenkhahn&Spruit02}
{Drenkhahn} G.,  {Spruit} H.~C.,  2002, \mn@doi [\aap]
  {10.1051/0004-6361:20020839}, \href
  {https://ui.adsabs.harvard.edu/abs/2002A&A...391.1141D} {391, 1141}

\bibitem[\protect\citeauthoryear{{Duffell} \& {MacFadyen}}{{Duffell} \&
  {MacFadyen}}{2013}]{Duffell_MacFadyen_13}
{Duffell} P.~C.,  {MacFadyen} A.~I.,  2013, \mn@doi [\apj]
  {10.1088/0004-637X/775/2/87}, \href
  {https://ui.adsabs.harvard.edu/abs/2013ApJ...775...87D} {775, 87}

\bibitem[\protect\citeauthoryear{{Duffell} \& {MacFadyen}}{{Duffell} \&
  {MacFadyen}}{2014}]{Duffell_14}
{Duffell} P.~C.,  {MacFadyen} A.~I.,  2014, \mn@doi [\apjl]
  {10.1088/2041-8205/791/1/L1}, \href
  {https://ui.adsabs.harvard.edu/abs/2014ApJ...791L...1D} {791, L1}

\bibitem[\protect\citeauthoryear{{Farah} et~al.,}{{Farah}
  et~al.}{2018}]{Farah_18}
{Farah} W.,  et~al., 2018, \mn@doi [\mnras] {10.1093/mnras/sty1122}, \href
  {https://ui.adsabs.harvard.edu/abs/2018MNRAS.478.1209F} {478, 1209}

\bibitem[\protect\citeauthoryear{{Fern{\'a}ndez} \& {Metzger}}{{Fern{\'a}ndez}
  \& {Metzger}}{2016}]{Fernandez&Metzger16}
{Fern{\'a}ndez} R.,  {Metzger} B.~D.,  2016, \mn@doi [Annu. Rev. Nucl. Part.
  Sci.] {10.1146/annurev-nucl-102115-044819}, \href
  {http://adsabs.harvard.edu/abs/2016ARNPS..66...23F} {66, 23}

\bibitem[\protect\citeauthoryear{{Gajjar} et~al.,}{{Gajjar}
  et~al.}{2018}]{Gajjar_18}
{Gajjar} V.,  et~al., 2018, \mn@doi [\apj] {10.3847/1538-4357/aad005}, \href
  {https://ui.adsabs.harvard.edu/abs/2018ApJ...863....2G} {863, 2}

\bibitem[\protect\citeauthoryear{{Giannios}, {Mimica}  \& {Aloy}}{{Giannios}
  et~al.}{2008}]{Giannios_08}
{Giannios} D.,  {Mimica} P.,   {Aloy} M.~A.,  2008, \mn@doi [\aap]
  {10.1051/0004-6361:20078931}, \href
  {https://ui.adsabs.harvard.edu/abs/2008A&A...478..747G} {478, 747}

\bibitem[\protect\citeauthoryear{{Goldreich} \& {Julian}}{{Goldreich} \&
  {Julian}}{1970}]{Goldreich&Julian70}
{Goldreich} P.,  {Julian} W.~H.,  1970, \mn@doi [\apj] {10.1086/150486}, \href
  {https://ui.adsabs.harvard.edu/abs/1970ApJ...160..971G} {160, 971}

\bibitem[\protect\citeauthoryear{{Goldreich} \& {Lynden-Bell}}{{Goldreich} \&
  {Lynden-Bell}}{1969}]{Goldreich&LyndenBell69}
{Goldreich} P.,  {Lynden-Bell} D.,  1969, \mn@doi [\apj] {10.1086/149947},
  \href {http://adsabs.harvard.edu/abs/1969ApJ...156...59G} {156, 59}

\bibitem[\protect\citeauthoryear{{Goldstein} et~al.}{{Goldstein}
  et~al.}{2017}]{Goldstein+17}
{Goldstein} A.,  et~al., 2017, \mn@doi [\apjl] {10.3847/2041-8213/aa8f41},
  \href {http://adsabs.harvard.edu/abs/2017ApJ...848L..14G} {848, L14}

\bibitem[\protect\citeauthoryear{{Goodman}}{{Goodman}}{1986}]{Goodman86}
{Goodman} J.,  1986, \mn@doi [ApJ] {10.1086/184741}, \href
  {http://adsabs.harvard.edu/abs/1986ApJ...308L..47G} {308, L47}

\bibitem[\protect\citeauthoryear{{Gourdji}, {Rowlinson}, {Wijers}  \&
  {Goldstein}}{{Gourdji} et~al.}{2020}]{Gourdji+20}
{Gourdji} K.,  {Rowlinson} A.,  {Wijers} R. A.~M.~J.,   {Goldstein} A.,  2020,
  arXiv e-prints, \href {https://ui.adsabs.harvard.edu/abs/2020arXiv200302706G}
  {p. arXiv:2003.02706}

\bibitem[\protect\citeauthoryear{{Gourgouliatos} \&
  {Lynden-Bell}}{{Gourgouliatos} \& {Lynden-Bell}}{2019}]{Gourgouliatos+19}
{Gourgouliatos} K.~N.,  {Lynden-Bell} D.,  2019, \mn@doi [\mnras]
  {10.1093/mnras/sty2766}, \href
  {https://ui.adsabs.harvard.edu/abs/2019MNRAS.482.1942G} {482, 1942}

\bibitem[\protect\citeauthoryear{{Granot}, {Komissarov}  \&
  {Spitkovsky}}{{Granot} et~al.}{2011}]{Granot+11}
{Granot} J.,  {Komissarov} S.~S.,   {Spitkovsky} A.,  2011, \mn@doi [\mnras]
  {10.1111/j.1365-2966.2010.17770.x}, \href
  {https://ui.adsabs.harvard.edu/abs/2011MNRAS.411.1323G} {411, 1323}

\bibitem[\protect\citeauthoryear{{Hansen} \& {Lyutikov}}{{Hansen} \&
  {Lyutikov}}{2001}]{Hansen&Lyutikov01}
{Hansen} B.~M.~S.,  {Lyutikov} M.,  2001, \mn@doi [MNRAS]
  {10.1046/j.1365-8711.2001.04103.x}, \href
  {http://adsabs.harvard.edu/abs/2001MNRAS.322..695H} {322, 695}

\bibitem[\protect\citeauthoryear{{Harry} \& {LIGO Scientific
  Collaboration}}{{Harry} \& {LIGO Scientific Collaboration}}{2010}]{Harry+10}
{Harry} G.~M.,  {LIGO Scientific Collaboration} 2010, \mn@doi [Classical and
  Quantum Gravity] {10.1088/0264-9381/27/8/084006}, \href
  {https://ui.adsabs.harvard.edu/abs/2010CQGra..27h4006H} {27, 084006}

\bibitem[\protect\citeauthoryear{{Heintz} et~al.,}{{Heintz}
  et~al.}{2020}]{Heintz+20}
{Heintz} K.~E.,  et~al., 2020, arXiv e-prints, \href
  {https://ui.adsabs.harvard.edu/abs/2020arXiv200910747H} {p. arXiv:2009.10747}

\bibitem[\protect\citeauthoryear{{Hessels} et~al.,}{{Hessels}
  et~al.}{2019}]{Hessels+19}
{Hessels} J.~W.~T.,  et~al., 2019, \mn@doi [\apjl] {10.3847/2041-8213/ab13ae},
  \href {https://ui.adsabs.harvard.edu/abs/2019ApJ...876L..23H} {876, L23}

\bibitem[\protect\citeauthoryear{{Inoue}}{{Inoue}}{2004}]{Inoue_2004}
{Inoue} S.,  2004, \mn@doi [\mnras] {10.1111/j.1365-2966.2004.07359.x}, \href
  {https://ui.adsabs.harvard.edu/abs/2004MNRAS.348..999I} {348, 999}

\bibitem[\protect\citeauthoryear{{Ioka}}{{Ioka}}{2003}]{Ioka_2003}
{Ioka} K.,  2003, \mn@doi [\apjl] {10.1086/380598}, \href
  {https://ui.adsabs.harvard.edu/abs/2003ApJ...598L..79I} {598, L79}

\bibitem[\protect\citeauthoryear{{Ioka} \& {Taniguchi}}{{Ioka} \&
  {Taniguchi}}{2000}]{Ioka&Taniguchi00}
{Ioka} K.,  {Taniguchi} K.,  2000, \mn@doi [\apj] {10.1086/309004}, \href
  {https://ui.adsabs.harvard.edu/abs/2000ApJ...537..327I} {537, 327}

\bibitem[\protect\citeauthoryear{{Iwamoto}, {Amano}, {Hoshino}  \&
  {Matsumoto}}{{Iwamoto} et~al.}{2017}]{Iwamoto_17}
{Iwamoto} M.,  {Amano} T.,  {Hoshino} M.,   {Matsumoto} Y.,  2017, \mn@doi
  [\apj] {10.3847/1538-4357/aa6d6f}, \href
  {https://ui.adsabs.harvard.edu/abs/2017ApJ...840...52I} {840, 52}

\bibitem[\protect\citeauthoryear{{James}, {Anderson}, {Wen}, {Bosveld}, {Chu},
  {Kovalam}, {Slaven-Blair}  \& {Williams}}{{James} et~al.}{2019}]{James+19}
{James} C.~W.,  {Anderson} G.~E.,  {Wen} L.,  {Bosveld} J.,  {Chu} Q.,
  {Kovalam} M.,  {Slaven-Blair} T.~J.,   {Williams} A.,  2019, \mn@doi [\mnras]
  {10.1093/mnrasl/slz129}, \href
  {https://ui.adsabs.harvard.edu/abs/2019MNRAS.489L..75J} {489, L75}

\bibitem[\protect\citeauthoryear{{Kaplan} et~al.,}{{Kaplan}
  et~al.}{2015}]{Kaplan+15}
{Kaplan} D.~L.,  et~al., 2015, \mn@doi [\apjl] {10.1088/2041-8205/814/2/L25},
  \href {https://ui.adsabs.harvard.edu/abs/2015ApJ...814L..25K} {814, L25}

\bibitem[\protect\citeauthoryear{{Karastergiou} et~al.,}{{Karastergiou}
  et~al.}{2015}]{Karastergiou_2015}
{Karastergiou} A.,  et~al., 2015, \mn@doi [\mnras] {10.1093/mnras/stv1306},
  \href {https://ui.adsabs.harvard.edu/abs/2015MNRAS.452.1254K} {452, 1254}

\bibitem[\protect\citeauthoryear{{Kasliwal} et~al.}{{Kasliwal}
  et~al.}{2017}]{Kasliwal+17}
{Kasliwal} M.~M.,  et~al., 2017, \mn@doi [Science] {10.1126/science.aap9455},
  \href {https://ui.adsabs.harvard.edu/abs/2017Sci...358.1559K} {358, 1559}

\bibitem[\protect\citeauthoryear{{Kathirgamaraju}, {Tchekhovskoy}, {Giannios}
  \& {Barniol Duran}}{{Kathirgamaraju} et~al.}{2019}]{Kathirgamaraju+19}
{Kathirgamaraju} A.,  {Tchekhovskoy} A.,  {Giannios} D.,   {Barniol Duran} R.,
  2019, \mn@doi [\mnras] {10.1093/mnrasl/slz012}, \href
  {https://ui.adsabs.harvard.edu/abs/2019MNRAS.484L..98K} {484, L98}

\bibitem[\protect\citeauthoryear{{Kennel} \& {Coroniti}}{{Kennel} \&
  {Coroniti}}{1984}]{Kennel&Coroniti84}
{Kennel} C.~F.,  {Coroniti} F.~V.,  1984, \mn@doi [\apj] {10.1086/162357},
  \href {https://ui.adsabs.harvard.edu/abs/1984ApJ...283..710K} {283, 710}

\bibitem[\protect\citeauthoryear{{Khangulyan}, {Aharonian}, {Bogovalov}  \&
  {Rib{\'o}}}{{Khangulyan} et~al.}{2012}]{Khangulyan+12}
{Khangulyan} D.,  {Aharonian} F.~A.,  {Bogovalov} S.~V.,   {Rib{\'o}} M.,
  2012, \mn@doi [\apjl] {10.1088/2041-8205/752/1/L17}, \href
  {https://ui.adsabs.harvard.edu/abs/2012ApJ...752L..17K} {752, L17}

\bibitem[\protect\citeauthoryear{{Kramer} \& {Stairs}}{{Kramer} \&
  {Stairs}}{2008}]{Kramer&Stairs08}
{Kramer} M.,  {Stairs} I.~H.,  2008, \mn@doi [\araa]
  {10.1146/annurev.astro.46.060407.145247}, \href
  {https://ui.adsabs.harvard.edu/abs/2008ARA&A..46..541K} {46, 541}

\bibitem[\protect\citeauthoryear{{Kumar} \& {Bo{\v{s}}njak}}{{Kumar} \&
  {Bo{\v{s}}njak}}{2020}]{Kumar&Bosnjak20}
{Kumar} P.,  {Bo{\v{s}}njak} {\v{Z}}.,  2020, \mn@doi [\mnras]
  {10.1093/mnras/staa774}, \href
  {https://ui.adsabs.harvard.edu/abs/2020MNRAS.494.2385K} {494, 2385}

\bibitem[\protect\citeauthoryear{{Lai}}{{Lai}}{2012}]{Lai12}
{Lai} D.,  2012, \mn@doi [The Astrophysical Journal Letters]
  {10.1088/2041-8205/757/1/L3}, \href
  {http://adsabs.harvard.edu/abs/2012ApJ...757L...3L} {757, L3}

\bibitem[\protect\citeauthoryear{{Lipunov} \& {Panchenko}}{{Lipunov} \&
  {Panchenko}}{1996}]{Lipunov&Panchenko96}
{Lipunov} V.~M.,  {Panchenko} I.~E.,  1996, \aap, \href
  {http://adsabs.harvard.edu/abs/1996A%26A...312..937L} {312, 937}

\bibitem[\protect\citeauthoryear{{Lorimer}, {Bailes}, {McLaughlin}, {Narkevic}
  \& {Crawford}}{{Lorimer} et~al.}{2007}]{Lorimer+07}
{Lorimer} D.~R.,  {Bailes} M.,  {McLaughlin} M.~A.,  {Narkevic} D.~J.,
  {Crawford} F.,  2007, \mn@doi [Science] {10.1126/science.1147532}, \href
  {http://adsabs.harvard.edu/abs/2007Sci...318..777L} {318, 777}

\bibitem[\protect\citeauthoryear{{Lu}, {Piro}  \& {Waxman}}{{Lu}
  et~al.}{2020}]{Lu+20}
{Lu} W.,  {Piro} A.~L.,   {Waxman} E.,  2020, arXiv e-prints, \href
  {https://ui.adsabs.harvard.edu/abs/2020arXiv200312581L} {p. arXiv:2003.12581}

\bibitem[\protect\citeauthoryear{{Lyubarsky}}{{Lyubarsky}}{2008}]{Lyubarsky08}
{Lyubarsky} Y.,  2008, \mn@doi [\apj] {10.1086/589435}, \href
  {http://adsabs.harvard.edu/abs/2008ApJ...682.1443L} {682, 1443}

\bibitem[\protect\citeauthoryear{{Lyubarsky}}{{Lyubarsky}}{2019}]{Lyubarsky19}
{Lyubarsky} Y.,  2019, \mn@doi [\mnras] {10.1093/mnras/sty3233}, \href
  {https://ui.adsabs.harvard.edu/abs/2019MNRAS.483.1731L} {483, 1731}

\bibitem[\protect\citeauthoryear{{Lyubarsky}}{{Lyubarsky}}{2020}]{Lyubarsky20}
{Lyubarsky} Y.,  2020, \mn@doi [\apj] {10.3847/1538-4357/ab97b5}, \href
  {https://ui.adsabs.harvard.edu/abs/2020ApJ...897....1L} {897, 1}

\bibitem[\protect\citeauthoryear{{Lyubarsky} \& {Kirk}}{{Lyubarsky} \&
  {Kirk}}{2001}]{Lyubarsky&Kirk01}
{Lyubarsky} Y.,  {Kirk} J.~G.,  2001, \mn@doi [\apj] {10.1086/318354}, \href
  {https://ui.adsabs.harvard.edu/abs/2001ApJ...547..437L} {547, 437}

\bibitem[\protect\citeauthoryear{{Lyutikov}}{{Lyutikov}}{2019}]{Lyutikov19}
{Lyutikov} M.,  2019, \mn@doi [\mnras] {10.1093/mnras/sty3303}, \href
  {https://ui.adsabs.harvard.edu/abs/2019MNRAS.483.2766L} {483, 2766}

\bibitem[\protect\citeauthoryear{{Margalit} \& {Metzger}}{{Margalit} \&
  {Metzger}}{2019}]{Margalit&Metzger19}
{Margalit} B.,  {Metzger} B.~D.,  2019, arXiv e-prints, \href
  {https://ui.adsabs.harvard.edu/abs/2019arXiv190411995M} {p. arXiv:1904.11995}

\bibitem[\protect\citeauthoryear{{Margalit}, {Berger}  \& {Metzger}}{{Margalit}
  et~al.}{2019}]{Margalit+19}
{Margalit} B.,  {Berger} E.,   {Metzger} B.~D.,  2019, \mn@doi [\apj]
  {10.3847/1538-4357/ab4c31}, \href
  {https://ui.adsabs.harvard.edu/abs/2019ApJ...886..110M} {886, 110}

\bibitem[\protect\citeauthoryear{{Margutti} et~al.,}{{Margutti}
  et~al.}{2018}]{Margutti+18}
{Margutti} R.,  et~al., 2018, \mn@doi [\apjl] {10.3847/2041-8213/aab2ad}, \href
  {https://ui.adsabs.harvard.edu/abs/2018ApJ...856L..18M} {856, L18}

\bibitem[\protect\citeauthoryear{{Mart{\'\i}} \& {M{\"u}ller}}{{Mart{\'\i}} \&
  {M{\"u}ller}}{2003}]{Marti_03}
{Mart{\'\i}} J.~M.,  {M{\"u}ller} E.,  2003, \mn@doi [Living Reviews in
  Relativity] {10.12942/lrr-2003-7}, \href
  {https://ui.adsabs.harvard.edu/abs/2003LRR.....6....7M} {6, 7}

\bibitem[\protect\citeauthoryear{{McWilliams} \& {Levin}}{{McWilliams} \&
  {Levin}}{2011}]{McWilliams&Levin11}
{McWilliams} S.~T.,  {Levin} J.,  2011, \mn@doi [Astrophys. J.]
  {10.1088/0004-637X/742/2/90}, \href
  {http://adsabs.harvard.edu/abs/2011ApJ...742...90M} {742, 90}

\bibitem[\protect\citeauthoryear{{Metzger} \& {Berger}}{{Metzger} \&
  {Berger}}{2012}]{Metzger&Berger12}
{Metzger} B.~D.,  {Berger} E.,  2012, \mn@doi [Astrophys. J.]
  {10.1088/0004-637X/746/1/48}, \href
  {http://adsabs.harvard.edu/abs/2012ApJ...746...48M} {746, 48}

\bibitem[\protect\citeauthoryear{{Metzger} \& {Zivancev}}{{Metzger} \&
  {Zivancev}}{2016}]{Metzger&Zivancev16}
{Metzger} B.~D.,  {Zivancev} C.,  2016, \mn@doi [\mnras]
  {10.1093/mnras/stw1800}, \href
  {https://ui.adsabs.harvard.edu/abs/2016MNRAS.461.4435M} {461, 4435}

\bibitem[\protect\citeauthoryear{{Metzger} et~al.,}{{Metzger}
  et~al.}{2010}]{Metzger+10}
{Metzger} B.~D.,  et~al., 2010, \mn@doi [Mon. Not. R. Astron. Soc.]
  {10.1111/j.1365-2966.2010.16864.x}, \href
  {http://adsabs.harvard.edu/abs/2010MNRAS.406.2650M} {406, 2650}

\bibitem[\protect\citeauthoryear{{Metzger}, {Thompson}  \&
  {Quataert}}{{Metzger} et~al.}{2018}]{Metzger+18}
{Metzger} B.~D.,  {Thompson} T.~A.,   {Quataert} E.,  2018, \mn@doi [\apj]
  {10.3847/1538-4357/aab095}, \href
  {https://ui.adsabs.harvard.edu/abs/2018ApJ...856..101M} {856, 101}

\bibitem[\protect\citeauthoryear{{Metzger}, {Margalit}  \& {Sironi}}{{Metzger}
  et~al.}{2019}]{Metzger+19}
{Metzger} B.~D.,  {Margalit} B.,   {Sironi} L.,  2019, \mn@doi [\mnras]
  {10.1093/mnras/stz700}, \href
  {https://ui.adsabs.harvard.edu/abs/2019MNRAS.485.4091M} {485, 4091}

\bibitem[\protect\citeauthoryear{{Mignone} \& {Bodo}}{{Mignone} \&
  {Bodo}}{2006}]{Mignone_Bodo_06}
{Mignone} A.,  {Bodo} G.,  2006, \mn@doi [\mnras]
  {10.1111/j.1365-2966.2006.10162.x}, \href
  {https://ui.adsabs.harvard.edu/abs/2006MNRAS.368.1040M} {368, 1040}

\bibitem[\protect\citeauthoryear{{Mimica} \& {Aloy}}{{Mimica} \&
  {Aloy}}{2010}]{Mimica_10}
{Mimica} P.,  {Aloy} M.~A.,  2010, \mn@doi [\mnras]
  {10.1111/j.1365-2966.2009.15669.x}, \href
  {https://ui.adsabs.harvard.edu/abs/2010MNRAS.401..525M} {401, 525}

\bibitem[\protect\citeauthoryear{{Mingarelli}, {Levin}  \&
  {Lazio}}{{Mingarelli} et~al.}{2015}]{Mingarelli+15}
{Mingarelli} C. M.~F.,  {Levin} J.,   {Lazio} T. J.~W.,  2015, \mn@doi [\apjl]
  {10.1088/2041-8205/814/2/L20}, \href
  {https://ui.adsabs.harvard.edu/abs/2015ApJ...814L..20M} {814, L20}

\bibitem[\protect\citeauthoryear{{Mizuno}, {Zhang}, {Giacomazzo}, {Nishikawa},
  {Hardee}, {Nagataki}  \& {Hartmann}}{{Mizuno} et~al.}{2009}]{Mizuno_09}
{Mizuno} Y.,  {Zhang} B.,  {Giacomazzo} B.,  {Nishikawa} K.-I.,  {Hardee}
  P.~E.,  {Nagataki} S.,   {Hartmann} D.~H.,  2009, \mn@doi [\apjl]
  {10.1088/0004-637X/690/1/L47}, \href
  {https://ui.adsabs.harvard.edu/abs/2009ApJ...690L..47M} {690, L47}

\bibitem[\protect\citeauthoryear{{Moortgat} \& {Kuijpers}}{{Moortgat} \&
  {Kuijpers}}{2003}]{Moortgat&Kuijpers03}
{Moortgat} J.,  {Kuijpers} J.,  2003, \mn@doi [\aap]
  {10.1051/0004-6361:20030271}, \href
  {https://ui.adsabs.harvard.edu/abs/2003A&A...402..905M} {402, 905}

\bibitem[\protect\citeauthoryear{{Most} \& {Philippov}}{{Most} \&
  {Philippov}}{2020}]{Most&Philippov20}
{Most} E.~R.,  {Philippov} A.~A.,  2020, arXiv e-prints, \href
  {https://ui.adsabs.harvard.edu/abs/2020arXiv200106037M} {p. arXiv:2001.06037}

\bibitem[\protect\citeauthoryear{{Nakar} \& {Piran}}{{Nakar} \&
  {Piran}}{2011}]{Nakar&Piran11}
{Nakar} E.,  {Piran} T.,  2011, \mn@doi [Nature] {10.1038/nature10365}, \href
  {http://adsabs.harvard.edu/abs/2011Natur.478...82N} {478, 82}

\bibitem[\protect\citeauthoryear{{Nakar}, {Gottlieb}, {Piran}, {Kasliwal}  \&
  {Hallinan}}{{Nakar} et~al.}{2018}]{Nakar+18}
{Nakar} E.,  {Gottlieb} O.,  {Piran} T.,  {Kasliwal} M.~M.,   {Hallinan} G.,
  2018, \mn@doi [\apj] {10.3847/1538-4357/aae205}, \href
  {https://ui.adsabs.harvard.edu/abs/2018ApJ...867...18N} {867, 18}

\bibitem[\protect\citeauthoryear{{Nimmo} et~al.,}{{Nimmo}
  et~al.}{2020}]{Nimmo_20}
{Nimmo} K.,  et~al., 2020, arXiv e-prints, \href
  {https://ui.adsabs.harvard.edu/abs/2020arXiv201005800N} {p. arXiv:2010.05800}

\bibitem[\protect\citeauthoryear{{Paczynski}}{{Paczynski}}{1986}]{Paczynski86}
{Paczynski} B.,  1986, \mn@doi [Astrophys. J. Lett.] {10.1086/184740}, \href
  {http://adsabs.harvard.edu/abs/1986ApJ...308L..43P} {308, L43}

\bibitem[\protect\citeauthoryear{{Palaniswamy}, {Wayth}, {Trott}, {McCallum},
  {Tingay}  \& {Reynolds}}{{Palaniswamy} et~al.}{2014}]{Palaniswamy+14}
{Palaniswamy} D.,  {Wayth} R.~B.,  {Trott} C.~M.,  {McCallum} J.~N.,  {Tingay}
  S.~J.,   {Reynolds} C.,  2014, \mn@doi [\apj] {10.1088/0004-637X/790/1/63},
  \href {https://ui.adsabs.harvard.edu/abs/2014ApJ...790...63P} {790, 63}

\bibitem[\protect\citeauthoryear{{Palenzuela}, {Lehner}, {Ponce}, {Liebling},
  {Anderson}, {Neilsen}  \& {Motl}}{{Palenzuela} et~al.}{2013}]{Palenzuela+13}
{Palenzuela} C.,  {Lehner} L.,  {Ponce} M.,  {Liebling} S.~L.,  {Anderson} M.,
  {Neilsen} D.,   {Motl} P.,  2013, \mn@doi [Phys. Rev. Lett.]
  {10.1103/PhysRevLett.111.061105}, \href
  {http://adsabs.harvard.edu/abs/2013PhRvL.111f1105P} {111, 061105}

\bibitem[\protect\citeauthoryear{{Paschalidis} \& {Ruiz}}{{Paschalidis} \&
  {Ruiz}}{2019}]{Paschalidis&Ruiz19}
{Paschalidis} V.,  {Ruiz} M.,  2019, \mn@doi [\prd]
  {10.1103/PhysRevD.100.043001}, \href
  {https://ui.adsabs.harvard.edu/abs/2019PhRvD.100d3001P} {100, 043001}

\bibitem[\protect\citeauthoryear{{Philippov}, {Uzdensky}, {Spitkovsky}  \&
  {Cerutti}}{{Philippov} et~al.}{2019}]{Philippov+19}
{Philippov} A.,  {Uzdensky} D.~A.,  {Spitkovsky} A.,   {Cerutti} B.,  2019,
  \mn@doi [\apjl] {10.3847/2041-8213/ab1590}, \href
  {https://ui.adsabs.harvard.edu/abs/2019ApJ...876L...6P} {876, L6}

\bibitem[\protect\citeauthoryear{{Piro}}{{Piro}}{2012}]{Piro12}
{Piro} A.~L.,  2012, \mn@doi [\apj] {10.1088/0004-637X/755/1/80}, \href
  {http://adsabs.harvard.edu/abs/2012ApJ...755...80P} {755, 80}

\bibitem[\protect\citeauthoryear{{Plotnikov} \& {Sironi}}{{Plotnikov} \&
  {Sironi}}{2019}]{Plotnikov&Sironi19}
{Plotnikov} I.,  {Sironi} L.,  2019, \mn@doi [\mnras] {10.1093/mnras/stz640},
  \href {https://ui.adsabs.harvard.edu/abs/2019MNRAS.485.3816P} {485, 3816}

\bibitem[\protect\citeauthoryear{{Ponce}, {Palenzuela}, {Lehner}  \&
  {Liebling}}{{Ponce} et~al.}{2014}]{Ponce+14}
{Ponce} M.,  {Palenzuela} C.,  {Lehner} L.,   {Liebling} S.~L.,  2014, \mn@doi
  [\prd] {10.1103/PhysRevD.90.044007}, \href
  {http://adsabs.harvard.edu/abs/2014PhRvD..90d4007P} {90, 044007}

\bibitem[\protect\citeauthoryear{{Porth}, {Komissarov}  \& {Keppens}}{{Porth}
  et~al.}{2013}]{Porth+13}
{Porth} O.,  {Komissarov} S.~S.,   {Keppens} R.,  2013, \mn@doi [\mnras]
  {10.1093/mnrasl/slt006}, \href
  {https://ui.adsabs.harvard.edu/abs/2013MNRAS.431L..48P} {431, L48}

\bibitem[\protect\citeauthoryear{{Prochaska} \& {Zheng}}{{Prochaska} \&
  {Zheng}}{2019}]{Prochaska_Zheng_2019}
{Prochaska} J.~X.,  {Zheng} Y.,  2019, \mn@doi [\mnras] {10.1093/mnras/stz261},
  \href {https://ui.adsabs.harvard.edu/abs/2019MNRAS.485..648P} {485, 648}

\bibitem[\protect\citeauthoryear{{Ramirez-Ruiz}, {Andrews}  \&
  {Schr{\o}der}}{{Ramirez-Ruiz} et~al.}{2019}]{Ramirez-Ruiz+19}
{Ramirez-Ruiz} E.,  {Andrews} J.~J.,   {Schr{\o}der} S.~L.,  2019, \mn@doi
  [\apjl] {10.3847/2041-8213/ab3f2c}, \href
  {https://ui.adsabs.harvard.edu/abs/2019ApJ...883L...6R} {883, L6}

\bibitem[\protect\citeauthoryear{{Ravi}}{{Ravi}}{2019}]{Ravi_19}
{Ravi} V.,  2019, \mn@doi [Nature Astronomy] {10.1038/s41550-019-0831-y}, \href
  {https://ui.adsabs.harvard.edu/abs/2019NatAs...3..928R} {3, 928}

\bibitem[\protect\citeauthoryear{{Rowlinson} \& {Anderson}}{{Rowlinson} \&
  {Anderson}}{2019}]{Rowlinson&Anderson19}
{Rowlinson} A.,  {Anderson} G.~E.,  2019, \mn@doi [\mnras]
  {10.1093/mnras/stz2295}, \href
  {https://ui.adsabs.harvard.edu/abs/2019MNRAS.489.3316R} {489, 3316}

\bibitem[\protect\citeauthoryear{{Rowlinson} et~al.,}{{Rowlinson}
  et~al.}{2020}]{Rowlinson+20}
{Rowlinson} A.,  et~al., 2020, arXiv e-prints, \href
  {https://ui.adsabs.harvard.edu/abs/2020arXiv200812657R} {p. arXiv:2008.12657}

\bibitem[\protect\citeauthoryear{{Schnittman}, {Dal Canton}, {Camp}, {Tsang}
  \& {Kelly}}{{Schnittman} et~al.}{2018}]{Schnittman+18}
{Schnittman} J.~D.,  {Dal Canton} T.,  {Camp} J.,  {Tsang} D.,   {Kelly} B.~J.,
   2018, \mn@doi [\apj] {10.3847/1538-4357/aaa08b}, \href
  {https://ui.adsabs.harvard.edu/abs/2018ApJ...853..123S} {853, 123}

\bibitem[\protect\citeauthoryear{{Schutz}}{{Schutz}}{2011}]{Schutz11}
{Schutz} B.~F.,  2011, \mn@doi [Class. Quantum Grav.]
  {10.1088/0264-9381/28/12/125023}, \href
  {http://adsabs.harvard.edu/abs/2011CQGra..28l5023S} {28, 125023}

\bibitem[\protect\citeauthoryear{{Sironi} \& {Spitkovsky}}{{Sironi} \&
  {Spitkovsky}}{2011}]{Sironi&Spitkovsky11}
{Sironi} L.,  {Spitkovsky} A.,  2011, \mn@doi [\apj]
  {10.1088/0004-637X/741/1/39}, \href
  {https://ui.adsabs.harvard.edu/abs/2011ApJ...741...39S} {741, 39}

\bibitem[\protect\citeauthoryear{{Somiya}}{{Somiya}}{2012}]{Somiya12}
{Somiya} K.,  2012, \mn@doi [Classical and Quantum Gravity]
  {10.1088/0264-9381/29/12/124007}, \href
  {https://ui.adsabs.harvard.edu/abs/2012CQGra..29l4007S} {29, 124007}

\bibitem[\protect\citeauthoryear{{Spitler} et~al.,}{{Spitler}
  et~al.}{2016}]{Spitler+16}
{Spitler} L.~G.,  et~al., 2016, \mn@doi [\nat] {10.1038/nature17168}, \href
  {https://ui.adsabs.harvard.edu/abs/2016Natur.531..202S} {531, 202}

\bibitem[\protect\citeauthoryear{{The CHIME/FRB Collaboration} et~al.,}{{The
  CHIME/FRB Collaboration} et~al.}{2020}]{CHIME+20}
{The CHIME/FRB Collaboration} et~al., 2020, arXiv e-prints, \href
  {https://ui.adsabs.harvard.edu/abs/2020arXiv200510324T} {p. arXiv:2005.10324}

\bibitem[\protect\citeauthoryear{{The LIGO Scientific Collaboration} \& {the
  Virgo Collaboration}}{{The LIGO Scientific Collaboration} \& {the Virgo
  Collaboration}}{2020}]{LVC2020population}
{The LIGO Scientific Collaboration} {the Virgo Collaboration} 2020, Population
  Properties of Compact Objects from the Second LIGO-Virgo Gravitational-Wave
  Transient Catalog (\mn@eprint {arXiv} {2010.14533})

\bibitem[\protect\citeauthoryear{{The LIGO Scientific Collaboration}, {the
  Virgo Collaboration}, {Abbott}  et~al.}{{The LIGO Scientific Collaboration}
  et~al.}{2020}]{LIGO+20a}
{The LIGO Scientific Collaboration} {the Virgo Collaboration} {Abbott} B.~P.,
  et~al., 2020, arXiv e-prints, \href
  {https://ui.adsabs.harvard.edu/abs/2020arXiv200101761T} {p. arXiv:2001.01761}

\bibitem[\protect\citeauthoryear{{Thornton} et~al.,}{{Thornton}
  et~al.}{2013}]{Thornton+13}
{Thornton} D.,  et~al., 2013, \mn@doi [Science] {10.1126/science.1236789},
  \href {https://ui.adsabs.harvard.edu/abs/2013Sci...341...53T} {341, 53}

\bibitem[\protect\citeauthoryear{{Timokhin} \& {Harding}}{{Timokhin} \&
  {Harding}}{2019}]{Timokhin&Harding19}
{Timokhin} A.~N.,  {Harding} A.~K.,  2019, \mn@doi [\apj]
  {10.3847/1538-4357/aaf050}, \href
  {https://ui.adsabs.harvard.edu/abs/2019ApJ...871...12T} {871, 12}

\bibitem[\protect\citeauthoryear{{Totani}}{{Totani}}{2013}]{Totani13}
{Totani} T.,  2013, \mn@doi [\pasj] {10.1093/pasj/65.5.L12}, \href
  {https://ui.adsabs.harvard.edu/abs/2013PASJ...65L..12T} {65, L12}

\bibitem[\protect\citeauthoryear{{Tsang}}{{Tsang}}{2013}]{Tsang13}
{Tsang} D.,  2013, \mn@doi [Astrophys. J.] {10.1088/0004-637X/777/2/103}, \href
  {http://adsabs.harvard.edu/abs/2013ApJ...777..103T} {777, 103}

\bibitem[\protect\citeauthoryear{{Tsang}, {Read}, {Hinderer}, {Piro}  \&
  {Bondarescu}}{{Tsang} et~al.}{2012}]{Tsang+12}
{Tsang} D.,  {Read} J.~S.,  {Hinderer} T.,  {Piro} A.~L.,   {Bondarescu} R.,
  2012, \mn@doi [\prl] {10.1103/PhysRevLett.108.011102}, \href
  {https://ui.adsabs.harvard.edu/abs/2012PhRvL.108a1102T} {108, 011102}

\bibitem[\protect\citeauthoryear{{Usov}}{{Usov}}{1992}]{Usov92}
{Usov} V.~V.,  1992, \mn@doi [\nat] {10.1038/357472a0}, \href
  {http://adsabs.harvard.edu/abs/1992Natur.357..472U} {357, 472}

\bibitem[\protect\citeauthoryear{{Usov} \& {Katz}}{{Usov} \&
  {Katz}}{2000}]{Usov&Katz00}
{Usov} V.~V.,  {Katz} J.~I.,  2000, \aap, \href
  {https://ui.adsabs.harvard.edu/abs/2000A&A...364..655U} {364, 655}

\bibitem[\protect\citeauthoryear{{Vietri}}{{Vietri}}{1996}]{Vietri96}
{Vietri} M.,  1996, \mn@doi [\apjl] {10.1086/310340}, \href
  {http://adsabs.harvard.edu/abs/1996ApJ...471L..95V} {471, L95}

\bibitem[\protect\citeauthoryear{{Wada}, {Shibata}  \& {Ioka}}{{Wada}
  et~al.}{2020}]{Wada+20}
{Wada} T.,  {Shibata} M.,   {Ioka} K.,  2020, arXiv e-prints, \href
  {https://ui.adsabs.harvard.edu/abs/2020arXiv200804661W} {p. arXiv:2008.04661}

\bibitem[\protect\citeauthoryear{{Wang}, {Dai}, {Liu}  \& {Wu}}{{Wang}
  et~al.}{2016}]{Wang+16}
{Wang} L.-J.,  {Dai} Z.-G.,  {Liu} L.-D.,   {Wu} X.-F.,  2016, \mn@doi
  [Astrophys. J.] {10.3847/0004-637X/823/1/15}, \href
  {http://adsabs.harvard.edu/abs/2016ApJ...823...15W} {823, 15}

\bibitem[\protect\citeauthoryear{{Wang}, {Peng}, {Wu}  \& {Dai}}{{Wang}
  et~al.}{2018}]{Wang+18}
{Wang} J.-S.,  {Peng} F.-K.,  {Wu} K.,   {Dai} Z.-G.,  2018, \mn@doi [\apj]
  {10.3847/1538-4357/aae531}, \href
  {https://ui.adsabs.harvard.edu/abs/2018ApJ...868...19W} {868, 19}

\bibitem[\protect\citeauthoryear{{Weibel}}{{Weibel}}{1959}]{Weibel_59}
{Weibel} E.~S.,  1959, \mn@doi [\prl] {10.1103/PhysRevLett.2.83}, \href
  {https://ui.adsabs.harvard.edu/abs/1959PhRvL...2...83W} {2, 83}

\bibitem[\protect\citeauthoryear{{Yuan}, {Beloborodov}, {Chen}  \&
  {Levin}}{{Yuan} et~al.}{2020}]{Yuan_2020}
{Yuan} Y.,  {Beloborodov} A.~M.,  {Chen} A.~Y.,   {Levin} Y.,  2020, arXiv
  e-prints, \href {https://ui.adsabs.harvard.edu/abs/2020arXiv200604649Y} {p.
  arXiv:2006.04649}

\bibitem[\protect\citeauthoryear{{Zhang}}{{Zhang}}{2013}]{Zhang13}
{Zhang} B.,  2013, \mn@doi [Astrophys. J. Lett.] {10.1088/2041-8205/763/1/L22},
  \href {http://adsabs.harvard.edu/abs/2013ApJ...763L..22Z} {763, L22}

\bibitem[\protect\citeauthoryear{{Zhang}}{{Zhang}}{2019}]{Zhang_2019}
{Zhang} B.,  2019, \mn@doi [\apjl] {10.3847/2041-8213/ab0ae8}, \href
  {https://ui.adsabs.harvard.edu/abs/2019ApJ...873L...9Z} {873, L9}

\bibitem[\protect\citeauthoryear{{Zhang}}{{Zhang}}{2020}]{Zhang20}
{Zhang} B.,  2020, \mn@doi [\apjl] {10.3847/2041-8213/ab7244}, \href
  {https://ui.adsabs.harvard.edu/abs/2020ApJ...890L..24Z} {890, L24}

\bibitem[\protect\citeauthoryear{{Zhang} \& {Kobayashi}}{{Zhang} \&
  {Kobayashi}}{2005}]{Zhang_Kobayashi_05}
{Zhang} B.,  {Kobayashi} S.,  2005, \mn@doi [\apj] {10.1086/429787}, \href
  {https://ui.adsabs.harvard.edu/abs/2005ApJ...628..315Z} {628, 315}

\bibitem[\protect\citeauthoryear{{Zrake} \& {Arons}}{{Zrake} \&
  {Arons}}{2017}]{Zrake&Arons_17}
{Zrake} J.,  {Arons} J.,  2017, \mn@doi [\apj] {10.3847/1538-4357/aa826d},
  \href {https://ui.adsabs.harvard.edu/abs/2017ApJ...847...57Z} {847, 57}

\bibitem[\protect\citeauthoryear{{Zrake} \& {MacFadyen}}{{Zrake} \&
  {MacFadyen}}{2012}]{Zrake_MacFadyen_12}
{Zrake} J.,  {MacFadyen} A.~I.,  2012, \mn@doi [\apj]
  {10.1088/0004-637X/744/1/32}, \href
  {https://ui.adsabs.harvard.edu/abs/2012ApJ...744...32Z} {744, 32}

\bibitem[\protect\citeauthoryear{{Zrake}, {Xie}  \& {MacFadyen}}{{Zrake}
  et~al.}{2018}]{Zrake+18}
{Zrake} J.,  {Xie} X.,   {MacFadyen} A.,  2018, \mn@doi [\apjl]
  {10.3847/2041-8213/aaddf8}, \href
  {https://ui.adsabs.harvard.edu/abs/2018ApJ...865L...2Z} {865, L2}

\makeatother
\end{thebibliography}

\end{document}